\documentclass[reprint,superscriptaddress,aps,pre,floatfix]{revtex4-2}
\usepackage{microtype}
\usepackage[utf8]{inputenc}
\usepackage[tbtags]{amsmath}
\usepackage{amsthm,amssymb,dsfont,mathrsfs,bbm,xfrac,tabularx,booktabs}
\usepackage[linesnumbered,ruled,vlined]{algorithm2e}
\usepackage{upgreek}
\usepackage{mathtools}
\newtheorem{theorem}{Theorem}
\newtheorem{prop}[theorem]{Proposition}
\newtheorem*{prop*}{Proposition}
\newcommand{\dd}[1]{\mathrm{d}#1}
\usepackage{comment}
\usepackage{tikz}
\usetikzlibrary{patterns,decorations,arrows,shapes.geometric,arrows,mindmap,backgrounds,positioning,positioning,fit,backgrounds,positioning,arrows,matrix,calc}
\usetikzlibrary{shapes,backgrounds,mindmap,shadows,shapes.geometric,topaths,arrows,pgfplots.groupplots,fadings,calc,decorations.markings,positioning,chains,fit}
\tikzfading[name=fade l,left color=transparent!100,right color=transparent!0]
\tikzfading[name=fade r,right color=transparent!100,left color=transparent!0]
\tikzfading[name=fade d,bottom color=transparent!100,top color=transparent!0]
\tikzfading[name=fade u,top color=transparent!100,bottom color=transparent!0]
\usepackage{braket}
\usepackage{pgfplotstable}

\usepackage{ifthen}

\pgfplotsset{compat=1.16}

\newdimen\nodeSize
\newdimen\nodeDist
\tikzset{position/.style args={#1:#2 from #3}{at=(#3.#1), anchor=#1+180, shift=(#1:#2)}}

\newcommand{\arrowIn}{
\tikz \draw[-stealth] (-1pt,0) -- (1pt,0);
}

\begin{document}
\title{Thermodynamics of the Fredrickson-Andersen Model on the Bethe Lattice}
\author{Gianmarco Perrupato}
\affiliation{Department of Computing Sciences, Bocconi University, Via Roentgen 1, 20136 Milan, Italy}
\author{Tommaso Rizzo}
\affiliation{Dipartimento di Fisica, Sapienza Universit\`a di Roma, P.le A. Moro 5, 00185 Rome, Italy}
\affiliation{Institute of Complex Systems (ISC) - CNR, Rome unit, P.le A. Moro 5, 00185 Rome, Italy}
\begin{abstract}
The statics of the Fredrickson-Andersen model (FAM) of the liquid-glass transition is solved on the Bethe lattice (BL). The kinetic constraints of the FAM imply on the BL an ergodicity-breaking transition to a (glassy) phase where a fraction of spins of the system is permanently blocked, and the remaining ``free'' spins become non-trivially correlated. We compute several observables of the ergodicity-broken phase, such as the self-overlap, the configurational entropy and the spin-glass susceptibility, and we compare the analytical predictions with numerical experiments. The cavity equations that we obtain allow to define algorithms for fast equilibration and accelerated dynamics. We find that at variance with spin-glass models, the correlations inside a state do not exhibit a critical behavior. 
\end{abstract}
\maketitle

\section{Introduction}

Kinetically constrained models (KCM)s are lattice spin models of the liquid-glass transition \cite{ritort2003glassy,garrahan2011kinetically}. The main property of KCMs is their capability of reproducing some features of glasses, notably two-step relaxation and stretched exponential decay, in the absence of thermodynamic singularities. A prototypical example is the Fredrickson-Andersen model (FAM) \cite{fredrickson1984kinetic}, defined as follows. An initial configuration of the system is generated from a factorized Boltzmann-Gibbs distribution: with each site $i$ of a given lattice is associated a spin down ($s_i=-1$) with probability $p$, or up ($s_i=1$) with probability $1-p$. Afterwards the systems is evolved with a single-spin dynamics satisfying detailed balance w.r.t.\ the factorized distribution of the initial configuration. The key ingredient is that the dynamics is kinetically constrained, namely in order for a spin $s_i$ to flip, at least $f$ of its neighbors, with $f$ a site-independent external parameter, should be in the up state. The parameter $f$ is called \emph{facilitation}, and one usually refers to spins up as \emph{facilitating} spins. These constraints try to mimic the cooperative nature of the dynamics in supercooled liquids, where the ability to move of a particle can be dramatically restricted by its neighborhood \cite{ritort2003glassy}. 

The FAM has been studied extensively on the Bethe lattice (BL), \emph{i.e.}\ on finite-connectivity random graphs enjoying the so-called locally-tree-like property, meaning that the neighborhood of a site taken at random is typically a tree up to a distance that is diverging in the thermodynamic limit. On \emph{regular} BLs, in which each node has the same number $z$ of neighbors, the kinetic constraint implies for $1<f<z$ an ergodicity-breaking transition at a certain critical occupation probability $p_c$. Ergodicity breaking occurs at $p_c$ with the formation of a cluster of ``blocked'' spins that cannot be flipped by the dynamics, causing them to retain the same state as in the initial condition at all times. The appearance of this blocked cluster is related with the bootstrap percolation (BP) problem on graphs, also called $k$-core. A $k$-core is a subset of the nodes of a graph such that each node of the $k$-core: i) should satisfy a given property $\mathcal{A}$, ii) should have at least $k$ neighbors satisfying $\mathcal{A}$. If $\mathcal{A}$ is the property of a spin of being down-oriented, the set of blocked-down spins forms a $k$-core with $k=z-f+1$. 

A natural order parameter for the problem is defined in terms of the so-called local \emph{persistence} $\phi_i(t)$ as a function of time $t$. The persistence may have slightly different definitions in the literature. A possibility is to consider the persistence of the down spins. More precisely $\phi_i(t)$ at site $i$ is equal to one if $s_i(t')=-1$ for all $0 \leq t' \leq t$, and zero otherwise. Therefore the averaged persistence $\phi(t)=1/N\sum_i\phi_i(t)$ counts the number of negative sites that have never flipped up to time $t$, divided by the total number of spins, and it represents an order parameter for the theory. Indeed for $p<p_c$, $\phi(t)$ goes to zero in the large time limit, while at $p_c$, for $1<f<z-1$, it reaches a non-zero plateau value $\phi_{plat}$, signaling the presence of the cluster of blocked spins. As discussed in section \ref{sec:statSol} both $p_c$ and $\phi_{plat}$ can be exactly computed on the BL by means of the connection with BP, and the locally-tree-likeness of the BL. 


The critical dynamics of the FAM on the BL displays peculiar features relating this model with mean-field spin glasses and supercooled liquids in infinite dimensions. Indeed it has been shown analytically \cite{perrupato2022exact} that it exhibits a Mode-Coupling-Theory (MCT) nature, consistent with findings from earlier numerical simulations \cite{sellitto2005facilitated,sellitto2015crossover,ikeda2017fredrickson,rizzo2020solvable}. This means that for $p$ close to $p_c$, there is a timescale $\tau_\beta$, defining the so-called $\beta$-regime, where the deviation $g(t)=\phi(t)-\phi_{plat}$ of the persistence function $\phi(t)$ from its plateau value $\phi_{plat}$ at the critical point is small, and satisfies the following MCT equation \cite{gotze2008complex}: 
\begin{equation}
\sigma = - \lambda \, g^2(t) + \frac{d}{dt} \, \int_0^t  \, g(t')\, g(t-t')\, dt' \ ,
\label{MCTcrit}
\end{equation}
where $\sigma$ is a linear function of $p-p_c$, and $\lambda$ is the so-called parameter exponent. Eq.~\eqref{MCTcrit} implies that $g(t)$ is $O(\sqrt{|\sigma}|)$ in the $\beta$-regime, and obeys the following scaling laws:
\begin{equation}
\label{eq:scalinglaw}
g(t) = \sqrt{|\sigma|} \, f_{\pm}(t/\tau_\beta),
\end{equation}
where the master functions $f_{\pm}$ are the solutions of Eq.~\eqref{MCTcrit} with $\sigma=\pm 1$. In the liquid phase, the $\beta$-regime is followed by a so-called $\alpha$-regime, during which the persistence decays from $\phi_{plat}$ to zero. Eq.~\eqref{eq:scalinglaw} implies that in the liquid phase $g(t)$ approaches zero with a $t^{-a}$ law, and then it becomes negative (i.e. $\phi(t)$ becomes smaller than its plateau value at the critical point) with a $-t^b$ law. The model-dependent exponents $a$ and $b$ are determined by $\lambda$ through 
\begin{equation}
\label{lambdaMCT}
\lambda=\frac{\Gamma^2(1-a)}{\Gamma(1-2a)}=\frac{\Gamma^2(1+b)}{\Gamma(1+2b)}.
\end{equation}
Interestingly one finds from Eq.~\eqref{MCTcrit} that the $\beta$ time-scale $\tau_\beta$ diverges with $\sigma$ as $\tau_\beta \propto |\sigma|^{-1/(2\,a)}$ both in the glassy ($\sigma\geq 0$) and in the liquid phase ($\sigma< 0$). From Eq.~\eqref{MCTcrit} it can be deduced also the time-scale of the $\alpha$-regime, that diverges like $\tau_\alpha \propto |\sigma|^{-\gamma}$ with $\gamma=1/(2 a)+1/(2 b)$. At the critical point ($\sigma=0$), $g(t)$ has a power-law behaviour: 
\begin{equation}
\label{eq:scalinglaw-1}
g(t)=\phi(t)-\phi_{plat} \propto 1/t^a.
\end{equation}
For $z=4$ and $f=2$ one has in particular $\phi_{plat}=21/32$, and $a\approx 0.340356$ \cite{perrupato2022exact}. In the glassy phase the master function displays arrest for times larger than $\tau_{\beta}$. In \cite{perrupato2022exact} the parameter exponent $\lambda$, as well as the gauge coefficient relating $\sigma$ with $p-p_c$ are computed as function of $z$ and $f$ for different KCMs, including the FAM. 

Remarkably some of the results obtained on the BL allow to study the behaviour of the FAM in finite dimension through a renormalization group approach, the so-called $M$-layer expansion \cite{altieri2017loop,angelini2022unexpected}. In this framework it has been shown that the BP transition disappears in finite dimension \cite{rizzo2019fate}. This implies that in finite $d$ ergodicity is restored. Interestingly it turns out that the transition observed on the BL becomes a crossover from power-law to exponential increase of the relaxation time, as widely observed in actual supercooled liquids and spin-glass (SG) models. Indeed in supercooled liquids in $d\rightarrow\infty$, and mean-field (MF) SGs with one step of Replica-Symmetry-Breaking, one finds a dynamical transition that turns out to be an artifact of the MF nature of the models, and becomes a crossover in finite $d$ \cite{parisi2020theory}.  The disappearance of the sharp transition is a consequence of the fluctuations that are not taken into account at the MF level. In SGs such fluctuations are proved to be described in the $\beta$-regime by a modification of the MCT equations called Stochastic-Beta-Relaxation (SBR) \cite{rizzo2014long,rizzo2016dynamical}. The parameters defining the SBR equations were computed for the FAM by leveraging its solution on the BL. This allowed to demonstrate that the SBR predictions are consistently confirmed also in the FAM \cite{rizzo2020solvable}. 

Despite all these results, the theory on the BL is still not completely solved. A piece that up to now was still missing is the analytical computation of the plateau value of several observables characterizing the ergodicity-broken (glassy) phase, e.g.\ the self-overlap, the configurational entropy and the spin-glass susceptibility. In the glassy phase the phase space of the problem divides into an exponential number of disconnected components (states). Ergodicity breaking implies the onset of non-trivial correlations between the spins, that are absent in the total Boltzmann-Gibbs measure. In other words, while the total Boltzmann-Gibbs distribution is completely factorized, the distribution of the microstates of the system conditioned to a state is highly non-trivial. Consider for example a system of two independent spins that in order to move need a facilitation greater than zero, \emph{i.e.}\ a spin in order to flip needs the other to be in the up state. In this case the $(-1,-1)$ configuration is a separate blocked component, while the remaining configurations form another component over which the magnetizations are different from the typical ones, implying a correlation between the two spins. In the context of disordered systems, the computation of observables of models defined on the BL are usually addressed  through the cavity method. The starting point of the cavity method is the Boltzmann-Gibbs distribution of the problem from which, under certain assumptions, a set of closed (cavity) equations for the local marginal probabilities (or their probability distribution in case of replica symmetry breaking) is written down \cite{Mezard2001,Mezard2003,parisi2020random,perrupato2022ising}. However this procedure is not feasible in KCMs since, as already said, the Boltzmann-Gibbs distribution in this case is trivial, and one has to find a way to implement the kinetic constraint in the cavity description of the statics. In this work we provide a solution to this problem, and we compute several quantities that are compared with numerical simulations. As a result of the analysis we found that correlations inside a state are not diverging at the critical point in the FAM, at variance with mean-field SGs, in which the susceptibility inside a state is critical. This difference is somehow unexpected, given the similarities between SGs and KCMs. Indeed in both problems: i) the state-to-state fluctuations are critical, and they are both described by a double-pole singularity in momentum space \cite{Franz2011,rizzo2019fate}; ii) close to the critical point the order parameter is proved to be described by the same MCT equation of motion \cite{perrupato2022exact}. The finiteness of the SG susceptibility at the critical point has deep physical implications. Indeed it is possible to make an argument \cite{unpublished} showing that this property implies that the critical behavior of the overlap, $q(t)=1/N\sum_is_i(0)s_i(t)$, that is another order parameter for the problem often studied in the literature, is fully controlled by that of the persistence $\phi(t)$. This means that they both have the same critical behaviour (see Fig.~\ref{shifted_overlap}), satisfying the same MCT equation \eqref{MCTcrit} with the same value of $\lambda$. 

Another physical quantity we are able to access with the formalism presented here is related with the off-equilibrium dynamics of the FA model. As shown by Sellitto et al. \cite{sellitto2005facilitated}, FA exhibits aging behavior similar to that observed in mean-field spin glasses with one-step Replica Symmetry Breaking (1RSB).  More precisely initializing an instance of FA at $p_c^+$, and performing a quench above $p_c$, the energy $E(t)$ approaches at large times an asymptotic value $E(\infty)$ larger than its equilibrium value. In Sec.~\ref{sec:thresholdE} we give an analytical prediction of $E(\infty)$, in agreement with earlier numerical simulations \cite{sellitto2005facilitated}.

Interestingly, the cavity solution of the FAM suggests a message-passing algorithm for the simulation of a physical dynamics with accelerated thermalization, that could aid in examining the critical dynamics in KCMs. We discuss this idea in the following, leaving the implementation to a future work.
 
The manuscript is organised as follows. In section \ref{sec:statSol} we present the argument for the computation of the statics of the problem. The final result is conceptually simple: we end up with a set of equations that can be studied numerically. We warn the reader that the derivation in section \ref{sec:statSol} is somewhat technical, however to grasp the essence of the work it is not necessary to follow all the intermediate steps leading to the cavity equations.
The subsequent sections are devoted to the computation of different observables. For each observable we compare the analytical cavity prediction with numerical simulations. We compute the self-overlap (Sec.~\ref{sec:Overlap}), the marginal distribution of two neighboring spins (Sec.~\ref{sec:EdgeMarg}), the configurational entropy (Sec.~\ref{sec:ConfEntro}), the threshold energy (Sec.~\ref{sec:thresholdE}), the SG susceptibility (Sec.~\ref{sec:IntroSGSusc}) and the specific heat (Sec.~\ref{sec:SpecHeat}). In section \ref{sec:Conclusions} we present the conclusions. In App.~\ref{app:ITfunctions} we write the definition of the iterative functions appearing in the cavity equations. In App.~\ref{sec:appLowT} we compute the low temperature expansion of the overlap and the entropy, and in Appendices \ref{sec:AnalyticalCompSGSusc} and \ref{sec:NumericalCompSGSusc} we discuss the details about the derivation and the numerical measure of the SG susceptibility. 

\section{The Static Solution}
\label{sec:statSol}
As discussed in the introduction, in the FAM on the BL there is an ergodicity breaking transition that is signaled by the persistence function $\phi(t)$. In the ergodicity-broken (glassy) phase ($p\geq p_c$) the configuration space divides into an exponential number of states, that are characterized by non trivial (non-factorized) distributions. As we will see this implies that the complete order parameter of the theory becomes a probability distribution. 

Let us state more precisely the definition of the problem. An instance of the FAM is defined by a graph and an initial condition of the spins. As already said, in our case the graph is a BL, defined as a random graph drawn from the uniform measure on the set of all graphs with a certain number $N$ of nodes, such that each node has the same number $z$ of neighbors (random regular graphs) \footnote{The procedure used in the present work to generate BLs with $z=4$ is the following. The first step is to construct $M$ identical replicas of an elementary square cell $\mathcal{C}$ with $L^2$ sites and periodic boundary conditions. In this way each site $i$ of $\mathcal{C}$ has $M$ replicas, that we denote by $s_i^a$, with $a=1,\dots,M$. We call $N=L^2M$ the total number of sites. At this point we apply the following ``rewiring'' procedure. For each edge $(i,j)$ of $\mathcal{C}$, we generate a random permutation $\mathcal{P}$ of $(1,2,\dots,M)$, and, for each $a$, we replace the edge connecting $s_i^{a}$ to $s_j^{a}$ with an edge connecting $s_i^{a}$ to $s_j^{\mathcal{P}(a)}$. Note that this procedure, the so-called $M$-layer construction \cite{altieri2017loop}, does not change the coordination of the nodes and, as shown in \cite{altieri2017loop}, for large $M$ it produces an instance of Bethe lattice (the density of cycles of fixed length is vanishing for $M\rightarrow\infty$).}. We also define the cavity coordination $c=z-1$. Such ensemble of graphs enjoys the so-called \emph{locally-tree-like} property, namely a finite-size neighborhood of a node drawn at random on a BL is typically a tree in the limit $N\rightarrow\infty$. Once the graph is generated  the initial condition is drawn by assigning \emph{independently} to each site a spin up with probability $1-p$, or down with probability $p$. It is useful to define a temperature $T=1/\beta$ through $p=e^{\beta}/(e^{\beta}+e^{-\beta})$. In this way the energy function associated with the total Boltzmann-Gibbs distribution of the problem takes the simple form $H=\sum_is_i$. At this point the system is evolved according to a single-spin dynamics that satisfies detailed balance w.r.t.\ the factorized distribution. At each step a site $i$ is selected, and the clock is updated by one. If $s_i$ is facilitated, \emph{i.e.}\ if 
\begin{equation}
\label{eq:faccc}
\frac{1}{2}\sum_{j\in\partial i}(s_j+1)\geq f,
\end{equation}
then spin $s_i$ is flipped with a probability $w(s_i\rightarrow-s_i)$ that is given by the Metropolis rule:
\begin{equation}
\label{eq:MetropRule}
w(s_i\rightarrow-s_i)=\min\{e^{s_i\beta},1\}.
\end{equation}
If condition \eqref{eq:faccc} is not fulfilled, \emph{i.e.}\ if $s_i$ is not facilitated, $s_i$ is automatically left unchanged. In the following we study the FAM in the case of discontinuous transition ($1<f<z-1$). We will explicitly refer to the case $z=4$ and $f=2$, that we denote by ``$(4,2)$'', but our discussion can be easily extended to generic discontinuous models $(z,f)$. 

Let us consider a site $s_0$ on the Bethe lattice, that we call root. We have three possibilities: i) the root is blocked to minus one, ii) it is blocked to one, iii) it is free to move. The first two cases correspond, respectively, to magnetizations minus one and one. Let us start from the case in which $s_0$ is not blocked, and therefore its magnetization $m_0$ is such that $|m_0|<1$. Let us firstly introduce some definitions. Given two generic configurations $C$ and $C'$, we say that $C'$ is visitable from $C$ if there exists an allowed trajectory connecting them. An allowed trajectory $\tau$ is a collection of configurations $\tau=\{C^{1},C^{2},\dots,C^{d}\}$, with $C^{1}=C$ and $C^{d}=C'$, such that $C^{i+1}$ differs from $C^{i}$ only for the orientation of a single spin $s_k$, and $s_k$ is facilitated in $C^{i}$, namely there are at least $f$ neighbors of $s_k$ that are in the up state. In the following we denote by $\tau^{-1}=\{C^{d},\dots,C^{2},C^{1}\}$ the trajectory in which the moves are inverted w.r.t.\ $\tau$. Clearly if $C'$ is visitable from $C$, then $C$ is visitable from $C'$ because the dynamics is \emph{reversible}, \emph{i.e.}\ $\tau$ is allowed if and only if $\tau^{-1}$ is allowed. In the following when we say that a configuration is visitable without specifying, we mean that it is visitable from the initial condition. Note that due to the facilitation constraint, the minimum number of changes allowing to go from a visitable configuration to another is at least equal to the Hamming distance between them, and in general not all configurations are visitable. The fundamental observations at the basis of our analysis are enclosed in the following proposition:
\begin{prop}
\label{prop:prop}
Consider a non-blocked spin $s_0$ (the root). Independently of the value of the root on the initial condition, the following properties hold.
\begin{enumerate}
    \item Consider a visitable configuration $C_+$ in which $s_0=1$, and an allowed trajectory $\tau$ connecting the initial condition with $C_+$. The trajectory $\tau_+$, equal to $\tau$ except at most for $s_0$, that is conditioned to stay up ($s_0=1$) in all the moves, is an allowed trajectory; 
    \item for each visitable configuration $C$, the configuration $C_+$ in which $s_0=1$, and the other spins assume the same value as in $C$ is visitable;
    \item given a visitable configuration $C_{+}$ in which $s_0=1$, the configuration $C_{-}$ obtained from $C_{+}$ by reversing $s_0$ is visitable if and only if $s_0$ is not blocked in $C_{-}$. 
\end{enumerate}
\end{prop}
\textit{Proof}:
\begin{enumerate}
\item If a neighbor of the root can flip (is facilitated) if the root is down, it can flip also if the root is up. This proves that $\tau_+$ is allowed;
\item If $s_0=1$ in $C$ then $C_+=C$, that is visitable by hypothesis. Let us suppose that $s_0=-1$ in $C$. Since $s_0$ is a free spin, there exists a visitable configuration $C^{\star}$ in which $s_0=1$. Now consider an allowed trajectory $\tau$ connecting $C$ to $C^{\star}$. At the last point we proved that $\tau_+$ is allowed. This implies that $C_+$ is visitable, because it can be reached from a visitable configuration $C^{\star}$ through the allowed trajectory $\tau^{-1}_+$.
\item Suppose to start the dynamics from $C_{-}$. A visitable configuration $C'_{+}$ with $s_0=1$ exists if and only if $s_0$ is not blocked in $C_{-}$. In the case in which $s_0$ is not blocked, let us call $\tau$ an allowed trajectory connecting $C_{-}$ to $C'_{+}$. Since $\tau_{+}$ is an allowed trajectory, we have that $C_{-}$ is visitable from $C_{+}$ according to the following path:
\begin{equation}
C_{+}\xrightarrow[]{\tau_{+}}C_{+}'\xrightarrow[]{\tau^{-1}}C_{-}.
\end{equation}
\end{enumerate}
As we are going to show, Prop.~\ref{prop:prop} together with the local properties of the BL allows to compute self-consistently the statistics of the free spins. Suppose to start from an initial condition $C$ in which the root is free. Let us call $\mathcal{C}_+(C)$ ($\mathcal{C}_-$(C)) the set of all visitable configurations from $C$ in which the root is up (down). In the following we do not explicit the dependence of $\mathcal{C}_+,\mathcal{C}_-$ on the initial condition. It is useful to introduce the conditioned partition functions:  
\begin{equation}
\label{eq:condPartFunct}
Z(s_0=1)=\sum_{\underline{s}\in\mathcal{C}_+}e^{-\beta\sum_is_i},\,  Z(s_0=-1)=\sum_{\underline{s}\in\mathcal{C}_-}e^{-\beta\sum_is_i},
\end{equation}
The magnetization of the root can be expressed in terms of \eqref{eq:condPartFunct}:
\begin{equation}
m_0=\frac{1-Z(-1)/Z(1)}{1+Z(-1)/Z(1)},
\end{equation}
and thus we see that naturally we need only to characterize the ratio of partition functions. Exploiting Prop.~\ref{prop:prop}, these ratios can be computed self-consistently on the tree. Let us define $\tilde{Z}$ as the partition function of the visitable configurations in which the root is up with no energy term on the root:
\begin{equation}
\label{eq:partfunctConstr1}
Z(1)=e ^{-\beta} \tilde{Z}.    
\end{equation}
From Prop.~\ref{prop:prop} it follows that  
\begin{equation}
\label{eq:partfunctConstr2}
Z(-1)=e^{\beta}(\tilde{Z}-\hat{Z}),    
\end{equation}
where $\hat{Z}$ is the partition function (again with no energy term associated with the root) of the visitable up configurations such that the root is blocked if it is reversed. Using Eqs.~\eqref{eq:partfunctConstr1} and \eqref{eq:partfunctConstr2} we can write:
\begin{equation}
\label{eq:magnFreeSpins}
m_0=\frac{e^{-2\beta}-1+R}{e^{-2\beta}+1-R},\quad R \equiv \frac{\hat{Z}}{\tilde{Z}}.
\end{equation}
Let us define the $i$-th branch entering the root, $T_{i\rightarrow 0}$, as the sub-tree composed by all the nodes that can be reached from $s_0$ without visiting the sites in $\partial 0\setminus i$. Due to the tree structure of the graph, we have
\begin{equation}
\tilde{Z}=\prod_{i=1}^{z}\tilde{Z}_i,    
\end{equation}
where $\tilde{Z}_i$ is the partition function of the sub-system on branch $i$ computed on the up configurations $\mathcal{C}_{+}^{i\rightarrow 0}$, namely all the configuration that can be visited by the spins in $T_{i\rightarrow 0}$ when the root $s_0$ is conditioned in the up state. In order to evaluate $\hat{Z}$ in \eqref{eq:magnFreeSpins} we have to determine the configurations where the root is blocked once reversed. The latter can happen i.f.f.\ in more than $z-f$ branches the neighbor is blocked down once the root is reversed. Let us call $\hat{Z}_i$ the partition function (without energy term associated with the root) of the sub-system on branch $i$, computed on the subset of configurations belonging to $\mathcal{C}_{+}^{i\rightarrow 0}$ such that $s_i$ is blocked down if the root is reversed.

In the case $z=4$ and $f=2$ we need at least three such branches in order for the root to be blocked if reversed. It follows that the ratio $R$ is given by:
\begin{widetext}
\begin{equation}
\label{eq:Rsito}
\begin{split}
R=R_{site}(R_1,R_2,R_3,R_4)\equiv R_1  R_2 R_3+R_1  R_2 R_4  +R_1  R_3 R_4+R_2  R_3 R_4- 3 \, R_1  R_2 R_3 R_4,\quad R_i \equiv  \frac{\hat{Z}_i}{\tilde{Z}_i}.
\end{split}
\end{equation}
\end{widetext}
In this way we expressed $R$ in terms of quantities on the branches that, as we will see, can be computed iteratively. At this point let us define two variables  $\eta_i$ and $\mu_i$ associated with branch $i$:
\begin{equation}
\eta_i=\left\{
\begin{alignedat}{2}
&1\,\,\, \text{if $s_i$ is blocked down if $s_0$ is initialized down} \\
& 0\,\,\, \text{otherwise},
\end{alignedat}
\right.
\end{equation}
\begin{equation}
\mu_i=\left\{
\begin{alignedat}{2}
&1\,\,\, \text{if $s_i$ is blocked down if $s_0$ is initialized up} \\
& 0\,\,\, \text{otherwise}.
\end{alignedat}
\right.
\end{equation}
Note that all the quantities on the branches are defined on directed edges, e.g.\ $\eta_i$ should be written as $\eta_{i\rightarrow 0}$. In the following we are going to use both notations depending on the context. With these definitions, if the root is initialized down, it is blocked i.f.f. 
\begin{equation}
\sum_{i\in\partial 0} \eta_i > z-f,    
\end{equation}
while if the root is initialized up, it is blocked i.f.f. 
\begin{equation}
\sum_{i\in\partial 0} \mu_i > z-f.    
\end{equation}
The knowledge of the $z$ triplets $(\eta_{i\rightarrow 0},\mu_{i\rightarrow 0},R_{i\rightarrow 0})$, $i\in\partial 0$, is all we need in order to compute $m_0$. Indeed $(\eta_{i\rightarrow 0},\mu_{i\rightarrow 0})$ determine whether $s_0$ is blocked up, down or free. In case $s_0$ is blocked up or down its magnetization is, respectively, plus one or minus one. If it is free the $R_{i\rightarrow 0}$'s allow to compute $R$ (Eq.~\eqref{eq:Rsito}), that determines $m_0$ through Eq.~\eqref{eq:magnFreeSpins}. For this reason $P(\eta,\mu,R)$, that we compute in the next section, represents the order parameter of the theory. 
To conclude the section let us discuss how $P(\eta,\mu,R)$ is connected to the plateau value $\phi_{plat}=\lim_{t\rightarrow\infty}\phi(t)$ of the persistence function. In the case $(4,2)$ the probability that the root is blocked down can be written as:
\begin{equation}
\label{eq:PsiteFrozendown}
P_{site}^-=  4 p P^3-3 p P^4,\quad P \equiv \sum_{\mu}\int dR P(1,\mu,R),
\end{equation}
where $P$ is the probability that the neighbor is blocked down if the root is constrained down, and $P_{site}^-=\phi_{plat}$. The probability that the root is blocked up can be written as:
\begin{equation}
\label{eq:PsiteFrozenup}
P_{site}^+= 4 (1-p)\, D^3 -3 (1-p)\, D^4,  
\end{equation}
where $D$,
\begin{equation}
D \equiv \sum_{\eta}\int dR P(\eta,1,R),
\end{equation}
is the probability that the neighbor is blocked down if the root is constrained up. 
\subsection{The Iterative Equation}
\begin{figure}
\centering
\scalebox{0.95}{
\begin{tikzpicture}
\node[label=above:{},shape=circle, scale=1] (0) at (0,0) {};
\foreach \i in {0}
{
\node[label=above:{},shape=circle, semithick,position=\i:{2\nodeDist} from 0, scale=0.7, fill=black!28,draw] (A\i) {};
\foreach \j in {\i-55,\i,\i+55}
{
\node[label=above:{},shape=circle, semithick,position=\j:{\nodeDist} from A\i] (B\j) {};
\pgfmathparse{int(\j)};
\ifthenelse{\pgfmathresult<90 \AND \pgfmathresult>-90 \OR \pgfmathresult>270)}{\draw [semithick,path fading=east] (A\i) -- (B\j);}{\draw [semithick,path fading=west] (A\i) -- (B\j);}
}
}
\node[label=above:{},shape=circle, semithick,position=180:{2\nodeDist} from 0, scale=0.7,thick,densely dashed,draw] (A180) {};
\draw [semithick,gray] (A0) -- (A180);
\node[label=above:{},shape=circle,semithick,position=-100:{-0.2} from A0, scale=1.5] (0spin) {$s_n$};
\node[label=above:{},shape=circle,semithick,position=-80:{-0.2} from A180, scale=1.5] (180spin) {$s_0$};
\node[label=below:{},shape=circle,position=0:{0.3\nodeDist} from A180] (Arrr) {};
\node[label=below:{},shape=circle,position=0:{3\nodeDist} from Arrr] (Arrr1) {};
\draw[] (Arrr1) -- (Arrr) node[label={[label distance=-1.5mm]-175:{$(\eta,\mu,R)_{n\rightarrow 0}$}},sloped,pos=1,scale=2,allow upside down]{\arrowIn};
\end{tikzpicture}
}
\caption{Representation of a cavity graph. The cavity spin (root) $s_0$ is dashed in order to highlight that it is conditioned up or down. For the triplet of cavity fields we used the notation $(\eta,\mu,R)_{n\rightarrow 0}\equiv(\eta_{n\rightarrow 0},\mu_{n\rightarrow 0},R_{n\rightarrow 0})$. \label{fig:cavGraph1}}
\end{figure}
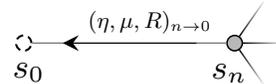
At this point we can discuss how to write down a closed equation for the distribution $P(\eta,\mu,R)$. Consider a cavity graph, or cavity branch, in which each spin has coordination $z$ except for $s_0$ (the root), that has just one neighbor $s_n$ (see Fig.~\ref{fig:cavGraph1}). If the root is conditioned down, $s_n$ is blocked down i.f.f.\ it is initialized down, and
\begin{equation}
\sum_{i=1}^{c} \eta_i > z-f-1,    
\end{equation}
while if the root is conditioned up, $s_n$ is blocked down i.f.f.\ it is initialized down, and 
\begin{equation}
\sum_{i=1}^{c} \eta_i > z-f.    
\end{equation}
Therefore in the case $(4,2)$ we find:
\begin{equation}
\label{eq:BPeqZ4m2}
P= 3 p P^2 -2 p P^3,\quad D=p\,P^3,  
\end{equation}
that are closed equations for $P$ and $D$ (see  Eqs.~\eqref{eq:PsiteFrozendown} and \eqref{eq:PsiteFrozenup}). The self-consistent equation for $P$ (and analogously that for $D$) can be visualised as follows: 
\begin{equation}
P=
\begin{gathered}
\scalebox{0.85}{
\begin{tikzpicture}
\node[label=above:{},shape=circle, scale=1] (0) at (0,0) {};
\foreach \i in {0}
{
\node[label=above:{},shape=circle, semithick,position=\i:{0.6\nodeDist} from 0, scale=0.7, fill=black!28,draw] (A\i) {};
\foreach \j in {\i-55,\i,\i+55}
{
\node[label=above:{},shape=circle, semithick,position=\j:{1.5\nodeDist} from A\i] (B\j) {};
\pgfmathparse{int(\j)};
\ifthenelse{\pgfmathresult<90 \AND \pgfmathresult>-90 \OR \pgfmathresult>270)}{\draw [semithick,path fading=east] (A\i) -- (B\j);}{\draw [semithick,path fading=west] (A\i) -- (B\j);}
}
}
\node[label=above:{},shape=circle, semithick,position=180:{0.6\nodeDist} from 0, scale=0.7,thick,densely dashed,draw] (A180) {};
\draw [semithick,gray] (A0) -- (A180);
\node[label=above:{},shape=circle,semithick,position=130:{-0.1} from A0, scale=1] (0spin) {$p$};
\node[label=above:{},shape=circle,semithick,position=55:{0.6} from A0, scale=1] (0spin) {$P$};
\node[label=above:{},shape=circle,semithick,position=6:{0.7} from A0, scale=1] (0spin) {$P$};
\node[label=above:{},shape=circle,semithick,position=-50:{0.4} from A0, scale=1] (0spin) {$1-P$};
\node[label=above:{},shape=circle,semithick,position=-80:{-0.2} from A180, scale=1.5] (180spin) {};
\node[label=below:{},shape=circle,position=0:{0.3\nodeDist} from A180] (Arrr) {};
\node[label=below:{},shape=circle,position=0:{3\nodeDist} from Arrr] (Arrr1) {};
\end{tikzpicture}
}
\end{gathered}
\,\,+\,\,
\begin{gathered}
\scalebox{0.85}{
\begin{tikzpicture}
\node[label=above:{},shape=circle, scale=1] (0) at (0,0) {};
\foreach \i in {0}
{
\node[label=above:{},shape=circle, semithick,position=\i:{0.6\nodeDist} from 0, scale=0.7, fill=black!28,draw] (A\i) {};
\foreach \j in {\i-55,\i,\i+55}
{
\node[label=above:{},shape=circle, semithick,position=\j:{1.5\nodeDist} from A\i] (B\j) {};
\pgfmathparse{int(\j)};
\ifthenelse{\pgfmathresult<90 \AND \pgfmathresult>-90 \OR \pgfmathresult>270)}{\draw [semithick,path fading=east] (A\i) -- (B\j);}{\draw [semithick,path fading=west] (A\i) -- (B\j);}
}
}
\node[label=above:{},shape=circle, semithick,position=180:{0.6\nodeDist} from 0, scale=0.7,thick,densely dashed,draw] (A180) {};
\draw [semithick,gray] (A0) -- (A180);
\node[label=above:{},shape=circle,semithick,position=130:{-0.1} from A0, scale=1] (0spin) {$p$};
\node[label=above:{},shape=circle,semithick,position=55:{0.6} from A0, scale=1] (0spin) {$P$};
\node[label=above:{},shape=circle,semithick,position=6:{0.7} from A0, scale=1] (0spin) {$P$};
\node[label=above:{},shape=circle,semithick,position=-90:{1.2} from A0, scale=1] (0spin) {};
\node[label=above:{},shape=circle,semithick,position=-45:{0.5} from A0, scale=1] (0spin) {$P$};
\node[label=above:{},shape=circle,semithick,position=-80:{-0.2} from A180, scale=1.5] (180spin) {};
\node[label=below:{},shape=circle,position=0:{0.3\nodeDist} from A180] (Arrr) {};
\node[label=below:{},shape=circle,position=0:{0.8\nodeDist} from Arrr] (Arrr1) {};
\end{tikzpicture}
}
\end{gathered} 
,
\end{equation}
where the cavity spin $s_0$ is dashed. For a low value $p$ of the density of up spins $P=D=0$. A non-zero solution $P_c=3/4$ appears discontinuously at the critical probability $p_c=8/9$. As anticipated in the introduction, the cluster of spins that are blocked down forms a $k$-core. The transition from zero to a finite volume of the $k$-core has a mixed-nature first order-second order. Indeed there is a jump in the order parameter $P$ as in a first order phase transition, however the difference between $P$ and its critical value $P_c$ is singular as a function of $p-p_c$, $P-P_c\approx \sqrt{p-p_c}$, as usually found in second order phase transitions. 

In order to write down a self-consistent equation for $P(\eta,\mu,R)$ let us condition the root $s_0=1$. If $s_n$ is blocked down when the root is up, all three $\eta_{j\rightarrow n}$'s entering $s_n$, $j\in\partial n\setminus 0$, are equal to one. In this case if we reverse the root, $s_n$ obviously remains blocked down, implying that $R=1$. Therefore we can write: 
\begin{eqnarray}
P(1,1,R)= D \, \delta (R-1).
\end{eqnarray}
If the neighbor is blocked up when the root is up, that's because all three $\mu_{j\rightarrow n}$'s, with $j\in\partial n\setminus 0$, are equal to one, and then the ratio $\hat{Z}/\tilde{Z}$ is zero: $R=0$. In all the other cases $s_n$ is free to move, and then the cavity partition function $Z_{n\rightarrow 0}(s_0=1,s_n=1)$ of the configurations with the root up and the neighbor up (without energy term associated with the root) is given by the product of partition functions of the up configurations on the branches entering $s_n$:
\begin{eqnarray}
\label{eq:CavPartPos}
Z_{n\rightarrow 0}(s_0=1,s_n=1)=e^{-\beta} \prod_{i\in\partial n\setminus 0} \tilde{Z}_i.
\end{eqnarray}
The cavity partition function $Z_{n\rightarrow 0}(s_0=1,s_n=-1)$ of the configurations in which the root is up and the neighbor is down is obtained from the difference between the cavity partition function of the visitable configurations with $s_n=1$, and the cavity partition function of the visitable configurations with $s_n=1$ in which $s_n$ is blocked if it is reversed (see the previous section). For $(4,2)$ the neighbor $s_n$ is blocked if reversed i.f.f.\ all three spins in $\partial n\setminus 0$ are blocked down when $s_n$ is reversed (we are still considering the case in which the root is up). Therefore $Z_{n\rightarrow 0}(s_0=1,s_n=-1)$ is given by:
\begin{equation}
\label{eq:CavPartNeg}
Z_{n\rightarrow 0}(s_0=1,s_n=-1)=e^{\beta}\left(\prod_{i\in\partial n\setminus 0}  \tilde{Z}_i-\prod_{i\in\partial n\setminus 0}  \hat{Z}_i\right).
\end{equation}
Combining Eqs.~\eqref{eq:CavPartPos} and \eqref{eq:CavPartNeg} we find the total cavity partition function $\tilde{Z}_{n\rightarrow 0}$ of the visitable configurations in which the root $s_0$ is up and the neighbor $s_n$ is not blocked:
\begin{equation}
\label{eq:ricZtil}
\tilde{Z}_{n\rightarrow 0}=e^{-\beta} \prod_{i\in\partial n\setminus 0} \tilde{Z}_i+e^{\beta}\left(\prod_{i\in\partial n\setminus 0}  \tilde{Z}_i-\prod_{i\in\partial n\setminus 0}  \hat{Z}_i\right).
\end{equation}
In order to find the iteration rule for $R$, we have to compute the cavity partition function $\hat{Z}_{n\rightarrow 0}$, that counts the visitable configurations such that $s_n$ is not blocked down when $s_0$ is up, and becomes blocked down when $s_0$ is reversed. For $(4,2)$ those are the configurations in which two spins in $\partial n\setminus 0$ are blocked down when the root is down, and one is not. Their partition function is given by:
\begin{equation}
\label{eq:ricZhat}
\hat{Z}_{n\rightarrow 0}=e^{\beta}(\hat{Z}_1 \hat{Z}_2 \tilde{Z}_3+\hat{Z}_1 \tilde{Z}_2 \hat{Z}_3+\tilde{Z}_1 \hat{Z}_2 \hat{Z}_3- 3\hat{Z}_1 \hat{Z}_2 \hat{Z}_3),
\end{equation}
where the subscripts $i=1,2,3$ are abbreviations for $i\rightarrow n$, and refer to the elements of $\partial n\setminus 0$. Dividing Eq.~\eqref{eq:ricZtil} by Eq.~\eqref{eq:ricZhat} we find $R_{n\rightarrow 0}$ (see the previous section):
\begin{equation}
\label{eq:Rnroz}
R_{n\rightarrow 0}=\frac{e^{\beta}(\hat{Z}_1 \hat{Z}_2 \tilde{Z}_3+\hat{Z}_1 \tilde{Z}_2 \hat{Z}_3+\tilde{Z}_1 \hat{Z}_2 \hat{Z}_3- 3 \hat{Z}_1 \hat{Z}_2 \hat{Z}_3)}{(e^{-\beta}+e^{\beta})\tilde{Z}_1\tilde{Z}_2\tilde{Z}_3-e^{\beta}\hat{Z}_1 \hat{Z}_2 \hat{Z}_3}.
\end{equation}
Dividing Eq.~\eqref{eq:Rnroz} by $\tilde{Z}_1\tilde{Z}_2\tilde{Z}_3$ we finally obtain an iteration rule for the cavity field $R$:
\begin{equation}
\label{eq:iterRule}
\begin{split}
R_{n\rightarrow 0}&=\frac{R_1 R_2 +R_1 R_3+R_2 R_3- 3 R_1 R_2 R_3}{e^{-2\beta}+1-R_1 R_2 R_3}=\\&\equiv R_{iter}(R_1,R_2,R_3).
\end{split}
\end{equation}
Note that $R_{iter}$ also depends explicitly on the inverse temperature $\beta=1/T$. In the following we omit sometimes this dependence for brevity. 

To summarize, the cavity distribution $P(\eta,\mu,R)$ can be computed self-consistently by: i) drawing $c$ triplets 
\begin{equation}
(\eta_1,\mu_1,R_1),\dots,(\eta_{c},\mu_{c},R_{c})   
\end{equation}
according to $P(\eta,\mu,R)$; ii) drawing an initial state for the spin $s_n$; iii) applying the iteration rules for the triplet. In formulas we have:
\begin{widetext}
\begin{multline}
\label{eq:Equatriple}
P(\eta,\mu,R) = \mathds{E}_{s_n}\left(\prod_{i=1}^{c}\,\sum_{\eta_i\mu_i} \int dR_i\,P(\eta_i,\mu_i,R_i)\right)\times\\\times\delta^{(k)}\Big(\eta-T_{s_n}(\{\eta_i,\mu_i\}_i^c)\Big)\,\delta^{(k)}\Big(\mu-M_{s_n}(\{\eta_i,\mu_i\}_i^c)\Big)\,\delta\Big(R-F_{s_n}(\{\eta_i,\mu_i,R_i\}_i^c)\bigg),
\end{multline}
\end{widetext}
where $\delta^{(k)}(\cdot)$ represents the Kronecker delta-function, and $\{\bullet_i\}_i^n\equiv \{\bullet_1,\bullet_2,\dots,\bullet_{n}\}$. The symbol $\mathds{E}_{s_n}$ is the average over the initial state of $s_n$, that is extracted down with probability $p$, or up with probability $1-p$. It is interesting to note that this operation has the same form of a quenched disorder average in SG models \cite{Mezard2001}. In other words the choice of the initial condition in the FAM plays the same role of the definition of an instance of quenched disorder. See App.~\ref{app:ITfunctions} for the definition of the iteration functions $T_{s_n},M_{s_n},F_{s_n}$.

From the cavity distribution of the triplet $P(\eta,\mu,R)$ it is possible to compute the distribution $P_{site}^{f}(R)$ of the total field $R$ (see Eq.~\eqref{eq:magnFreeSpins}) on a free site $s_0$ of the original graph: 
\begin{widetext}
\begin{equation}
\label{eq:Psite}
P_{site}^{f}(R) = \mathds{E}_{s_0}\left(\prod_{i=1}^{z} \,\sum_{\eta_i\mu_i}\int dP(\eta_i,\mu_i,R_i)\right)\Omega^{site}_{s_0}(\{\eta_i,\mu_i\}_i^z)\,\delta\big(R-R_{site}(\{R_i\}_i^z)\big),
\end{equation}
where the function $\Omega^{site}_{s_0}$ is given by:
\begin{equation}
\label{eq:FilterFreeSpins}
\Omega^{site}_{s_0}\big(\{\eta_i,\mu_i\}_i^z\big)=\delta_{s_0,1}\,\mathds{1}\bigg(\sum_i \mu_i \leq z-f\bigg)+\delta_{s_0,-1}\,\mathds{1}\bigg(\sum_i \eta_i \leq z-f\bigg).
\end{equation}
\end{widetext}
Note that with this definition the function $\Omega^{site}_{s_0}$ filters only those contributions resulting in a free site, and $P_{site}^{f}(R)$ is normalized to the probability of drawing a free site:
\begin{equation}
\int \dd P_{site}^{f}(R)=P^{f}_{site}=1-P^+_{site}-P^-_{site}.
\end{equation}
In the next sections we show that $P(\eta,\mu,R)$ provides a correct description of the local properties of the FAM in the glassy phase. It is important to note that we assumed the local distribution $P(\eta, \mu, R)$ of the cavity fields is the same across all the states, namely that it does not depend on the initial condition $C$. This corresponds to the hypothesis that all the ergodic components are \emph{equivalent}. We will test this assumption in the next section, with the comparison between the predictions of the cavity method with numerical simulations. Eq.~\eqref{eq:Equatriple} can be solved numerically by means of population dynamics algorithms (PDAs) \cite{mezard2009information,Mezard2003,Mezard2001,mezard2002random}. PDA consists in representing the order parameter $P(\eta,\mu,R)$ by a population of $M$ triplets:
\begin{equation}
\label{eq:RepPopDyn}
P(\eta,\mu,R)\approx \frac{1}{M}\sum_{i=1}^M \delta_{\eta,\eta_i}\,\delta_{\mu,\mu_i}\,\delta(R-R_i).
\end{equation}
Substituting \eqref{eq:RepPopDyn} into \eqref{eq:Equatriple} one obtains a self-consistent equation for the population of triplets, that is usually solved iteratively. Note that the quantities associated with the $k-$core are singular at the critical point. Therefore, in order to prevent the solution by iteration from a critical slowing-down, it can be particularly useful to adopt a ``protected'' implementation of PDA (see for example \cite{perrupato2022ising}). The basic idea is to try to decompose the measure $P(\eta,\mu,R)$ in such a way as to ``factorize'' the critical modes:
\begin{equation}
\begin{split}
\label{eq:decomp}
P(\eta,\mu,R)=&(1-P)\,\delta_{\eta,0}\delta_{\mu,0}P_{00}(R)+\\&+(P-D)\,\delta_{\eta,1}\delta_{\mu,0}P_{10}(R)+\\&+D\,\delta_{\eta,1}\delta_{\mu,1}\delta(R-1).
\end{split}
\end{equation}
The probabilities $P$ and $D$, that are singular at the transition, would cause a critical slowing-down if computed by iteration. Thanks to the ``factorised'' form of Eq.~\eqref{eq:decomp}, $P$ and $D$ are isolated, and can be determined analytically through the solution of the BP equation \eqref{eq:BPeqZ4m2}. The conditioned populations $P_{00}(R)$ and $P_{10}(R)$ can be represented again by populations of fields, and can be computed by iteration. 


\subsection{Algorithms for Finding Equilibrium Solutions and Accelerated Dynamics}
\label{sec:algosol}
Interestingly the cavity equations can be also used on a given graph to sample configurations of the free spins. This can be particularly convenient close to the critical point and in the low temperature regime, where the dynamics is slow due to the scarcity of facilitating spins. For a given instance of the problem one defines a triplet of fields $(\eta_{i\rightarrow j},\mu_{i\rightarrow j},R_{i\rightarrow j})$ for each directed edge $i\rightarrow j$ of the graph. Let us denote by $s_i$ the value taken by a generic site $i$ on the initial configuration. The triplets are updated according to a message-passing procedure (see \emph{Belief Propagation} \cite{mezard2009information}) in order to satisfy: 
\begin{equation}
\label{eq:BP1}
\eta_{i\rightarrow j}=T_{s_i}\big(\{(\eta,\mu)_{k\rightarrow i}\}_{k\in\partial i\setminus j}\big),
\end{equation}
\begin{equation}
\label{eq:BP2}
\mu_{i\rightarrow j}=M_{s_i}\big(\{(\eta,\mu)_{k\rightarrow i}\}_{k\in\partial i\setminus j}\big),
\end{equation}
\begin{equation}
\label{eq:BP3}
R_{i\rightarrow j}=F_{s_i}\big(\{(\eta,\mu,R)_{k\rightarrow i}\}_{k\in\partial i\setminus j}\big),
\end{equation}
that are the given-instance version of \eqref{eq:Equatriple}. Once the fields are fixed, one has access to the estimate of the marginal probability distribution of each site. In particular, from \eqref{eq:FilterFreeSpins}, we have that a site $i$ is free if 
\begin{equation}
\sum_{k\in\partial i}\mu_{k\rightarrow i}\leq z-f\,\,\textnormal{and}\,\,s_i=1
\end{equation}
or
\begin{equation}
\sum_{k\in\partial i}\eta_{k\rightarrow i}\leq z-f\,\,\textnormal{and}\,\,s_i=-1.
\end{equation}
If the site is free its magnetization is given by \eqref{eq:magnFreeSpins}, where 
\begin{equation}
R=R_{site}\left(\{R_{k\rightarrow i}\}_{k\in\partial i}\right).
\end{equation}
At this point a visitable configuration can be extracted with e.g.\ a decimation procedure.

Let us discuss now the algorithm for accelerated dynamics we mentioned in the introduction. In the single-spin dynamics \eqref{eq:MetropRule} a fraction of moves is rejected due to the facilitation constraint. The basic idea is that instead of equilibrating a single spin in the environment determined by its neighbors, it could be convenient to equilibrate instantaneously a region comprising all its neighbors at distance less than a given value $L$. In order to do so we can leverage the cavity method discussed before. The algorithm is the following. At each step one chooses a random site (seed) on the graph, and selects all the neighbors at distance at most $L$ from the seed. We call $\mathcal{V}'\subset \mathcal{V}$ the set of nodes selected by this procedure, and $\mathcal{E}'\subset \mathcal{E}$ the set of edges $(i,j)$ s.t.\ $i,j\in \mathcal{V}'$. Let us also define the frontier $\partial \mathcal{V}'$ as the set of spins not belonging to $\mathcal{V}'$ that have a neighbor in $\mathcal{V}'$, and the set $D_{\rightarrow G'}$ of all directed edges $i\rightarrow j$ satisfying $i\in \partial \mathcal{V}'$ and $j \in \mathcal{V}'$.
After the subgraph $G'(\mathcal{V}',\mathcal{E}')$ is fixed, the next step is to sample a configuration of the spins in $\mathcal{V}'$ while keeping fixed the spins in $\partial \mathcal{V}'$. In order to do that we can look for the solution of the given-instance cavity equations \eqref{eq:BP1},\eqref{eq:BP2} and \eqref{eq:BP3} in $G'$, at fixed messages in $D_{\rightarrow G'}$. The messages in $D_{\rightarrow G'}$ are determined by considering blocked the spins in $\partial \mathcal{V}'$. In particular for each edge $i\rightarrow j$ in $D_{\rightarrow G'}$ the boundary message is $\eta=0,\mu=0,R=0$ if $s_i=1$ and $\eta=1,\mu=1,R=1$ if $s_i=0$ (see the beginning of this section). Once the fixed point is reached by iteration of the triplets, one can sample by decimation a new configuration for the spins in $G'$. Note that this is equivalent to make an infinite number of dynamical moves for the spins in $\mathcal{V}'$ at fixed $\partial \mathcal{V}'$.
Therefore all the spins $\mathcal{V}'$ that are free for the given fixed configuration of the boundary $\partial \mathcal{V}'$ will move and we have to  update {\it all} the persistences of the free spins to zero, even if they have the same value in the old and new configuration. At this point the algorithm is iterated by extracting another random seed and applying the procedure that has just been described. 

\section{Computation of the Static Observables}
In this section we solve the cavity equations \eqref{eq:Equatriple}, compute different observables of the problem, and compare the predictions with numerical simulations. 

The kinetic constraint \eqref{eq:faccc} for $p_c\leq p<1 $ divides the configurations space into an exponential number of separate components (states). The total partition function $Z$ in this case is decomposed into a sum of the partition functions $Z_{\alpha}$ restricted to the single states:
\begin{equation}
Z=\sum_\alpha Z_\alpha.
\end{equation}
Note that the average according to the Boltzmann-Gibbs measure corresponds to an annealed average in which each state is weighted precisely according to its free energy. Averages inside a state of observables that depend on a \emph{single} configuration ({\it e.g.} the magnetization) can be written as averages over the Boltzmann-Gibbs measure, indeed we have:
\begin{equation}
\frac{1}{Z}\sum_{\alpha}  Z_\alpha  \sum_{s \in \alpha} P_\alpha(s)\,  O(s) = \frac{1}{Z} \sum_s e^{-\beta H(s)} O(s),
\end{equation}
where $P_{\alpha}(s)$ is the measure restricted to state $\alpha$. This is not possible for quantities that depend on more than one configuration inside the same state $O(s,s',s'',\dots)$, {\it e.g.} the overlap $q$ (see section \ref{sec:Overlap}).
Numerically those kind of averages can be measured through dynamics, using the fact that:
\begin{equation}
\lim_{t \rightarrow \infty} P(s',t|s,0) = P_{\alpha(s)}(s'),  
\end{equation}
where $\alpha(s)$ is the ergodic component of configuration $s$, and $P(s',t|s,0)$ is the probability of configuration $s'$ at time $t$ given configuration $s$ at time $0$. In this way the average of the observable $O$ can be written as
\begin{widetext}
\begin{multline}
\label{eq:calcObs}
O=\frac{1}{Z}\sum_{\alpha}  Z_\alpha  \sum_{s \in \,\alpha} P_\alpha(s) \sum_{s' \in\, \alpha} P_\alpha(s') \sum_{s'' \in \,\alpha} P_\alpha(s'') \dots   O(s,s',s'',\dots)=\\=\frac{1}{Z} \sum_s e^{-\beta H(s)} \sum_{s'} P(s',t|s,0) \sum_{s''}  P(s'',t|s,0) \dots O(s,s',s'',\dots). 
\end{multline}
\end{widetext}
Note that in the second line the Boltzmann-Gibbs measure corresponds to an average over the initial configuration, that is evolved according to different dynamical trajectories. The quantity 
\begin{multline}
\frac{1}{Z}\sum_s e^{-\beta H(s)}\sum_{s'} P(s',t|s,0) \times\\\times\sum_{s''}  P(s'',t|s,0) \dots O(s,s',s'',\dots),
\end{multline}
can be easily computed in numerical simulations, since the measure $e^{-\beta H(s)}/Z$ on the initial condition is factorized. The cavity calculation gives access to the probability $P_{\alpha}(s)$ of a typical state. In the large $N$ limit we expect all the states to be equivalent, implying that the average value of an observable over the whole system in a given state is independent of the state.
\subsection{Overlap}
\label{sec:Overlap}
\begin{figure}
\includegraphics[width=0.48\textwidth]{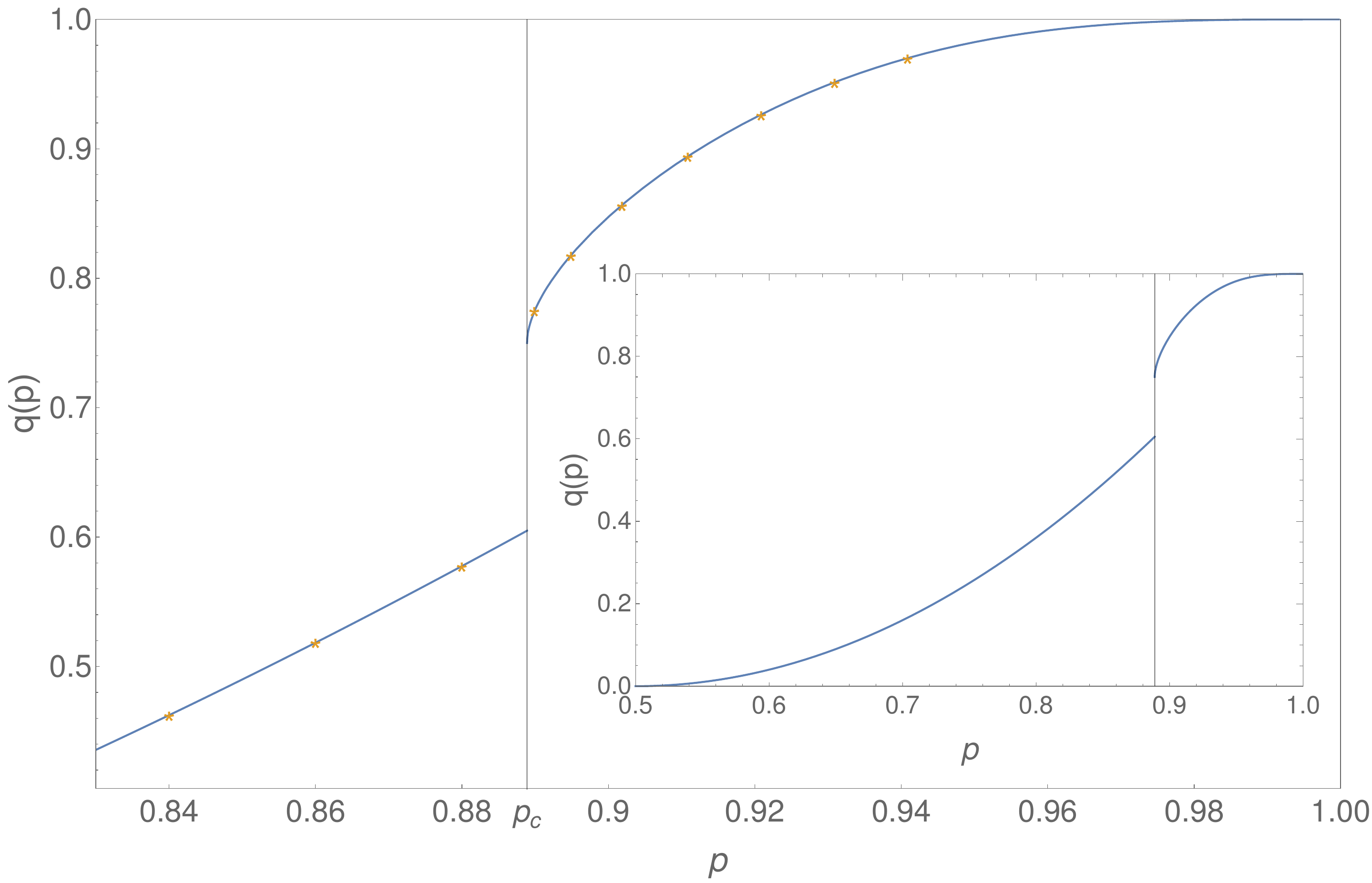}
\caption{Overlap as a function of $p$ for $z=4$ and $f=2$ ($p_c=8/9$). The continuous line is the cavity prediction, and the points correspond to numerical simulations on a system with size $N=9\times 10^6$. \emph{Inset}: Cavity prediction for the overlap on the whole domain $1/2\leq p\leq 1$.
\label{fig:datap115-1}}
\end{figure}

\begin{figure}
\includegraphics[width=0.48\textwidth]{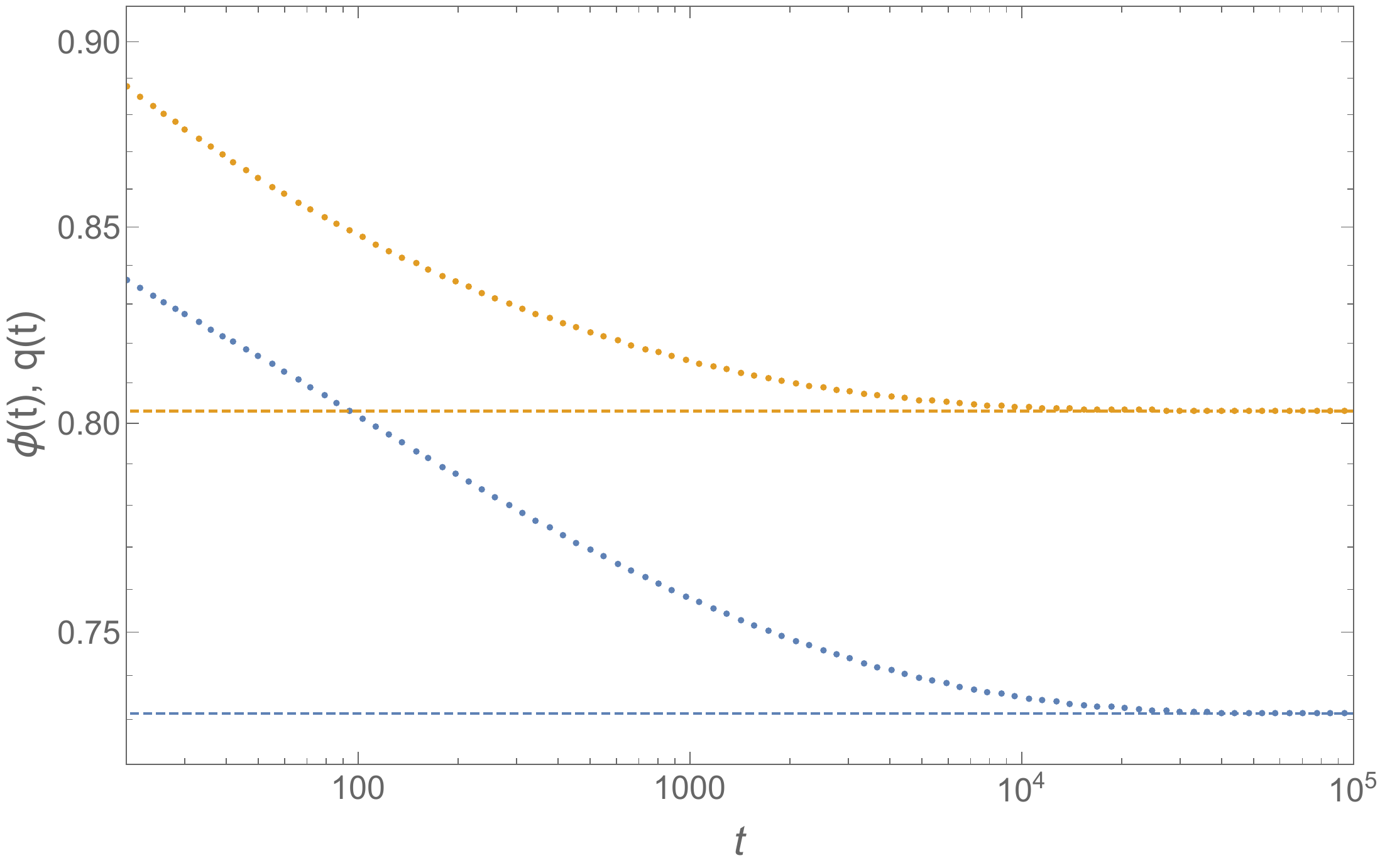}
\caption{Persistence $\phi(t)$ (bottom curve) and self-overlap $q(t)$ (top curve) for $e^{-2\beta}=0.12$ and facilitation $f=2$, on a Bethe lattice with coordination $z=4$ and number of nodes $N=4.5\cdot 10^5$. Numerical data are in perfect agreement with the analytical asymptotic predictions (horizontal lines).
\label{fig:datap115-2}}
\end{figure}

\begin{figure}
\includegraphics[width=0.48\textwidth]{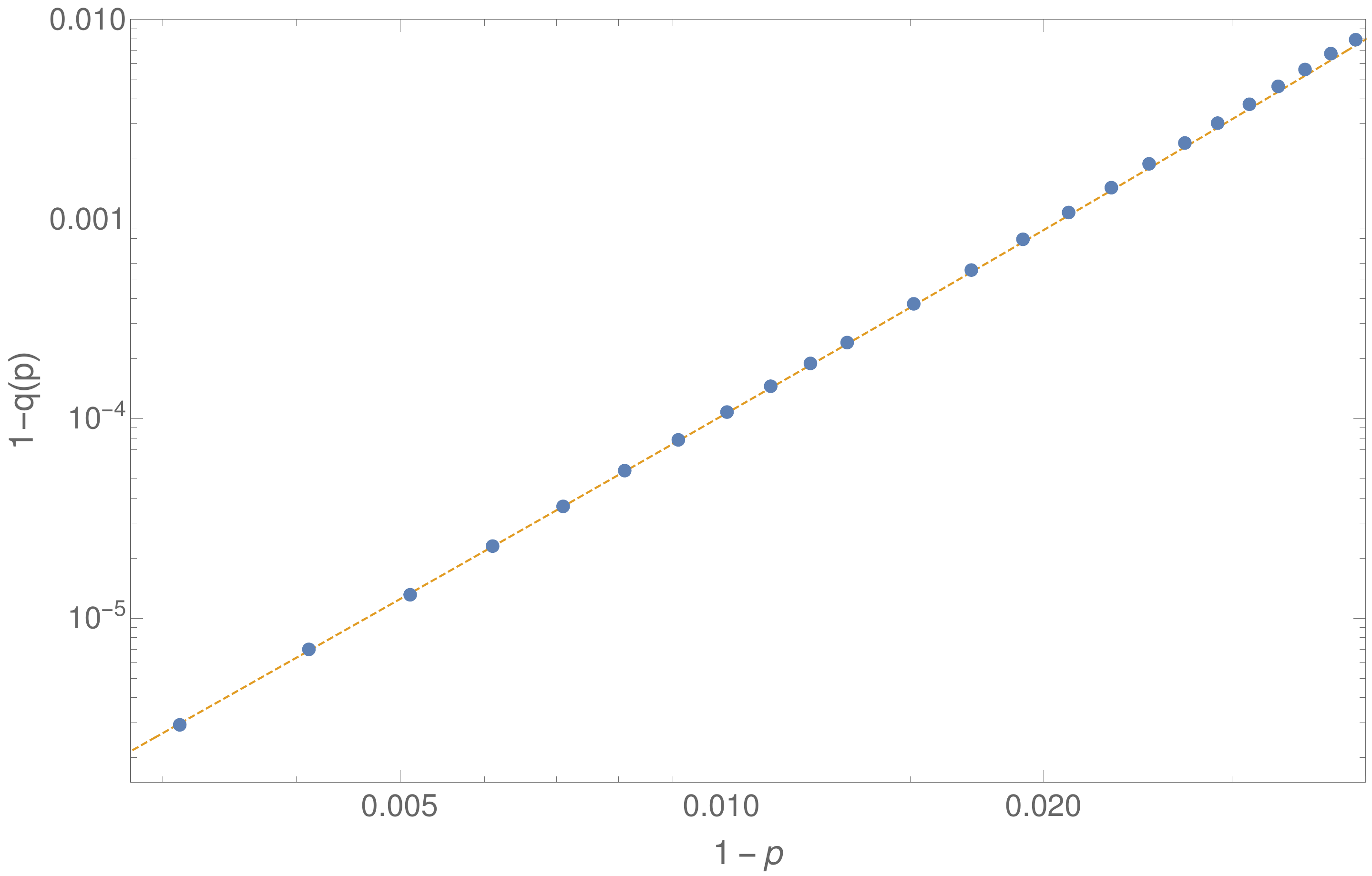}
\caption{Comparison between the cavity prediction (dots) of the overlap $q$ with the low temperature expansion (dashed line), $1-q\approx 96\,(1-p)^3+700\,(1-p)^4$ (see App.~\ref{sec:appLowT}). \label{fig:lowTExpOverlap}}
\end{figure}

The overlap $q(s,s')$ between two configurations $s,s'$ is defined by
\begin{equation}
q(s,s')=\frac{1}{N}\sum_{i=1}^N s_i\,s_i'.
\end{equation}
The average overlap $q$ inside a state is expressed by formula \eqref{eq:calcObs}:
\begin{equation}
q=\frac{1}{Z}\sum_{\alpha}  Z_\alpha  \sum_{s \in \,\alpha} P_\alpha(s) \sum_{s' \in \,\alpha} P_\alpha(s')\,   q(s,s'). 
\end{equation}
Explicitly we have:
\begin{equation}
q=\frac{1}{Z}\sum_{\alpha} Z_\alpha \frac{1}{N}\sum_{i=1}^N \left(m_i^{\alpha}\right)^2,
\end{equation}
where
\begin{equation}
m_i^{\alpha}=\sum_{s\in\alpha}P_{\alpha}(s)\,s_i.
\end{equation}
The average overlap $q$ has an important physical meaning, since it is an order parameter for the problem (see Fig.~\ref{fig:datap115-1}). Indeed in the liquid phase ($p<p_c$) the spins are all non-interacting, and the overlap is just $q=(1-2p)^2$. At $p_c$ a finite fraction of the spins becomes permanently blocked. This implies that the average overlap should have a jump at $p_c$, since for $p\geq p_c$, in a given state, the blocked spins have self-overlap one. Under the assumption of equivalence of the states, $1/N\sum_{i}(m_i^{\alpha})^2$ does not depend on $\alpha$, and it is given by the cavity computation of the previous sections: 
\begin{equation}
\label{eq:cavPredOverlap}
q=[m^2(R)]=P^{+}_{site}+P^{-}_{site}+\int dR\,P_{site}^f(R)\,m^2(R).
\end{equation}
In the previous formula we denoted by $[\cdot]$ the average w.r.t.\ the cavity distribution, and $m(R)$ is given by Eq.~\eqref{eq:magnFreeSpins}. The first two addends after the second equality are due to the blocked spins, that have self-overlap one. The cavity predictions obtained from Eq.~\eqref{eq:cavPredOverlap} are in perfect agreement with numerical simulations, as shown in Figs.~\ref{fig:datap115-1} and \ref{fig:datap115-2}, and with the low temperature expansion, see Fig.~\ref{fig:lowTExpOverlap} and App.~\ref{sec:appLowT}.

\begin{figure}
\includegraphics[width=0.48\textwidth]{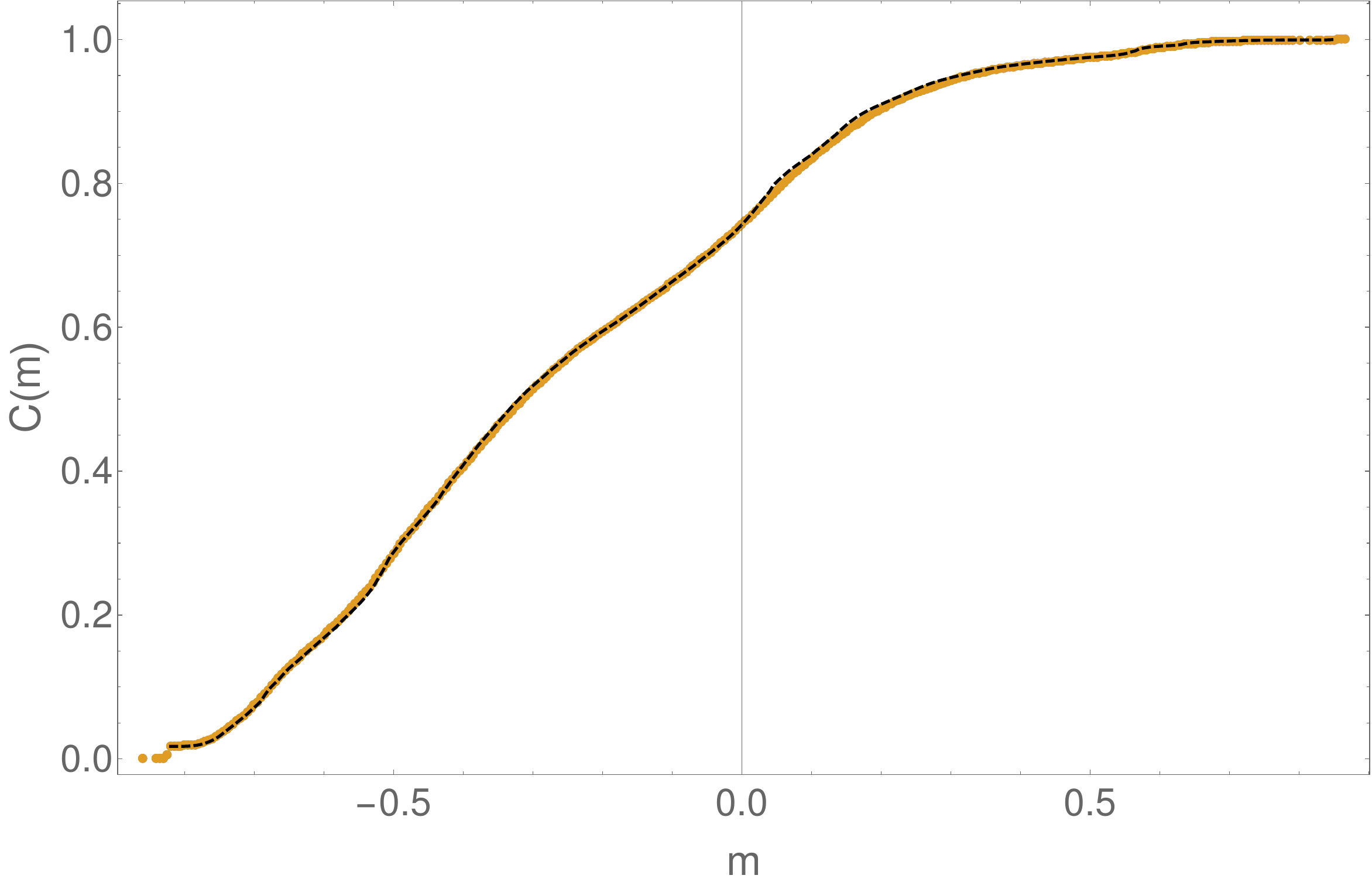}
\caption{Cumulative $C(m)$ of the probability distribution of the free spins' magnetization for $e^{-2\beta}=0.98$. The dashed line is the cavity prediction, while the points are obtained from numerical simulations with $N=9\times 10^5$.\label{histdistrmagn}
}
\end{figure}

A further comparison is presented in Fig.~\ref{histdistrmagn}, where we show the cavity prediction for the cumulative $C(m)$ of the probability distribution of the free spins magnetization:
\begin{equation}
C(m)=\int_{-1}^mdm'\int dR P_{site}^f(R)\, \delta\left(m'-m(R)\right),
\end{equation}
together with the numerical simulations. 

\begin{figure}
\includegraphics[width=0.48\textwidth]{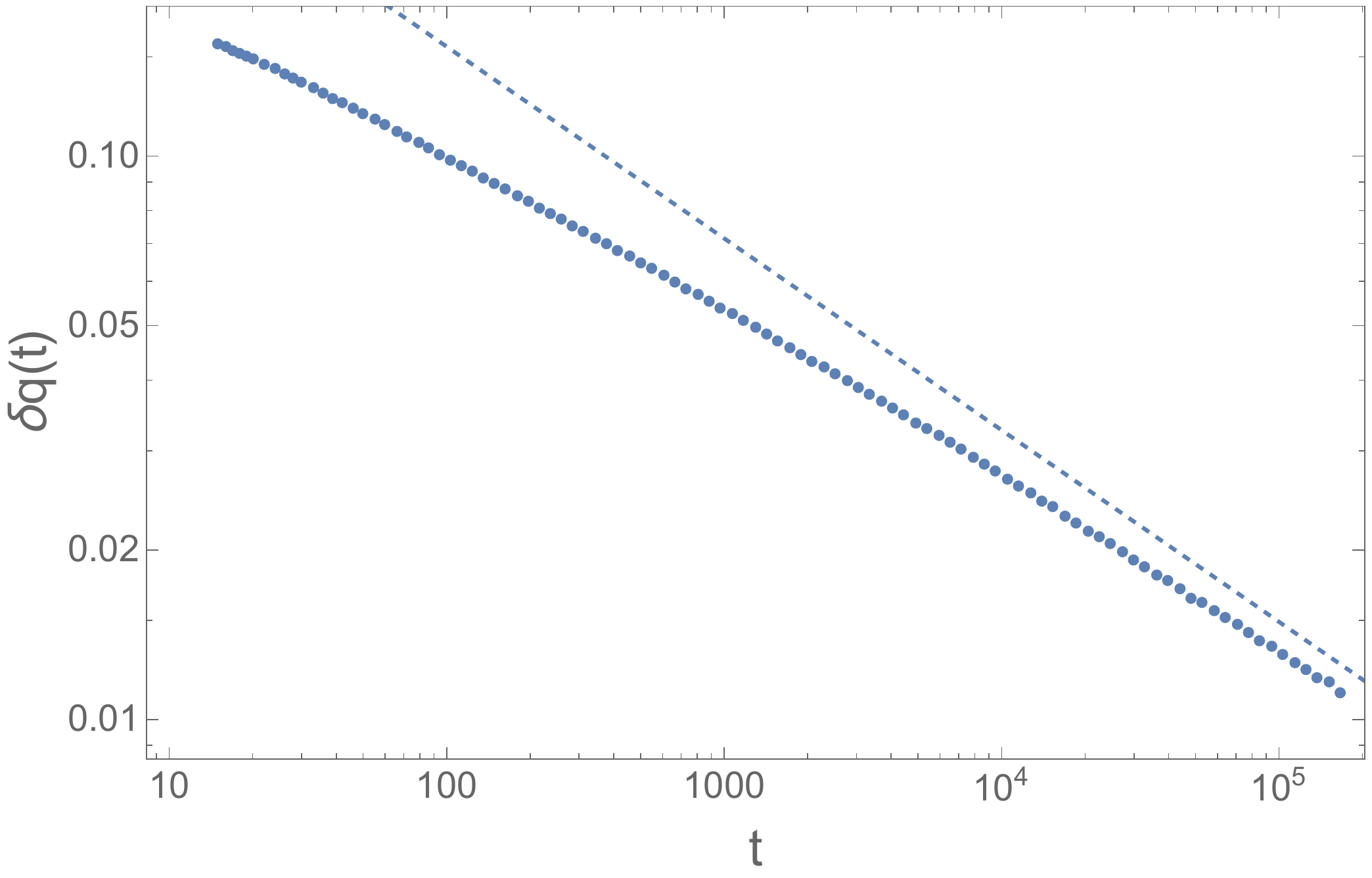}
\caption{Power-law decay of the overlap at the critical point. Dots represent $\delta q(t)=q(t)-q(\infty)$, where $q(t)$ is obtained from numerical simulations, and $q(\infty)\equiv q_c$, is the plateau value of the overlap at the critical point computed from the cavity method presented here. For $z=4$, $f=2$ we predict $q_c\approx 0.7494$. The dashed line is $\propto 1/t^a$, where $a=0.340356$ is computed in \cite{perrupato2022exact}. Only the slope of the dashed line is fixed by the theory, while there is no prediction for the microscopic time-scale that determines the coefficient of $1/t^a$. Numerical data are obtained on a system with size $N=9\times 10^6$.\label{shifted_overlap}
}
\end{figure}

In Fig.~\ref{shifted_overlap} we show the shifted overlap $\delta q(t)=q(t)-q(\infty)$ as a function of time at the critical point. The plateau $q(\infty)\equiv q_c\approx 0.7494$ is obtained from the cavity estimate. Interestingly the overlap shows the same critical behavior of the persistence. In particular the shifted overlap decays to zero with a power-law behavior $\propto t^{-a}$, where the MCT exponent (see the introduction) $a=0.340356$ is computed in \cite{perrupato2022exact}.   


\subsection{Edge Marginal}
\label{sec:EdgeMarg}
\begin{figure}
\includegraphics[width=0.48\textwidth]{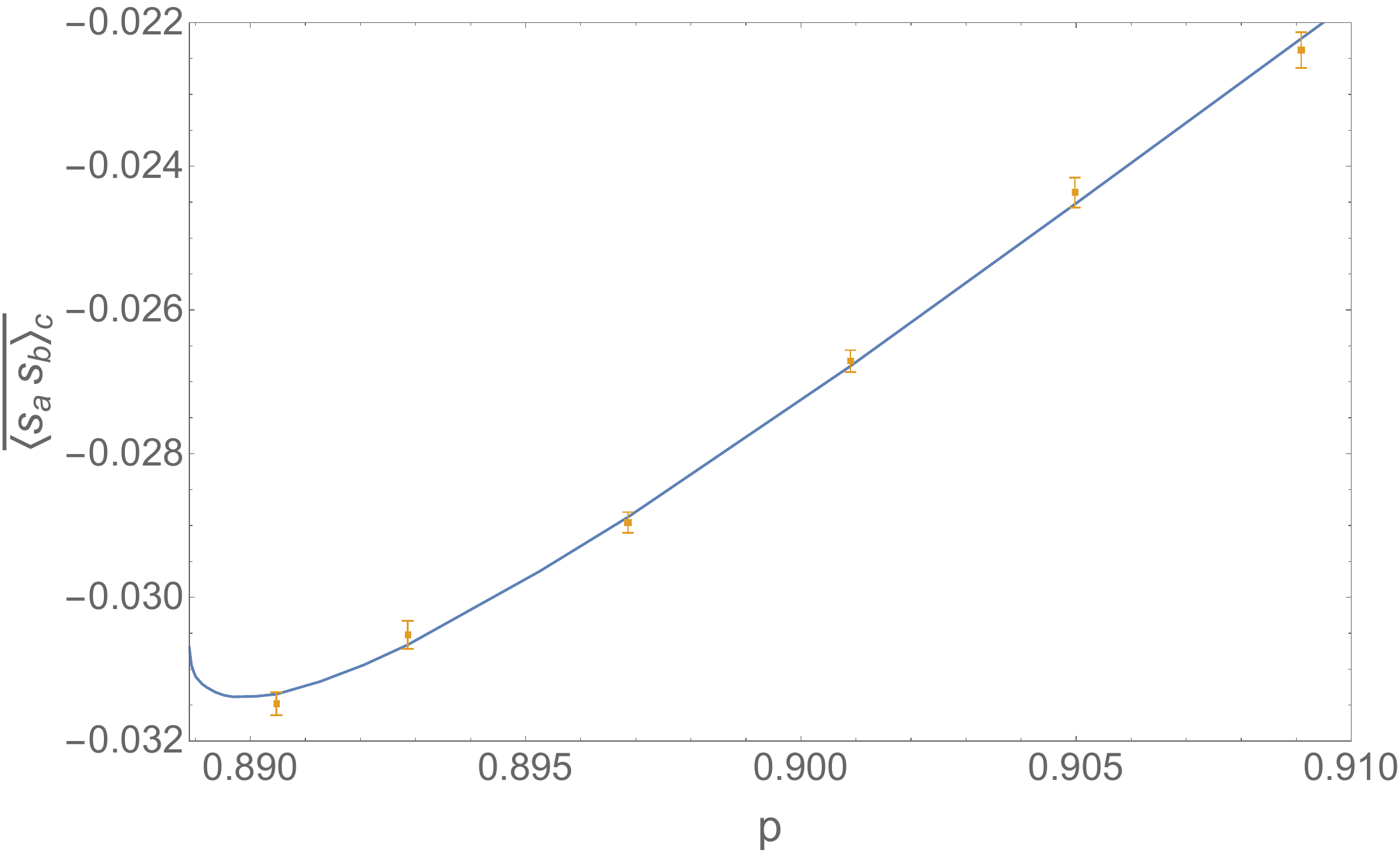}
\caption{Comparison between the cavity prediction (continuous line) of the connected correlation between two neighboring spins $\overline{\langle s_a s_b\rangle_c}$ (the bar represents the average over $a$ and $b$), and numerical simulations (points). Numerical simulation are performed with a system size $10^6$, and the cavity predictions are obtained with a PDA with population size $10^7$.\label{fig:ConnCorr}}
\end{figure}
As already discussed, in the glassy phase there are non-trivial correlations between the free sites of the lattice. These are due to the existence of a blocked cluster of spins. In order to clarify more this point we want to discuss an example that can be represented by the following diagram: 
\begin{equation*}
\begin{gathered}
\scalebox{0.6}{
\begin{tikzpicture}
\node[label=below:{}, shape=circle, scale=1,draw] (0) at (0,0) [right] {};
\node[label=below:{},shape=circle,position=0:{\nodeDist} from 0,scale=1,draw] (dx) {};
\node[label=below:{},shape=circle,position=-90:{\nodeDist} from 0,scale=1] (dw1) {};
\node[label=below:{},shape=circle,position=180:{\nodeDist} from 0,scale=1] (sx1) {};
\node[label=below:{},shape=circle,position=90:{\nodeDist} from 0,scale=1] (up) {};
\draw[black,thick,dashed] (0) -- (dw1);
\draw[black,thick,dashed] (0) -- (sx1);
\draw[black,thick] (0) -- (dx);
\draw[black,thick] (0) -- (up);
\node[label=below:{},shape=circle,position=0:{\nodeDist} from dx,scale=1] (dx1) {};
\node[label=below:{},shape=circle,position=90:{\nodeDist} from dx,scale=1] (up2) {};
\node[label=below:{},shape=circle,position=-90:{\nodeDist} from dx,scale=1] (dw2) {};
\draw[black,thick,dashed] (dx) -- (up2);
\draw[black,thick,dashed] (dx) -- (dw2);
\draw[black,thick] (dx) -- (dx1);
\end{tikzpicture}
}
\end{gathered}
\end{equation*}
where we represented with a continuous external leg a neighbor that is blocked up, and with a dashed external leg a neighbor that is blocked down. The empty circles represent spins that are initialized up. Since the neighbors are permanently blocked, the configuration $(-1,-1)$ in which both spins are down is impossible to reach with the kinetically constrained dynamics. This implies a correlation between the two spins (if one of them is down, the other must be up). In the liquid phase there is no blocked cluster, and the spins are uncorrelated. For this reason an interesting observable is the joint probability distribution of two free neighboring spins $s_a,s_b$ (edge marginal), that we compute in this section. This quantity should be become non-factorized in the glassy phase. 

Let us call $Z(s_a,s_b)$ the conditioned partition function of the couple $(s_a,s_b)$. The configurations with $s_a=1$ in the sub-tree whose nodes can be reached from $a$ without passing from $b$ have partition function:
\begin{equation}
\tilde{Z}^{\rightarrow a}=\prod_{i=1}^3\tilde{Z}^a_i,
\end{equation}
therefore
\begin{equation}
Z(1,1)=e^{-2\beta}\tilde{Z}^{\rightarrow a}\tilde{Z}^{\rightarrow b}.
\end{equation}
At this point, analogously to the single-spin observables of the last sections, the strategy is to write the partition functions conditioned to $(1,-1)$, $(-1,1)$ and $(-1,-1)$ in terms of the partition function of all reachable configurations with $(1,1)$ by subtracting from them all the cases in which at least one of the two spins is blocked if it is reversed. Let us start from the cases $(1,-1)$ and $(-1,1)$. When only one spin, say $s_a$, is reversed from $(1,1)$, $s_b$ is not blocked if and only if $s_a$ is not blocked. For this reason we have:
\begin{equation}
Z(-1,1)=\tilde{Z}^{\rightarrow b}\left(\tilde{Z}^{\rightarrow a}-\prod_{i=1}^3\hat{Z}^a_i\right),
\end{equation}
where we subtracted the configurations in which $s_a$ is blocked if it is reversed. The analogous expression also holds for $Z(1,-1)$. At this point let us study $Z(-1,-1)$. We should take into account the cases in which only one spin is blocked, and that in which both spins are blocked in $(-1,-1)$. We can write:
\begin{widetext}
\begin{equation}
\label{eq:mm}
\begin{split}
Z(-1,-1)=e^{2\beta}\Bigg\{&\tilde{Z}^{\rightarrow a}\tilde{Z}^{\rightarrow b}+\\
&-\prod_{i=1}^3\hat{Z}^a_i\left(\prod_{i=1}^3(\tilde{Z}_i^{b}-\hat{Z}_i^b)+\left((\tilde{Z}_1^b-\hat{Z}_1^b)(\tilde{Z}_2^b-\hat{Z}_2^b)\hat{Z}_3^b+\textnormal{perm.}\right)\right)+\\
&-\prod_{i=1}^3\hat{Z}^b_i\left(\prod_{i=1}^3(\tilde{Z}_i^a-\hat{Z}_i^a)+\left((\tilde{Z}_1^a-\hat{Z}_1^a)(\tilde{Z}_2^a-\hat{Z}_2^a)\hat{Z}_3^a+\textnormal{perm.}\right)\right)+\\
& -\left(\left(\hat{Z}^a_1\hat{Z}^a_2\tilde{Z}^a_3+\textnormal{perm.}\right)-2\hat{Z}^a_1\hat{Z}^a_2\hat{Z}^a_3\right)\times\left(\left(\hat{Z}^b_1\hat{Z}^b_2\tilde{Z}^b_3+\textnormal{perm.}\right)-2\hat{Z}^b_1\hat{Z}^b_2\hat{Z}^b_3\right)\Bigg\},
\end{split}
\end{equation}
where ``perm.'' means that one has to sum the terms in brackets with all permutations of the subscript indexes. The second and the third line of \eqref{eq:mm} subtract all the cases in which exactly one of the two spins is blocked if both $s_a$ and $s_b$ are reversed in the initial configuration. The fourth line subtracts all the cases in which both spins are blocked if reversed. At this point, dividing by $Z(1,1)$ we find:
\begin{equation}
\begin{split}
\frac{Z(-1,-1)}{Z(1,1)}=e^{4\beta}\Bigg\{1
&-\prod_{i=1}^3R^a_i\left(\prod_{i=1}^3(1-R_i^b)+\left((1-R_1^b)(1-R_2^b)R_3^b+\textnormal{perm.}\right)\right)+\\
&-\prod_{i=1}^3R^b_i\left(\prod_{i=1}^3(1-R_i^a)+\left((1-R_1^a)(1-R_2^a)R_3^a+\textnormal{perm.}\right)\right)+\\
& -\bigg(R^a_1R^a_2+R^a_1R^a_3+R^a_2R^a_3-2R^a_1R^a_2R^a_3\bigg)\bigg(R^b_1R^b_2+R^b_1R^b_3+R^b_2R^b_3-2R^b_1R^b_2R^b_3\bigg)\Bigg\}\\\equiv e^{4\beta} G_e&
\left(\{\underline{R}_{i\rightarrow a}\}_{i\in \partial a\setminus b},\{\underline{R}_{j\rightarrow a}\}_{j\in\partial b\setminus a}\right).
\end{split}
\end{equation}
\end{widetext}
Note that even if both spins are free, \emph{i.e.}\ both the single-spin marginal of $s_a$ and $s_b$ have support on $-1$ and $+1$, not all the configurations of the couple are in general allowed. In particular the state $s_a=-1,s_b=-1$ may have zero measure.

The previous expressions allow us to compute the correlation $C_{ab}=\langle s_a s_b\rangle$ of the couple. Necessary condition for $C_{ab}$ to be non trivial, \emph{i.e.}\ $C_{ab}\neq m_a\,m_b$, is that both spins are free. If this is the case we can write:
\begin{equation}
C_{ab}=\frac{Z(1,1)-Z(-1,1)-Z(1,-1)+Z(-1,-1)}{Z(1,1)+Z(-1,1)+Z(1,-1)+Z(-1,-1)},
\end{equation}
where the various terms are given by the previous expressions. Note that correlations can be also computed with the fluctuation-dissipation relation. Suppose to draw a configuration at a certain temperature, and \emph{afterwards} to add site-dependent fields $H_i$'s on the spins. In this way the blocked spins are unchanged. Given a free spin $s_i$, its magnetization depends on $H_i$ through:
\begin{equation}
\label{eq:addExtField}
m_i=\frac{e^{-2\beta(1-H_i)}-1+R}{e^{-2\beta(1-H_i)}+1-R},\quad R \equiv \frac{\hat{Z}}{\tilde{Z}}.
\end{equation}
Consider another free spin $s_j$. The connected correlation $\langle s_is_j\rangle_c$ is given by the fluctuation-dissipation relation:
\begin{equation}
\frac{1}{\beta}\frac{d m_i}{dH_j}=\langle s_is_j\rangle_c,
\end{equation}
where the derivative is computed at zero external fields. If $i=j$ we can immediately check that
\begin{equation}
\langle s_is_i\rangle_c=1-m_i^2.
\end{equation}
If $i\neq j$: 
\begin{equation}
\langle s_is_j\rangle_c=\frac{2e^{-2\beta}}{(e^{-2\beta}+1-R)^2}\frac{dR}{dH_j}.
\end{equation}
If $i$ and $j$ are neighboring sites we find
\begin{multline}
\frac{dR}{dH_j}=\frac{dR}{dR_{j\rightarrow i}}\frac{dR_{j\rightarrow i}}{dH_j}=\\=-\frac{2e^{-2\beta}R_{j\rightarrow i}}{e^{-2\beta}+1-\prod_{k\in\partial j\setminus i}R_{k\rightarrow j}}\frac{dR}{dR_{j\rightarrow i}}.
\end{multline}
The previous formulas will be particularly useful in Sec.~\ref{sec:IntroSGSusc} and App.~\ref{sec:AnalyticalCompSGSusc} for the computation of the SG susceptibility. Note that all the quantities discussed in this section are expressed in terms of the local cavity fields, and therefore they can be computed by solving the cavity equations \eqref{eq:Equatriple}. In Fig.~\ref{fig:ConnCorr} we show a comparison between the cavity prediction for the connected correlation function of two neighboring spins and numerical simulations.

\subsection{Configurational Entropy}
\label{sec:ConfEntro}

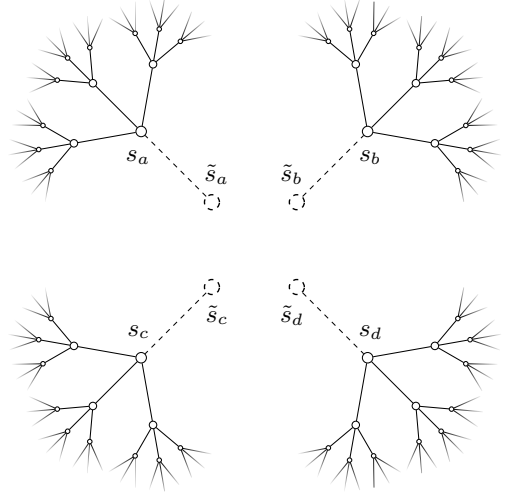
\begin{figure}
\centering
\scalebox{0.65}{
\begin{tikzpicture}
\node[label=above:{},shape=circle, scale=1] (0) at (0,0) {};
\foreach \i in {45,135,225,315}
{
\node[label=above:{},shape=circle, semithick,position=\i:{5\nodeDist} from 0, scale=0.7, draw] (A\i) {};
\foreach \j in {\i-55,\i,\i+55}
{
\node[label=above:{},shape=circle, semithick,position=\j:{2\nodeDist} from A\i, scale=0.5, draw] (B\j) {};
\draw [semithick] (A\i) -- (B\j);
\foreach \k in {\j-40,\j,\j+40}
{
\node[label=above:{},shape=circle, semithick,position=\k:{\nodeDist} from B\j, scale=0.3,draw] (C\k) {};
\draw [semithick] (C\k) -- (B\j);
\foreach \l in {\k-29,\k,\k+29}
{
\node[label=above:{},shape=circle, semithick,position=\l:{\nodeDist} from C\k] (D\l) {};
\pgfmathparse{int(\l)};
\ifthenelse{\pgfmathresult<90 \AND \pgfmathresult>-90 \OR \pgfmathresult>270 )}{\draw [semithick,path fading=east] (C\k) -- (D\l);}{\draw [semithick,path fading=west] (C\k) -- (D\l);}
}
}
}
}
\node[label=above:{},shape=circle, semithick,position=143:{3.8\nodeDist} from 0,scale=1.6] () {\small $s_a$};
\node[label=above:{},shape=circle, semithick,position=118:{1.5\nodeDist} from 0,scale=1.6] () {\small $\tilde{s}_a$};
\node[label=above:{},shape=circle, semithick,position=37:{3.8\nodeDist} from 0,scale=1.6] () {\small $s_b$};
\node[label=above:{},shape=circle, semithick,position=62:{1.5\nodeDist} from 0,scale=1.6] () {\small $\tilde{s}_b$};
\node[label=above:{},shape=circle, semithick,position=217:{3.8\nodeDist} from 0,scale=1.6] () {\small $s_c$};
\node[label=above:{},shape=circle, semithick,position=242:{1.5\nodeDist} from 0,scale=1.6] () {\small $\tilde{s}_c$};
\node[label=above:{},shape=circle, semithick,position=323:{3.8\nodeDist} from 0,scale=1.6] () {\small $s_d$};
\node[label=above:{},shape=circle, semithick,position=298:{1.5\nodeDist} from 0,scale=1.6] () {\small $\tilde{s}_d$};
\foreach \i in {45,135,225,315}
{
\node[label=above:{},shape=circle, semithick,position=\i:{1.5\nodeDist} from 0, scale=1, thick,dashed,draw] (CC\i) {};
\draw [dashed,semithick] (CC\i) -- (A\i);
}
\end{tikzpicture}
}
\caption{Cavity graph. The dashed nodes represent the so called cavity spins. The branches associated with the cavity spins are assumed as independent. 
\label{fig:cavGraph}}
\end{figure}
\begin{figure}
\centering
\scalebox{0.65}{
\begin{tikzpicture}
\node[label=above:{},shape=circle, scale=1] (0) at (0,0) {};
\foreach \i in {45,135,225,315}
{
\node[label=above:{},shape=circle, semithick,position=\i:{5\nodeDist} from 0, scale=0.7, draw] (A\i) {};
\foreach \j in {\i-55,\i,\i+55}
{
\node[label=above:{},shape=circle, semithick,position=\j:{2\nodeDist} from A\i, scale=0.5, draw] (B\j) {};
\draw [semithick] (A\i) -- (B\j);
\foreach \k in {\j-40,\j,\j+40}
{
\node[label=above:{},shape=circle, semithick,position=\k:{\nodeDist} from B\j, scale=0.3,draw] (C\k) {};
\draw [semithick] (C\k) -- (B\j);
\foreach \l in {\k-29,\k,\k+29}
{
\node[label=above:{},shape=circle, semithick,position=\l:{\nodeDist} from C\k] (D\l) {};
\pgfmathparse{int(\l)};
\ifthenelse{\pgfmathresult<90 \AND \pgfmathresult>-90 \OR \pgfmathresult>270 )}{\draw [semithick,path fading=east] (C\k) -- (D\l);}{\draw [semithick,path fading=west] (C\k) -- (D\l);}
}
}
}
}
\node[label=above:{},shape=circle, semithick,position=143:{3.8\nodeDist} from 0,scale=1.6] () {\small $s_a$};
\node[label=above:{},shape=circle, semithick,position=37:{3.8\nodeDist} from 0,scale=1.6] () {\small $s_b$};
\node[label=above:{},shape=circle, semithick,position=217:{3.8\nodeDist} from 0,scale=1.6] () {\small $s_c$};
\node[label=above:{},shape=circle, semithick,position=323:{3.8\nodeDist} from 0,scale=1.6] () {\small $s_d$};
\draw [thick] (A135) -- (A45);
\draw [thick] (A225) -- (A315);
\end{tikzpicture}
}
\caption{From the cavity graph of Fig.~\ref{fig:cavGraph} it is possible to obtain a Bethe lattice with $N$ nodes by the addition of two links, namely setting $\tilde{s}_a\equiv s_b$,$\tilde{s}_b\equiv s_a$,$\tilde{s}_c\equiv s_d$ and $\tilde{s}_d\equiv s_c$. \label{fig:cavGraphTonodeToEdge-1}}
\end{figure}
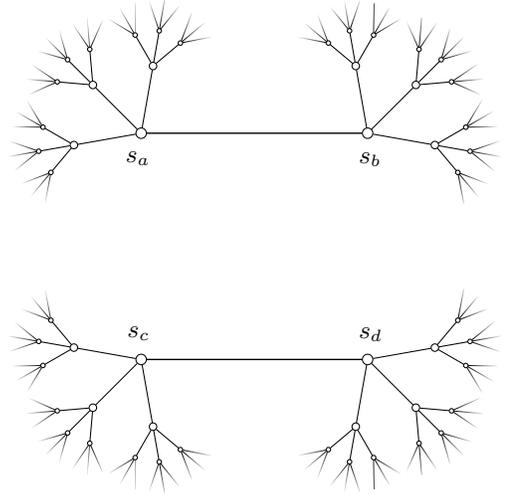

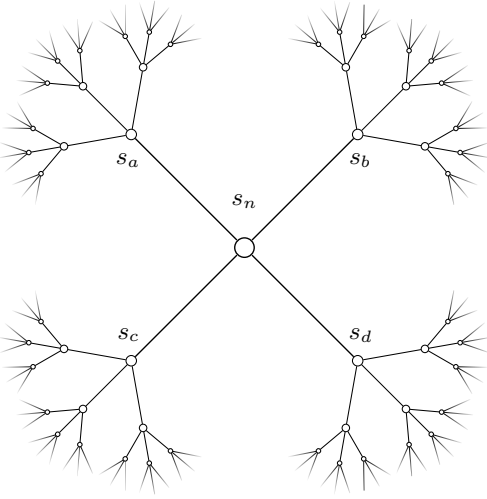
\begin{figure}
\centering
\scalebox{0.65}{
\begin{tikzpicture}
\node[label=above:{},shape=circle, scale=1] (0) at (0,0) {};
\foreach \i in {45,135,225,315}
{
\node[label=above:{},shape=circle, semithick,position=\i:{5\nodeDist} from 0, scale=0.7, draw] (A\i) {};
\foreach \j in {\i-55,\i,\i+55}
{
\node[label=above:{},shape=circle, semithick,position=\j:{2\nodeDist} from A\i, scale=0.5, draw] (B\j) {};
\draw [semithick] (A\i) -- (B\j);
\foreach \k in {\j-40,\j,\j+40}
{
\node[label=above:{},shape=circle, semithick,position=\k:{\nodeDist} from B\j, scale=0.3,draw] (C\k) {};
\draw [semithick] (C\k) -- (B\j);
\foreach \l in {\k-29,\k,\k+29}
{
\node[label=above:{},shape=circle, semithick,position=\l:{\nodeDist} from C\k] (D\l) {};
\pgfmathparse{int(\l)};
\ifthenelse{\pgfmathresult<90 \AND \pgfmathresult>-90 \OR \pgfmathresult>270 )}{\draw [semithick,path fading=east] (C\k) -- (D\l);}{\draw [semithick,path fading=west] (C\k) -- (D\l);}
}
}
}
}
\node[label=above:{},shape=circle, semithick,position=143:{3.8\nodeDist} from 0,scale=1.6] () {\small $s_a$};
\node[label=above:{},shape=circle, semithick,position=37:{3.8\nodeDist} from 0,scale=1.6] () {\small $s_b$};
\node[label=above:{},shape=circle, semithick,position=217:{3.8\nodeDist} from 0,scale=1.6] () {\small $s_c$};
\node[label=above:{},shape=circle, semithick,position=323:{3.8\nodeDist} from 0,scale=1.6] () {\small $s_d$};
\node[label=above:{},shape=circle, scale=1.3,thick,draw] (00) at (0,0) {};
\foreach \i in {45,135,225,315}
{
\draw [thick] (00) -- (A\i);
}
\node[label=above:{},shape=circle, semithick,position=90:{0.4\nodeDist} from 0,scale=1.6] () {\small $s_n$};
\end{tikzpicture}
}
\caption{From the cavity graph of Fig.~\ref{fig:cavGraph} it is possible to obtain a Bethe lattice with $N+1$ nodes by adding a spin $s_n$, namely by setting $\tilde{s}_a\equiv\tilde{s}_b\equiv\tilde{s}_c\equiv\tilde{s}_d\equiv s_n$. \label{fig:cavGraphTonodeToEdge-2}}
\end{figure}

In the glassy phase ($p\geq p_c$) the phase space breaks into an exponential number $\mathcal{N}$ of ergodic components:
\begin{equation}
\label{eq:numbErgComps}
\mathcal{N}\approx e^{N\Sigma}.
\end{equation}
In this section we compute the so-called configurational entropy $\Sigma$ of the FAM. 

\begin{figure}
\includegraphics[width=0.48\textwidth]{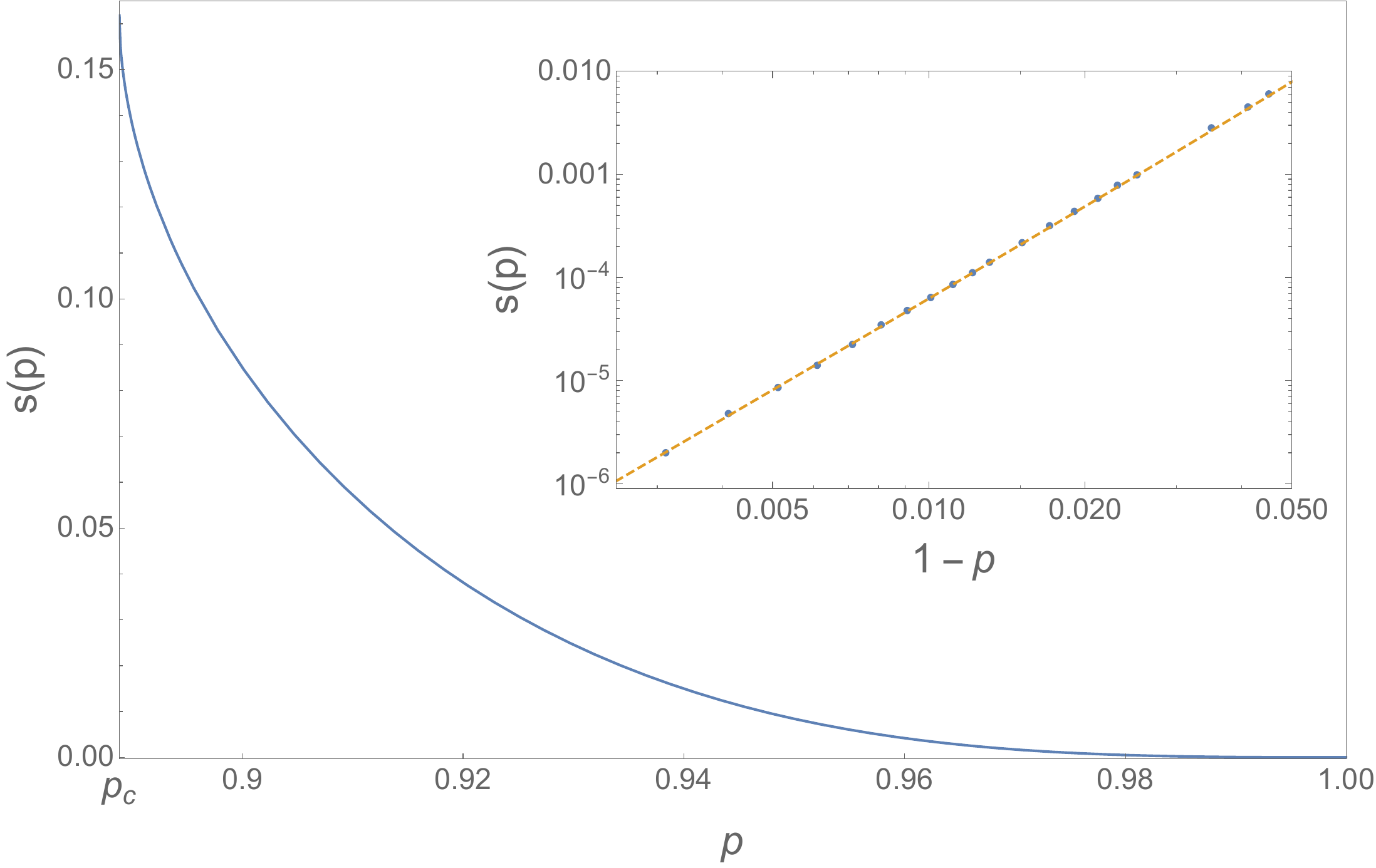}
\caption{Average entropy density $s$ of the state as a function of $p$. The continuous line is obtained by solving the cavity equation by a PDA with population size $N=10^7$. \emph{Inset}: the cavity estimates (dots) of the average entropy density are compared with the low temperature ($p\approx 1$) expansion (dashed line) discussed in App.~\ref{sec:appLowT} (see Eq.~\eqref{eq:expEntr}).\label{fig:EntroDensity}}
\end{figure}

\begin{figure}
\includegraphics[width=0.48\textwidth]{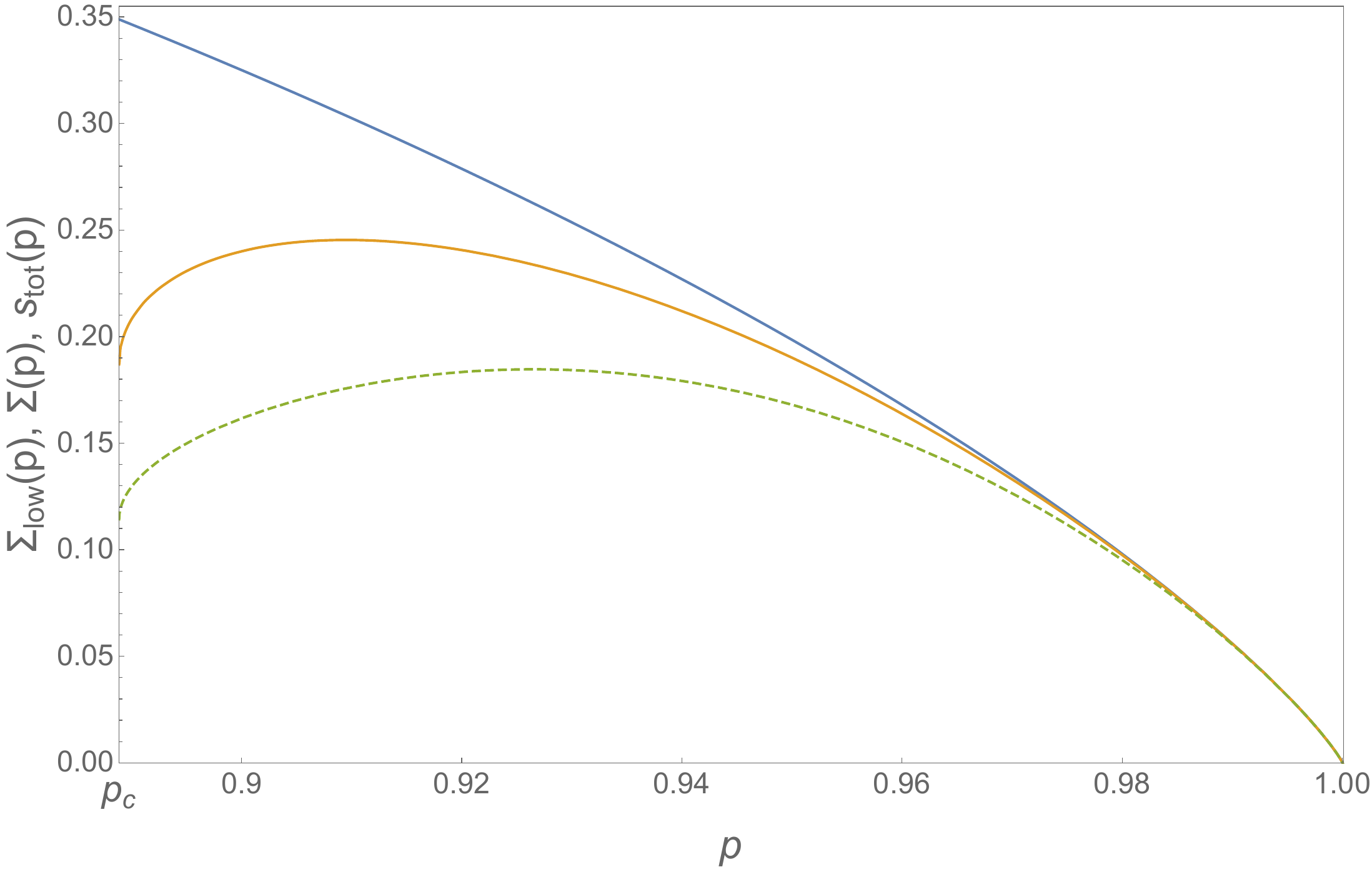}
\caption{\emph{From bottom to top}: lower bound $\Sigma_{low}(p)$ to the configurational entropy ($\Sigma_{low}(p)$ is obtained by using the upper bound for $\phi_f$ of Eq.~\eqref{eq:upperBound} into Eq.~\eqref{eq:confEntropy}), configurational entropy $\Sigma(p)$, and total entropy $s_{tot}(p)$.\label{fig:EntroConfigEntro}}
\end{figure}

Let us start from the computation of the {\it free entropy} $\Upphi$ (that is equal to the free energy up to a factor -$\beta$):
\begin{equation}
\label{eq:freeEntropy}
\Upphi=\mathds{E}_{C}\log{Z_{C}}.
\end{equation}
The fundamental idea \cite{Mezard2001,Mezard2003} is to define a cavity graph (see Fig.~\ref{fig:cavGraph}) in which there are $N$ spins with $z$ neighbors, and $z$ ``cavity'' spins $\tilde{s}_i$, $i=1,\dots,z$, with one neighbor only. Let us call $s_1,\dots,s_{z}$ the neighbors of, respectively, $\tilde{s}_1,\dots,\tilde{s}_{z}$. From this cavity graph we can define two new graphs: one by the addition of one node, that we call node-graph, and the other, that we call edge-graph, by the addition of $z/2$ edges (see Figs.~\ref{fig:cavGraphTonodeToEdge-1} and \ref{fig:cavGraphTonodeToEdge-2}). Note that if $z$ is odd one should start from a cavity graph with $2z$ spins. This can be done straightforwardly, and leads to the same formulas that we will find shortly. The node-graph has $N+1$ spins with $z$ neighbors, and zero cavity spins, and is obtained by setting $\tilde{s}_1\equiv\tilde{s}_2\equiv\dots\equiv\tilde{s}_z$. We call $s_n$ the new spin. The edge-graph has $N$ spins with $z$ neighbors, and zero cavity spins, and is obtained by setting $\tilde{s}_1\equiv s_2$ and $\tilde{s}_2\equiv s_1$, $\tilde{s}_3\equiv s_4$ and $\tilde{s}_4\equiv s_3$, etc. We introduce the free entropies  $\Upphi_C^{(n)}$ and $\Upphi_C^{(e)}$ of, respectively, the node-graph and the edge-graph, given the initial condition $C$:
\begin{equation}
\Upphi_C^{(n)}=\log Z_C^{(n)},\quad \Upphi_C^{(e)}=\log Z_C^{(e)},
\end{equation}
where $ Z_C^{(n)}$ and $Z_C^{(e)}$ are the respective partition functions. We call $\Upphi^{(n)}$ and $\Upphi^{(e)}$ the average of $\Upphi_C^{(n)}$ and $\Upphi_C^{(e)}$ over $C$. Thanks to extensivity, the difference of the two average free entropies equals the free entropy density $\phi$ in the thermodynamic limit:
\begin{equation}
\label{eq:DifferenceAvFreeEntropies}
\Upphi^{(n)}-\Upphi^{(e)}\approx \phi (N+1)-\phi N=\phi,
\end{equation}
where
\begin{equation}
\label{eq:freeEntropy-1}
\phi=\lim_{N\rightarrow \infty}\frac{1}{N}\,\mathds{E}_{C}\log{Z_{C}}.
\end{equation}
As we will see the difference $\Upphi^{(n)}-\Upphi^{(e)}$ can be easily written in terms of quantities that we can compute iteratively on the cavity graph. Let us start with the node-graph. We denote by $Z_C^{(n)}(\tau_n)$ the partition function conditioned to $s_n=\tau_n$. If, starting from configuration $C$, $s_n$ is blocked up we have:
\begin{equation}
\label{eq:frozUpNodGr}
\Upphi_C^{(n)}=\log{Z_C^{(n)}(1)}=-\beta+\sum_{i\in\partial n}\log{\tilde{Z}_i}, 
\end{equation}
else if it is blocked down:
\begin{equation}
\label{eq:frozDWNNodGr}
\Upphi_C^{(n)}=\log{Z_C^{(n)}(-1)}=\beta+\sum_{i\in\partial n}\log{Q_i}+\sum_{i\in\partial n}\log{\tilde{Z}_i}.
\end{equation}
In Eq.~\eqref{eq:frozDWNNodGr} we introduced a new field $Q_i\equiv \bar{Z}_i/\tilde{Z}_i$, that is the partition function on branch $i$ (as usual without Boltzmann term associated with the root) of all the configurations that can be reached when the root (in this case $s_n$) is conditioned down, divided by $\tilde{Z}_i$. If, starting from $C$, $s_n$ is free to move we find:
\begin{widetext}
\begin{multline}
\Upphi_C^{(n)}=\log{\Big(Z_C^{(n)}(1)+Z_C^{(n)}(-1)\Big)}=\log{\left(e^{-\beta}\prod_{i\in\partial n}\tilde{Z}_i+e^{\beta}\prod_{i\in\partial n}(\tilde{Z}_i-\hat{Z}_i)\right)}=\\=\log{\Big(e^{-\beta}+e^{\beta}(1-R_{site}\big(\{R_i\}_i^z)\big)\Big)}+\sum_{i\in\partial n}\log{\tilde{Z}_i},
\end{multline}
\end{widetext}
where $\{R_i\}_i^z=R_1,\dots,R_z$ are the cavity fields entering site $s_n$. Now let us consider the edge-graph. Let us take two spins $s_a$ and $s_b$ that have been linked in the edge-graph. We denote by $Z_C^{(e)}(\tau_a,\tau_b)$ the partition function conditioned to $s_a=\tau_a,s_b=\tau_b$. If, starting from $C$, both $s_a$ and $s_b$ are blocked up we have:
\begin{equation}
\Upphi_C^{(e)}=\log{Z_C^{(e)}(1,1)}=\log \tilde{Z}^a+\log \tilde{Z}^b,
\end{equation}
else if they are both blocked down:
\begin{multline}
\Upphi_C^{(e)}=\log{Z_C^{(e)}(-1,-1)}=\\=\log Q^a+\log Q^b+\log \tilde{Z}^a+\log \tilde{Z}^b,
\end{multline}
else if $s_a$ is blocked up and $s_b$ is blocked down:
\begin{equation}
\Upphi_C^{(e)}=\log{Z_C^{(e)}(1,-1)}=\log Q^a+\log \tilde{Z}^a+\log \tilde{Z}^b,
\end{equation}
else if $s_a$ is free and $s_b$ is blocked up:
\begin{equation}
\Upphi_C^{(e)}=\log{\Big(Z_C^{(e)}(1,1)+Z_C^{(e)}(-1,1)\Big)}=\log \tilde{Z}^a+\log \tilde{Z}^b,
\end{equation}
else if $s_a$ is free and $s_b$ is blocked down:
\begin{multline}
\Upphi_C^{(e)}(C)=\log{\Big(Z_C^{(e)}(1,-1)+Z_C^{(e)}(-1,-1)\Big)}=\\=\log Q^b+\log \tilde{Z}^a+\log \tilde{Z}^b.
\end{multline}
Lastly, if they are both free:
\begin{widetext}
\begin{multline}
\Upphi_C^{(e)}=\log{\Big(Z_C^{(e)}(1,1)+Z_C^{(e)}(1,-1)+Z_C^{(e)}(-1,1)+Z_C^{(e)}(-1,-1)\Big)}=\log{\tilde{Z}^a}+\log{\tilde{Z}^b}+\\+\log{\left(\frac{e^{-2\beta}+(1-\prod_{i\in\partial a\setminus b}R_{i\rightarrow a})+(1-\prod_{i\in\partial b\setminus a}R_{i\rightarrow b})+e^{2\beta}G_e\left(\{\underline{R}_{i\rightarrow a}\}_{i\in \partial a\setminus b},\{\underline{R}_{j\rightarrow a}\}_{j\in\partial b\setminus a}\right)}{\big(e^{-\beta}+e^{\beta}(1-\prod_{i\in\partial a\setminus b}R_{i\rightarrow a})\big)\big(e^{-\beta}+e^{\beta}(1-\prod_{i\in\partial b\setminus a}R_{i\rightarrow b})\big)}\right )},
\end{multline}
\end{widetext}
where the last term is obtained by adding and subtracting:
\begin{equation}
\log{\tilde{Z}^a\tilde{Z}^b}+\log{\tilde{Z}^{\rightarrow a}\tilde{Z}^{\rightarrow b}}.
\end{equation}
At this point by averaging $\Upphi_n(C)$ and $\Upphi_e(C)$ over $C$, we can compute the difference between the average free entropies, $\Upphi_n-\Upphi_e$, as a function of the fields on the branches. The crucial assumption is again the equivalence of the states, implying that the distribution of the cavity field does not depend on $C$. Using Eq.~\eqref{eq:DifferenceAvFreeEntropies} we find:
\begin{equation}
\phi=\phi_f-\beta (P^{+}_{site}-P^{-}_{site}),
\end{equation}
where $\phi_f$, the free entropy density of the free spins, is expressed in terms of local quantities:
\begin{equation}
\phi_f=\Delta \phi_f^{(n)}-\frac{z}{2}\Delta\phi_f^{(e)}.
\end{equation}
In particular the ``node'' free entropy shift $\Delta \phi_f^{(n)}$ is given by:
\begin{equation}
\label{eq:DPhin}
\Delta \phi_f^{(n)}=P_n\left[\log{\Big(e^{-\beta}+e^{\beta}\big(1-R_{site}(\{R_i\}_i^z)\big)\Big)}\right],
\end{equation}
where $P_n=1-P^{+}_{site}-P^{-}_{site}$ is the probability of a free spin, and the square brackets denote the average with respect to the cavity fields entering a free spin. The ``edge'' free entropy shift $\Delta \phi_f^{(e)}$ is
\begin{widetext}
\begin{equation}
\label{eq:DPhie}
\Delta \phi_f^{(e)}=P_e\left[\log{\left(\frac{e^{-2\beta}+(1-\prod_{i\in\partial a\setminus b}R_{i\rightarrow a})+(1-\prod_{i\in\partial b\setminus a}R_{i\rightarrow b})+e^{2\beta}G_e\left(\{\underline{R}_{i\rightarrow a}\}_{i\in \partial a\setminus b},\{\underline{R}_{j\rightarrow a}\}_{j\in\partial b\setminus a}\right)}{\big(e^{-\beta}+e^{\beta}(1-\prod_{i\in\partial a\setminus b}R_{i\rightarrow a})\big)\big(e^{-\beta}+e^{\beta}(1-\prod_{i\in\partial b\setminus a}R_{i\rightarrow b})\big)}\right)}\right],
\end{equation}
\end{widetext}
where the square brackets denote the average with respect to the cavity fields entering two neighboring free sites, and $P_e$, the probability that two neighboring spins are both free, can be easily computed from Eqs.~\eqref{eq:PsiteFrozendown} and \eqref{eq:PsiteFrozenup}. It is useful to introduce also the average entropy density $s$ of a state, that is given by 
\begin{equation}
\label{eq:entropyEq}
s=\beta u+\phi=\beta u_f+\phi_f,
\end{equation}
where $u$ is the energy density, and $u_f$ is the energy density of the free spins, that is given by:
\begin{equation}
\label{eq:EnDensFreeSpins}
u_f=\frac{e^{-2\beta}-1}{e^{-2\beta}+1}-(P^{+}_{site}-P^{-}_{site}).
\end{equation}
From Eq.~\eqref{eq:entropyEq} we obtain:
\begin{equation}
\label{eq:entroDens}
s=\beta\left(\frac{e^{-2\beta}-1}{e^{-2\beta}+1}-(P^{+}_{site}-P^{-}_{site})\right)+\phi_f.
\end{equation}
Note that by setting all the $R_i$'s to zero in Eqs.~\eqref{eq:DPhin} and \eqref{eq:DPhie}, corresponding to all free spins being independent, one obtains $\Delta\phi_f^{(e)}=0$, and $\phi_f=\Delta\phi_f^{(n)}=P_n\log{\left(e^{-\beta}+e^{\beta}\right)}$, that provides a simple upper bound to the free entropy of the free spins:
\begin{equation}
\label{eq:upperBound}
\phi_f=\Delta \phi_f^{(n)}-\frac{z}{2}\Delta \phi_f^{(e)}\leq P_n\log{\left(e^{-\beta}+e^{\beta}\right)}.
\end{equation}
At this point we can easily compute the configurational entropy $\Sigma$. Consider the total entropy $S_{tot}$ associated with the factorized Boltzmann-Gibbs measure $P(s)$ from which the initial condition is drawn:
\begin{multline}
\label{eq:totalEntropy}
S_{tot}=-\sum_{\alpha}\sum_{s\in\alpha}P(s)\log{P(s)}=\\=-\sum_{\alpha}\frac{Z_{\alpha}}{Z}\sum_{s\in\alpha}\frac{Z_{\alpha}(s)}{Z_{\alpha}}\log{\frac{Z_{\alpha}(s)}{Z_{\alpha}}}-\sum_{\alpha}\frac{Z_{\alpha}}{Z}\log{\frac{Z_{\alpha}}{Z}}.
\end{multline}
In Eq.~\eqref{eq:totalEntropy} we indexed with $\alpha$ the states of the system, we called $Z$ the partition function associated with $P(s)$, $Z_{\alpha}$ the partition function restricted to state $\alpha$, and $P_{\alpha}(s)$ the measure restricted to state $\alpha$. The first term on the second line of \eqref{eq:totalEntropy} corresponds to the average entropy inside a state. The last term of \eqref{eq:totalEntropy} divided by $N$ gives the configurational entropy $\Sigma$:
\begin{equation}
\Sigma=\lim_{N\rightarrow \infty}-\frac{1}{N}\sum_{\alpha}\frac{Z_{\alpha}}{Z}\log{\frac{Z_{\alpha}}{Z}}.
\end{equation}
Under the assumption of equivalence of the states, $\Sigma$ counts the number of ergodic components (Eq.~\eqref{eq:numbErgComps}). Since 
\begin{equation}
s_{tot}=\lim_{N\rightarrow\infty}\frac{S_{tot}}{N}=-p\log{(p)}-(1-p)\log{(1-p)},
\end{equation}
using Eqs.~\eqref{eq:entropyEq} and \eqref{eq:totalEntropy} we obtain:
\begin{multline}
\label{eq:confEntropy}
\Sigma(p)=-p\log{(p)}-(1-p)\log{(1-p)}+\\-\beta\left(\frac{e^{-2\beta}-1}{e^{-2\beta}+1}-(P^{+}_{site}-P^{-}_{site})\right)-\phi_f.
\end{multline}
In Figs.~\ref{fig:EntroDensity} and \ref{fig:EntroConfigEntro} we show, respectively, the average entropy density of a state $s(p)$, and the configurational entropy $\Sigma(p)$, that are computed in the case $z=4$, $f=2$ solving the iterative equation \eqref{eq:Equatriple} by means of a population dynamics algorithm, and then by using \eqref{eq:entroDens} and \eqref{eq:confEntropy}.

\begin{figure}
\includegraphics[width=0.48\textwidth]{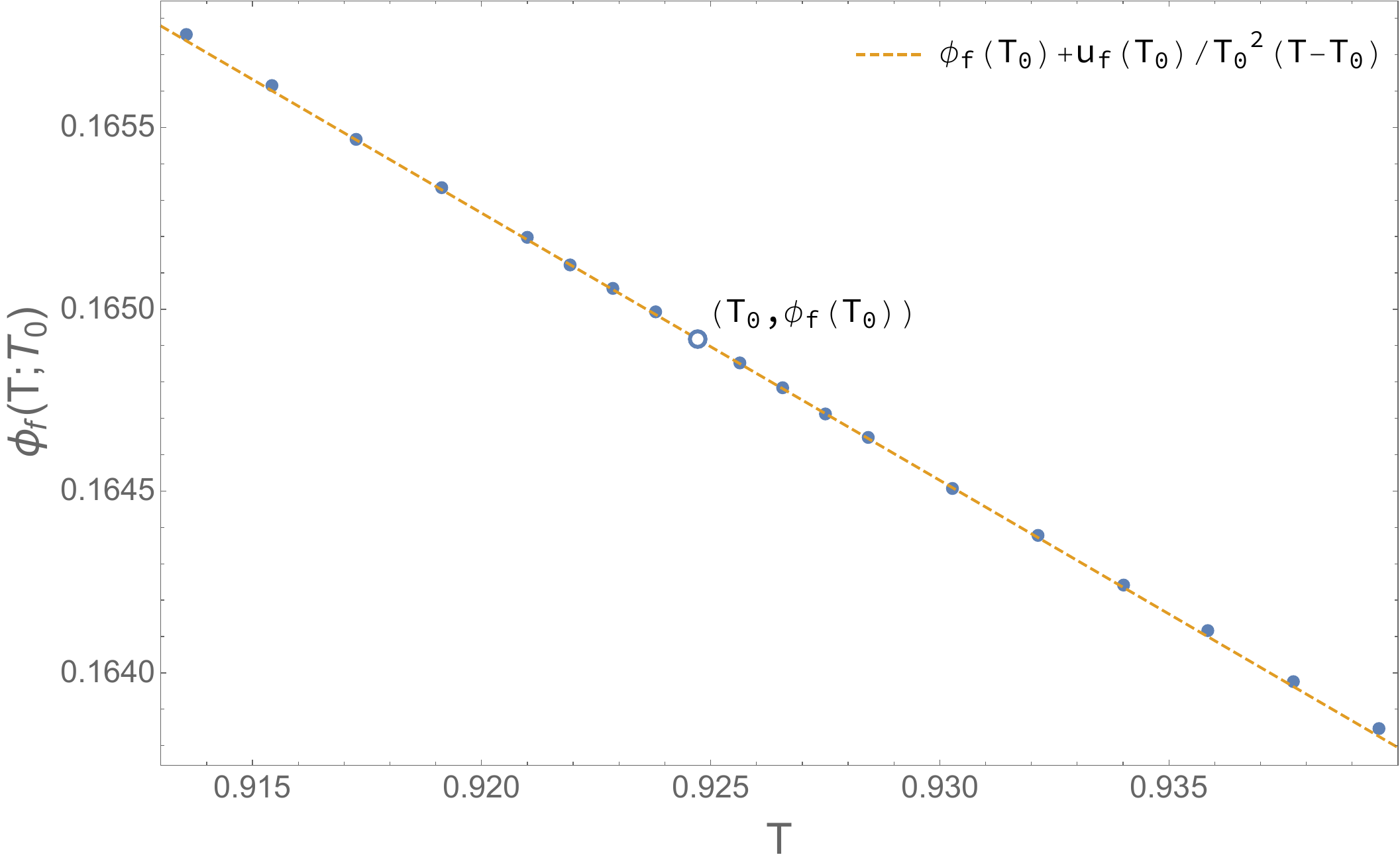}
\caption{$T_0$-conditioned free entropy of the free spins $\phi_f(T;T_0)$ as a function of $T$. The points correspond to the cavity estimates obtained with a PDA with a population of size $N=10^8$. The slope close to $T_0$ is consistent with the expectation $u_f(T_0)/T_0^2$ (see Eqs.~\eqref{eq:EnDensFreeSpins} and \eqref{eq:identFreeEntr}). In this example $T_0=0.924717$.\label{fig:checkFreeEntropy}}
\end{figure}

At this point we want to check the consistency of the cavity estimate for $s(p)$, verifying that:
\begin{equation}
\label{eq:identFreeEntr}
u_f(T)/T^2=\lim_{N\rightarrow\infty}\frac{1}{N}\mathds{E}\frac{d}{d T}\log{Z}.
\end{equation}
The energy density of the free spins $u_f(T)$ is given by formula \eqref{eq:EnDensFreeSpins}, while the quantity on the RHS can be computed starting from our cavity estimate of $\phi_f$ as follows. Note that the derivative on the RHS of Eq.~\eqref{eq:identFreeEntr} affects only the free spins, because it is taken \emph{inside} the state, \emph{i.e.}\ before averaging over the initial configuration, and therefore the RHS is different from the derivative of the free entropy $d\phi/dT$:
\begin{equation}
\label{eq:IdTot}
\begin{split}
\frac{d \phi}{d T}&=\lim_{N\rightarrow \infty}\frac{1}{N}\frac{d}{d T}\mathds{E}\log{Z}=\\&=\beta^2u_f-\lim_{N\rightarrow \infty}\frac{\beta^2}{N}\left(\sum_{\alpha}\frac{Z_{\alpha}}{Z}U_{\alpha}\log Z_{\alpha}-U\Upphi\right).
\end{split}
\end{equation}
In Eq.~\eqref{eq:IdTot} we denoted with $U_{\alpha}$ the internal energy in state $\alpha$, and by $U=\sum_{\alpha}Z_{\alpha}/Z\,U_{\alpha}$ the average internal energy. Note that there is one more term w.r.t.\ Eq.~\eqref{eq:identFreeEntr}, that comes out from the derivative of the measure over the initial configuration. In order to compute the derivative inside the state we should derive the free entropy keeping fixed the measure over the initial configuration. This can be easily done by a simple modification of the iterative scheme presented in section \ref{sec:statSol}. In particular one has to define two temperatures: $T_0$ and $T$. $T_0$ is the temperature at which the initial configuration is extracted, \emph{i.e.}\ it is the temperature that determines the size of the cluster of blocked spins, while $T$ is the temperature of the dynamics. In practice the couples $(\eta,\mu)$, that determine the probabilities P,D (see Eq.~\eqref{eq:BPeqZ4m2}), should be iterated with the temperature $T_0$, while the cavity field $R$, that determines the properties of the free spins, should be iterated using $T$. In this way we can compute by the iterative method of this section the conditioned free entropy $\phi(T;T_0)$ at temperature $T$, given an initial condition that is extracted at temperature $T_0$. We have that:
\begin{equation}
\label{eq:ufCond}
u_f(T_0)/T_0^2=\frac{d\phi(T;T_0)}{dT}\big|_{T=T_0}=\frac{d\phi_f(T;T_0)}{dT}\big|_{T=T_0},
\end{equation}
where the second equality comes from the fact that the derivative at fixed initial condition only affects the free spins. In Fig.~\ref{fig:checkFreeEntropy} we test the consistency of our cavity estimates checking that Eq.~\eqref{eq:ufCond} is verified. For a similar temperature conditioning applied to the calculation of another observable, the specific heat, we direct the reader to Sec.~\ref{sec:SpecHeat} (see Fig.~\ref{fig:checkFreeEntropy}).
\begin{figure}
\includegraphics[width=0.48\textwidth]{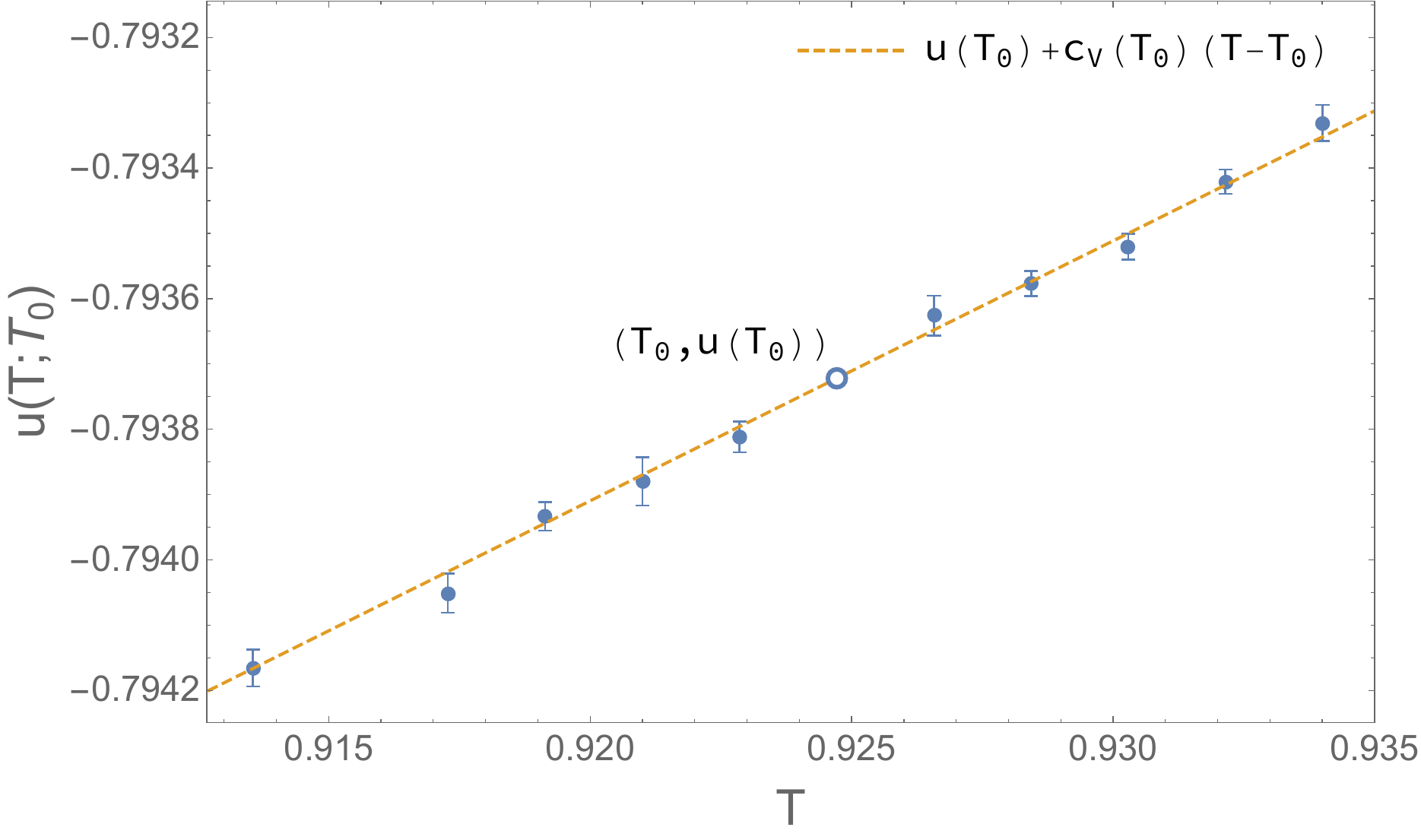}
\caption{$T_0$-conditioned energy density obtained from numerical simulations with a system of size $N=4.5\times 10^7$. Each point corresponds to the average over $10$ independent samples. The dashed line is the cavity estimate. In particular $u(T_0)=-\tanh{1/T_0}$, and the slope in $T_0$ is given by the specific heat $c_V(T_0)$ (see Eqs.~\eqref{eq:iterForRprime} and \eqref{eq:specificHeatCcomput}). In this example $T_0=0.924717$.\label{fig:specHeat}}
\end{figure}
\begin{figure}
\includegraphics[width=0.48\textwidth]{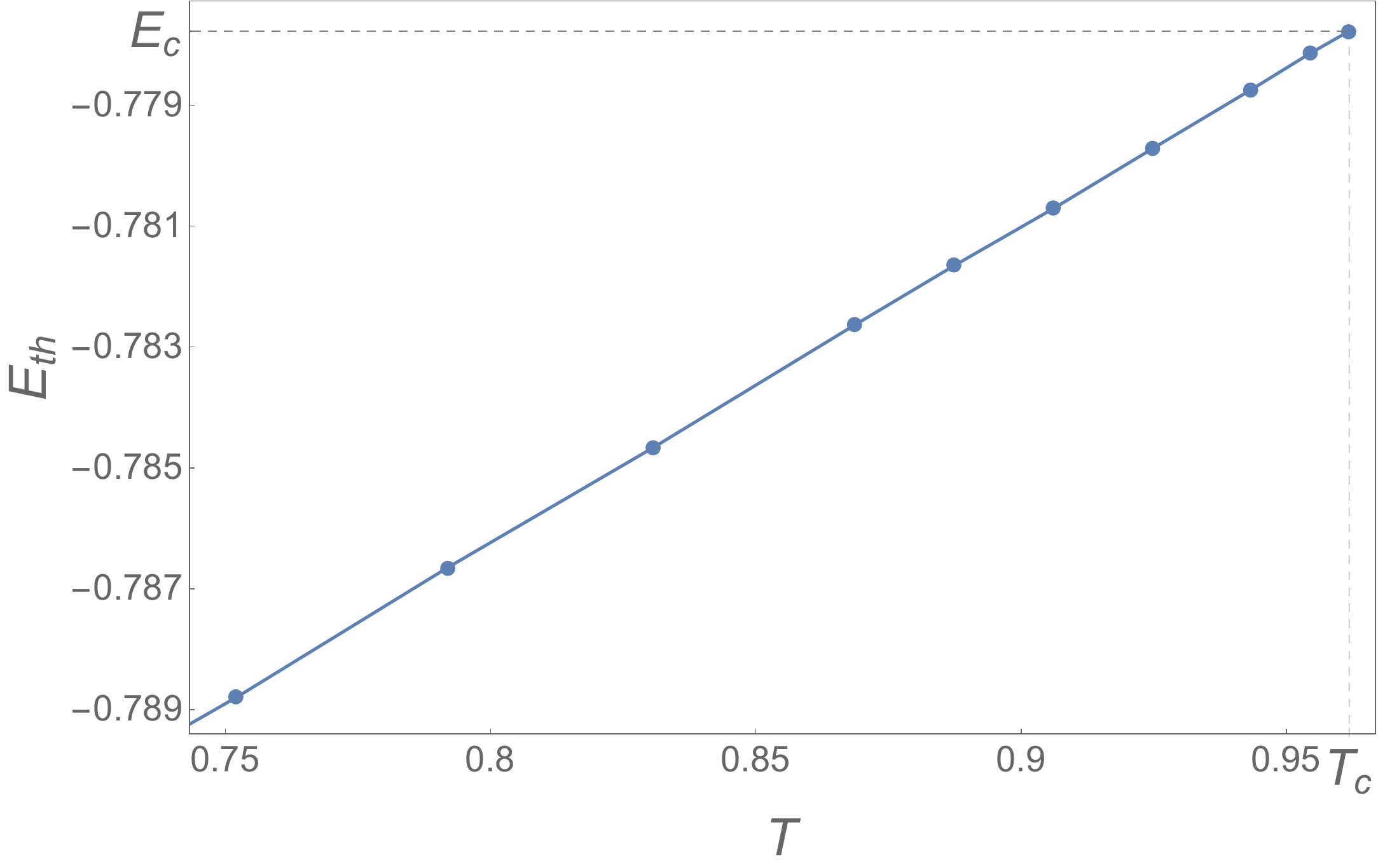}
\caption{Threshold energy as a function of temperature, computed with a population dynamics algorithm with a population of size $N=10^7$. The estimates are obtained by solving the cavity equations with two temperatures, one corresponding to the temperature of the dynamics of the free spins ($T$), and the other to the temperature at which the initial condition is extracted, that in this case corresponds to the critical temperature $T_c$ (see the text). The continuous line is an interpolation between the points. With our notations the critical temperature and the critical energy are, respectively, $T_c=2/log{(8)}$ and $E_c=-7/9$.\label{fig:thresholdE}}
\end{figure}

\subsection{The Threshold Energy}
\label{sec:thresholdE}
As we mentioned in the introduction, there are remarkable similarities between the FA model and mean-field SG models, more specifically those displaying one step of Parisi's Replica-Symmetry-Breaking (1RSB).
In \cite{sellitto2005facilitated} it has been shown that the similarities extend to off-equilibrium dynamics and more specifically to aging behavior. At equilibrium 1RSB-SG diplay a MCT-like transition at the so-called dynamical temperature $T_d$ \cite{kirkpatrick1987dynamics,kirkpatrick1987p,crisanti1993spherical}. Their off-equilibrium dynamics following a quench to a temperature below $T_d$ displays aging \cite{bouchaud1998out}: the system is unable to equilibrate and ages indefinitely. In particular, after a transient regime the energy $E(t)$ approaches at large times an asymptotic  value $E(\infty)$ larger than the equilibrium value. Cugliandolo and Kurchan \cite{cugliandolo1993analytical} noticed that in a notable case the asymptotic value coincides with the energy of the most numerous metastable states, the so-called threshold energy $E_{th}$ \footnote{The role of the threshold energy in off-equilibrium dynamics has been reassessed in recent literature \cite{folena2020rethinking,folena2023on}.}. 

Sellitto {\it et al.}  have shown  numerically \cite{sellitto2005facilitated} that off-equilibrium dynamics of the FA model after a quench below $T_c$ displays aging behavior qualitatively similar to that observed in mean-field Spin-Glasses. Furthermore they put forward the hypothesis that $E(\infty)$ can be identified with the threshold energy {\it defined} as the energy that an equilibrium configuration at temperature $T_c^-$ attains when it is cooled down to the quench temperature $T<T_c$. They also provided numerical evidences for the validity of this scenario by numerical simulations. Be as it may, in \cite{sellitto2005facilitated} the threshold energy itself had to be computed numerically by cooling equilibrium configurations, while the formalism we have developed here allows to compute it analytically by considering the $T_0$-conditioned setting used in the last section (see also Sec.~\ref{sec:SpecHeat}). The results are plotted in Fig.~\ref{fig:thresholdE}.
We note that to compare with their data (Fig.~6 in \cite{sellitto2005facilitated}) one should note that our definition of the energy differs by a factor two with respect to theirs, and thus the critical temperature is also two times larger.  
\subsection{The Spin-Glass Susceptibility}
\label{sec:IntroSGSusc}
\begin{figure}
\includegraphics[width=0.48\textwidth]{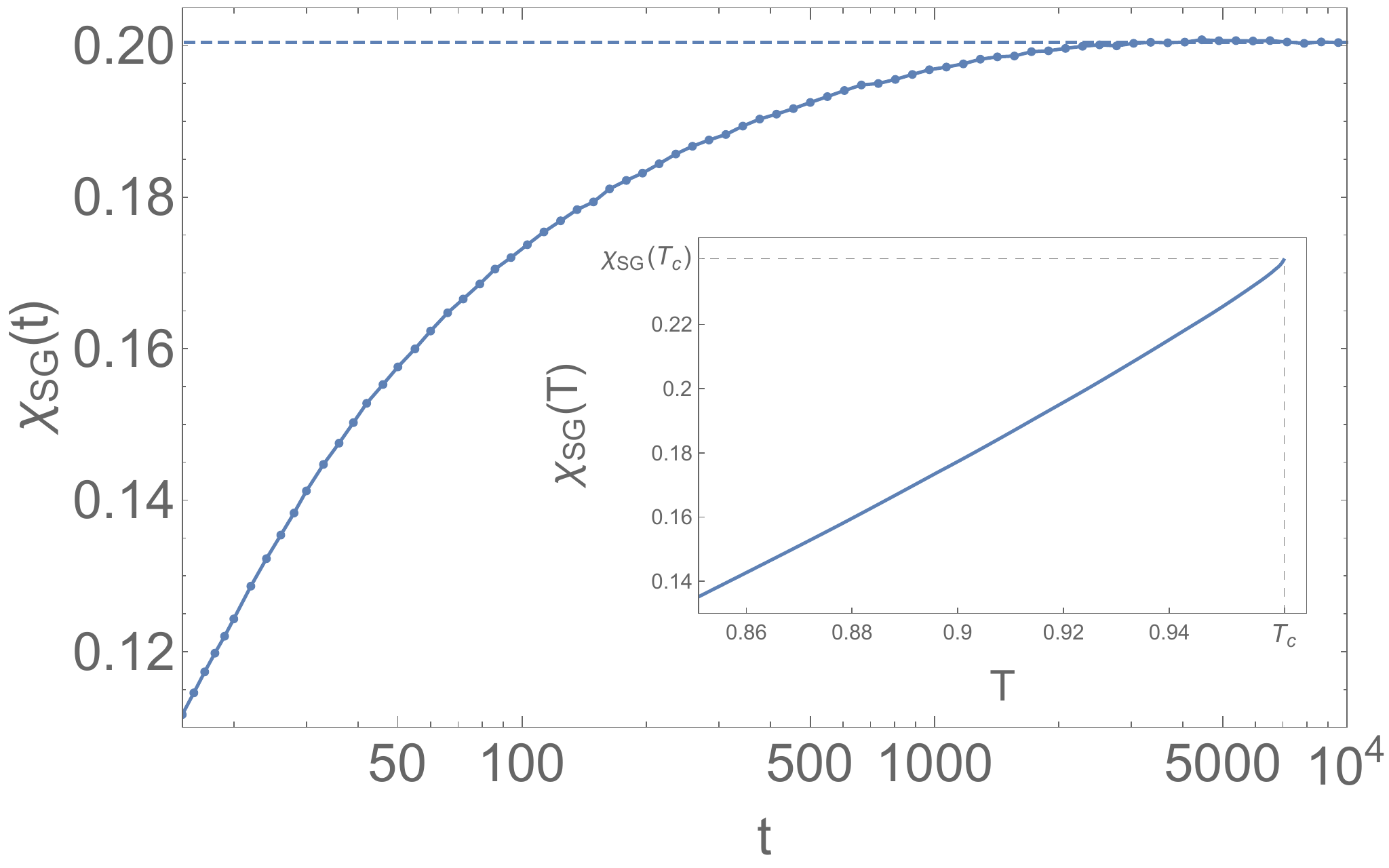}
\caption{Spin-glass susceptibility $\chi_{SG}$ as a function of time $t$, for $e^{-2\beta}=0.115$ and $z=4,f=2$. The dots (joined with a continuous line) correspond to numerical simulations with $N=4.5\cdot 10^5$, and are obtained by averaging over $7.5 \times 10^6$ samples. The dashed horizontal line is the cavity prediction. \emph{Inset}: cavity estimate of $\chi_{SG}$ as a function of temperature $T=1/\beta$. Note that $\chi_{SG}$ is finite at $T_c$.\label{fig:dataChiSG}}
\end{figure}

In this section we study the SG susceptibility $\chi_{SG}$,
\begin{equation}
\label{eq:defSpGlSus}
\chi_{SG}=\frac{1}{N}\,\mathds{E}\sum_{i,j\in\mathcal{V}}\langle s_is_j\rangle_{c}^2,
\end{equation}
leaving the details of the computation to App.~\ref{sec:AnalyticalCompSGSusc}. As usual, in Eq.~\eqref{eq:defSpGlSus} we denoted by $\langle\bullet\rangle$ the thermal average, and by $\mathds{E}$ the average over disorder (in our case the initial configuration). On a tree $\chi_{SG}$ can be expressed in terms of a conditioned susceptibility $\chi(\eta,\mu,R)$ defined on the cavity graph. In particular we have
\begin{equation}
\label{eq:LinearEqChiSg}
\chi_{SG}=\Gamma_0+\mathbf{G}\chi,
\end{equation}
where $\mathbf{G}$ is a linear operator that maps $\chi$ into a number. Both $\mathbf{G}$ and $\Gamma_0$ are determined by the order parameter $P(\eta,\mu,R)$ (see App.~\ref{sec:AnalyticalCompSGSusc}). The cavity susceptibility $\chi$ satisfies a fixed-point equation that takes the form:
\begin{equation}
\label{eq:LinearEqChi}
\chi=\xi_0+\mathbf{F}\chi,
\end{equation}
where $\mathbf{F}$ is a linear operator mapping $\chi$ into a function of $(\eta,\mu,R)$, and $\xi_0=\xi_0(\eta,\mu,R)$ is a function. Both $\mathbf{F}$ and $\xi_0$ are determined by the order parameter $P(\eta,\mu,R)$. If $(\mathds{1}-\mathbf{F})$ is invertible in the subspace to which $\xi_0$ belongs, we have
\begin{equation}
\label{eq:inversion}
\chi=(\mathds{1}-\mathbf{F})^{-1}\xi_0.
\end{equation}
In Fig.~\ref{fig:dataChiSG} we show a comparison between numerical simulations and the cavity predictions, obtained discretizing the operator $\mathbf{F}$, computing $\chi$ through \eqref{eq:inversion} and then computing $\chi_{SG}$ through \eqref{eq:LinearEqChiSg}. In App.~\ref{sec:NumericalCompSGSusc} we discuss the numerical technique we used for the measure of $\chi_{SG}$. 

Despite the overlap behaving similarly to the persistence function, \emph{i.e.}\ it has a jump at $p_c$ and a square root singularity above $p_c$, their fluctuations are different, as discussed in \cite{franz2013finite}. In particular $(\mathds{1}-\mathbf{F})$ is always invertible, implying that $\chi$ and $\chi_{SG}$ are finite at the transition (see Fig.~\ref{fig:dataChiSG}).

\subsection{Specific Heat}
\label{sec:SpecHeat}
In this section we compute the average thermal fluctuation inside a state. These are given by the specific heat $c_V$:
\begin{equation}
\label{eq:specHeat}
c_V=\lim_{N\rightarrow\infty}\frac{1}{N}\sum_{\alpha}\frac{Z_{\alpha}}{Z}\frac{d U_{\alpha}}{d T}.
\end{equation}
Equation \eqref{eq:specHeat} can be computed following the same observations given for the computation of the derivative of the free entropy at fixed initial condition (see Eq.~\eqref{eq:identFreeEntr}). The starting point is the average $T_0$-conditioned energy density $u(T,T_0)$, where as before $T_0$ is the temperature at which the initial condition is drawn, and $T$ is the temperature of the dynamics (the one entering Eq.~\eqref{eq:MetropRule}). The specific heat $c_V$ is simply given by:
\begin{equation}
\label{eq:compCV}
c_V(T_0)=\frac{d u(T;T_0)}{d T}\bigg|_{T=T_0}=\frac{d m(T;T_0)}{d T}\bigg|_{T=T_0}.
\end{equation}
The derivative on the RHS of Eq.~\eqref{eq:compCV} can be obtained either by computing $m(T;T_0)$ for different $T$, and then by estimating the slope at $T_0$ (like we did for $\phi_f$ in Fig.~\ref{fig:checkFreeEntropy}), or it can be computed self-consistently, as we want to show in this section. From Eq.~\eqref{eq:magnFreeSpins} we have that on a free site $0$:
\begin{equation}
\frac{d m_0}{d T}=\frac{\partial m_0}{\partial R}\left(\sum_{i\in\partial a}\frac{d R}{d R_{i\rightarrow a}}\frac{d R_{i\rightarrow a}}{d T}\right)+\frac{\partial m_0}{\partial T},
\end{equation}
from which we obtain
\begin{equation}
\label{eq:specificHeatCcomput}
c_V=z \left[\frac{\partial m_0}{\partial R} \frac{d R(\{R_i\}_i^z)}{d R_1}R_1'\right]+\left[\frac{\partial m_0}{\partial T}\right],
\end{equation}
where the square brackets is the average over all the free sites, and the cavity field $R_1'$, representing the perturbation of a cavity field $R$ w.r.t.\ an infinitesimal change of the temperature, can be computed self-consistently. Indeed differentiating the iterative rule of the cavity field (see Eq.~\eqref{eq:iterRule}) we find the following recursion:
\begin{equation}
\label{eq:RecursionPerturb}
\begin{split}
R'_{i\rightarrow j}&=\sum_{k\in\partial i\setminus j}\frac{dR_{i\rightarrow j}}{dR_{k\rightarrow i}}R_{k\rightarrow i}'+\frac{\partial R_{i\rightarrow j}}{\partial T}\\
&\equiv V\left(\{R_{k\rightarrow i},R_{k\rightarrow i}'\}_{k\in\partial i\setminus j}\right),
\end{split}
\end{equation}
where for $k\in\partial i\setminus j$:
\begin{equation}
\frac{dR_{i\rightarrow j}}{dR_{k\rightarrow i}}=\frac{d R_{iter}\left(\{R_{\ell\rightarrow i}\}_{\ell\in\partial i\setminus j},T\right)}{d R_{k\rightarrow i}},
\end{equation}
\begin{equation}
\frac{\partial R_{i\rightarrow j}}{\partial T}=\frac{\partial R_{iter}\left(\{R_{\ell\rightarrow i}\}_{\ell\in\partial i\setminus j},T\right)}{\partial T},
\end{equation}
and $R_{iter}$ is defined in Eq.~\eqref{eq:iterRule}. From Eq.~\eqref{eq:RecursionPerturb} we find the following iterative equation for the joint distribution of the cavity fields
\begin{widetext}
\begin{multline}
\label{eq:iterForRprime}
P(\eta,\mu,R,R') = \mathds{E}_{s_n}\left(\prod_{i=1}^{c} \int dP(\eta_i,\mu_i,R_i,R_i')\right)\,\delta\Big(R'-V\left(\{R_{i},R_{i}'\}_i^c\right)\Big)\times\\\times\delta^{(k)}\Big(\eta-T_{s_n}(\{\eta_i,\mu_i\}_i^c)\Big)\,\delta^{(k)}\Big(\mu-M_{s_n}(\{\eta_i,\mu_i\}_i^c)\Big)\,\delta\Big(R-F_{s_n}(\{\eta_i,\mu_i,R_i\}_i^c)\bigg).
\end{multline}
\end{widetext}
The solution of \eqref{eq:iterForRprime} allows to compute the specific heat $c_V$ via \eqref{eq:specificHeatCcomput}. In Fig.~\ref{fig:specHeat} we show a comparison between numerical simulations and the cavity prediction for $c_V$.

\section{Conclusions}
\label{sec:Conclusions}

The close relationship between bootstrap percolation and the FAM allows to characterize the properties of the blocked spins in the glassy phase but yields no information on the free spins.
In this work we have solved this problem deriving analytical equations that were then solved by standard population dynamics algorithms. 
We have thus been able to obtain quantitative predictions for several observables and we have shown their validity by comparing them with the outcome of numerical simulations. 

In particular we obtained the distribution of the magnetizations inside a given state and the overlap.
We also computed the entropy of a typical state and the configurational entropy, \emph{i.e.}\ the number of different blocked configuration (the states). The latter was obtained as the difference between the total entropy and the former, thus it would be interesting to obtain it from a direct computation.
The approach was extended to obtain predictions for various quantities of interest: the specific heat, the spin-glass susceptibility and the threshold energy, that is relevant for aging behavior. 

The analytical computation is based on the cavity method (CM), whose application to the FAM is non trivial. Indeed the CM is always applied to \emph{interacting} variables by leveraging the locally-tree-like property of the Bethe lattice. At contrast the Hamiltonian of the FAM is trivial, dynamics is essential and one could have wondered if any thermodynamic quantity could be computed in a purely static approach like the one presented here. 

We have also argued that the cavity equations yield an algorithm that could be useful in a variety of interesting problems.
For instance once an initial configuration is generated it is usually fast to determine the backbone of blocked spins but it is not easy to find a configuration of the free spins uncorrelated with the initial one, because dynamical evolution maybe very slow, especially close to the critical point. Instead the algorithm presented in Sec.~\ref{sec:algosol} allows to easily sample equilibrium configurations by a decimation procedure, thus avoiding dynamical slowing down.
As we mentioned before, the algorithm could be also used to define dynamical moves where, instead of equilibrating a single spin in the environment determined by its neighbors, one equilibrates instantaneously a {\it region} comprising all its neighbors at distance less than a given value $L$ in the environment determined by the region boundary. At fixed value of $L$ this algorithm accelerates the dynamics while keeping its local character, and this could be useful in contexts where the interesting physics occurs at very large times.

Apart from the technical and methodological aspects highlighted in the previous paragraphs we want to emphasize what we believe is the most interesting physical outcome of this study, \emph{i.e.}\ the fact that the Spin-Glass susceptibility remains finite at the critical point.
Indeed in the introduction and in the section on the threshold energy (Sec.~\ref{sec:thresholdE}) we mentioned the close similarity between the dynamics of the FAM and Spin-Glass models with one-step of Parisi's Replica Symmetry breaking. Remarkably this similarity extends to state-to-state fluctuations at criticality, that are essential to understand the behavior of these systems in physical dimensions \cite{Franz2011,rizzo2014long,rizzo2016dynamical,rizzo2019fate,rizzo2020solvable}. However at the level of {\it thermal} fluctuations, \emph{i.e.}\ inside a given state, there is a crucial difference: here we have shown that they remain finite at criticality, confirming earlier numerical observations by Franz and Sellitto \cite{franz2013finite}, while they diverge in the replica approach relevant not only for Spin-Glasses but also for supercooled liquids in infinite dimensions \cite{parisi2020theory}.  

\begin{acknowledgments}
We acknowledge the financial support of the Simons Foundation (Grant No. 454949, Giorgio Parisi). 
\end{acknowledgments}

\appendix 

\section{Iteration Functions}
\label{app:ITfunctions}
The iteration functions $T_{s_n},M_{s_n}$ and $F_{s_n}$ depend on the initial value of $s_n$, and are defined as follows. If $s_n$ is extracted in the up state we have:
\begin{equation}
\label{eq:T1M1}
T_1(\{\eta_i,\mu_i\}_i^c)=0,\quad M_1(\{\eta_i,\mu_i\}_i^c)=0,
\end{equation}
following from the fact that in this case, independently of the value of the cavity triplets entering $s_n$, we have $(\eta,\mu)=(0,0)$. Moreover we have that 
\begin{equation}
\begin{split}
&F_1(\{\eta_i,\mu_i,R_i\}_i^c)=\\&=F_1(\{\mu_i,R_i\}_i^c)=
\begin{cases}
R_{iter}(\{R_i\}_i^c)\quad\textnormal{if}\,\,\sum_{i}\mu_i\leq z-f\\
0 \quad\textnormal{otherwise},
\end{cases}
\end{split}
\end{equation}
where it is enforced that if the sum of the $\mu_i$'s is larger than $z-f$, $s_n$ is blocked up. Note that the function $R_{iter}$ is given by Eq.~\eqref{eq:iterRule} in the case $(4,2)$, and for generic $(z,f)$ can be easily obtained following the scheme discussed in Sec.~\ref{sec:statSol}. 

If $s_n$ is extracted in the down state one has the following cases depending on the values of the entering $\eta_i$'s. If the sum of the $\eta_i$'s is larger than $z-f$, $s_n$ is blocked in the down state, and $(\eta,\mu)=(1,1)$. Vice-versa $s_n$ cannot be blocked down if the root is up, and therefore $\mu=0$. In formulas we have:
\begin{equation}
M_{-1}(\{\eta_i,\mu_i\}_i^c)=M_{-1}(\{\eta_i\}_i^c)=
\begin{cases}
1\,\,\textnormal{if}\,\,\sum_{i}\eta_i>z-f\\
0\,\,\textnormal{otherwise}.
\end{cases}
\end{equation}
If $\sum_i\eta_i>z-f-1$, $s_n$ is blocked down if the root is in the down state, implying $\eta=1$, otherwise $\eta=0$:
\begin{equation}
\begin{split}
&T_{-1}(\{\eta_i,\mu_i\}_i^c)=\\&=T_{-1}(\{\eta_i\}_i^c)=
\begin{cases}
1\,\,\textnormal{if}\,\,\sum_{i}\eta_i>z-f-1\\
0\,\,\textnormal{otherwise},
\end{cases}
\end{split}
\end{equation}
\vspace{0.2cm}
\begin{equation}
\begin{split}
&F_{-1}(\{\eta_i,\mu_i,R_i\}_i^c)=\\&=F_{-1}(\{\eta_i,R_i\}_i^c)=
\begin{cases}
R_{iter}(\{R_i\}_i^c)\,\,\textnormal{if}\,\,\sum_{i}\eta_i\leq z-f\\
1 \,\,\textnormal{otherwise}.
\end{cases}
\end{split}
\end{equation}

\section{Low Temperature Expansion}
\label{sec:appLowT}
At zero temperature all the spins are blocked. If $T$ is small enough we can imagine that most of the spins are blocked, and that the remaining spins, the free ones, belong to small isolated clusters. With this idea in mind in this section we want to compute the low temperature expansion of some of the observables discussed in the previous sections. In the following we refer, as usual, to the case $z=4$ and $f=2$, but the results can be easily generalised to other models. We call $\delta=1-p$ our perturbative parameter. As we will see the leading contributions are order $\delta^2$, however the terms $\delta^2$ cancel out in the expansions of the observables. We will start with a discussion about the computation up to order $\delta^3$, and then we will extend the discussion to order $\delta^4$.

\subsection{The Expansion}
The leading contribution, resulting in a single free spin, is obtained if the spin is initialized down, it has two neighbors that are blocked down, and two neighbors that are blocked up. We represent this occurrence as follows:
\begin{equation}
\begin{gathered}
\scalebox{0.6}{
\begin{tikzpicture}
\node[label=below:{}, shape=circle, scale=1,pattern=north east lines,draw] (0) at (0,0) [right] {};
\node[label=below:{},shape=circle,position=0:{\nodeDist} from 0,scale=1] (dx) {};
\node[label=below:{},shape=circle,position=90:{\nodeDist} from 0,scale=1] (up) {};
\node[label=below:{},shape=circle,position=-90:{\nodeDist} from 0,scale=1] (dwn) {};
\node[label=below:{},shape=circle,position=180:{\nodeDist} from 0,scale=1] (sx) {};
\draw[black,thick] (0) -- (dx);
\draw[black,thick] (0) -- (up);
\draw[black,thick,dashed] (0) -- (sx);
\draw[black,thick,dashed] (0) -- (dwn);
\end{tikzpicture}
}
\end{gathered}
\end{equation}
Here we use the graphical convention of representing with a continuous external leg a neighbor that is blocked up, and with a dashed external leg a neighbor that is blocked down. We also represent as full circles the spins that are initialized down, and as empty circles the spins that are initialized up. Note that each external \emph{continuous} leg gives a contribution $D_{cl}=(1-p)D^3$, that expanded at small temperatures (small $\delta=1-p$) becomes 
\begin{equation}
\label{eq:dclexp}
D_{cl}=\delta-12\,\delta^2+39\,\delta^3-40\,\delta^4+\left(\delta^5\right),
\end{equation}
while each external dashed leg gives a contribution $P_{cl}=p P^3$, leading to:
\begin{equation}
\label{eq:pclexp}
P_{cl}=1-4\,\delta-3\,\delta^2-16\,\delta^3-95\,\delta^4+\left(\delta^5\right).
\end{equation}
The density $n_1$ of the number of clusters of one free spin takes the following form:
\begin{widetext}
\begin{equation}
\label{eq:contrib1spin}
n_1=
\,\, 6\,
\begin{gathered}
\scalebox{0.6}{
\begin{tikzpicture}
\node[label=below:{}, shape=circle, scale=1,pattern=north east lines,draw] (0) at (0,0) [right] {};
\node[label=below:{},shape=circle,position=0:{\nodeDist} from 0,scale=1] (dx) {};
\node[label=below:{},shape=circle,position=90:{\nodeDist} from 0,scale=1] (up) {};
\node[label=below:{},shape=circle,position=-90:{\nodeDist} from 0,scale=1] (dwn) {};
\node[label=below:{},shape=circle,position=180:{\nodeDist} from 0,scale=1] (sx) {};
\draw[black,thick] (0) -- (dx);
\draw[black,thick] (0) -- (up);
\draw[black,thick,dashed] (0) -- (dwn);
\draw[black,thick,dashed] (0) -- (sx);
\end{tikzpicture}
}
\end{gathered}
 +
\,\, 6\,
\begin{gathered}
\scalebox{0.6}{
\begin{tikzpicture}
\node[label=below:{}, shape=circle, scale=1,draw] (0) at (0,0) [right] {};
\node[label=below:{},shape=circle,position=0:{\nodeDist} from 0,scale=1] (dx) {};
\node[label=below:{},shape=circle,position=90:{\nodeDist} from 0,scale=1] (up) {};

\node[label=below:{},shape=circle,position=-90:{\nodeDist} from 0,scale=1] (dwn) {};
\node[label=below:{},shape=circle,position=180:{\nodeDist} from 0,scale=1] (sx) {};
\draw[black,thick] (0) -- (dx);
\draw[black,thick] (0) -- (up);
\draw[black,thick,dashed] (0) -- (dwn);
\draw[black,thick,dashed] (0) -- (sx);
\end{tikzpicture}
}
\end{gathered}
+
\,\, 4\,
\begin{gathered}
\scalebox{0.6}{
\begin{tikzpicture}
\node[label=below:{}, shape=circle, scale=1,pattern=north east lines,draw] (0) at (0,0) [right] {};
\node[label=below:{},shape=circle,position=0:{\nodeDist} from 0,scale=1] (dx) {};
\node[label=below:{},shape=circle,position=90:{\nodeDist} from 0,scale=1] (up) {};
\node[label=below:{},shape=circle,position=180:{\nodeDist} from 0,scale=1] (sx) {};
\node[label=below:{},shape=circle,position=-90:{\nodeDist} from 0,scale=1] (dwn) {};
\draw[black,thick] (0) -- (dx);
\draw[black,thick] (0) -- (up);
\draw[black,thick] (0) -- (sx);
\draw[black,thick,dashed] (0) -- (dwn);
\end{tikzpicture}
}
\end{gathered}
+\,\, 4\,
\begin{gathered}
\scalebox{0.6}{
\begin{tikzpicture}
\node[label=below:{}, shape=circle, scale=1,draw] (0) at (0,0) [right] {};
\node[label=below:{},shape=circle,position=0:{\nodeDist} from 0,scale=1] (dx) {};
\node[label=below:{},shape=circle,position=90:{\nodeDist} from 0,scale=1] (up) {};
\node[label=below:{},shape=circle,position=180:{\nodeDist} from 0,scale=1] (sx) {};
\node[label=below:{},shape=circle,position=-90:{\nodeDist} from 0,scale=1] (dwn) {};
\draw[black,thick] (0) -- (dx);
\draw[black,thick] (0) -- (up);
\draw[black,thick] (0) -- (sx);
\draw[black,thick,dashed] (0) -- (dwn);
\end{tikzpicture}
}
\end{gathered}
+\,
\begin{gathered}
\scalebox{0.6}{
\begin{tikzpicture}
\node[label=below:{}, shape=circle, scale=1,pattern=north east lines,draw] (0) at (0,0) [right] {};
\node[label=below:{},shape=circle,position=0:{\nodeDist} from 0,scale=1] (dx) {};
\node[label=below:{},shape=circle,position=90:{\nodeDist} from 0,scale=1] (up) {};
\node[label=below:{},shape=circle,position=180:{\nodeDist} from 0,scale=1] (sx) {};
\node[label=below:{},shape=circle,position=-90:{\nodeDist} from 0,scale=1] (dwn) {};
\draw[black,thick] (0) -- (dx);
\draw[black,thick] (0) -- (up);
\draw[black,thick] (0) -- (sx);
\draw[black,thick] (0) -- (dwn);
\end{tikzpicture}
}
\end{gathered}
+O\left(\delta^5\right).
\end{equation}
\end{widetext}
The weights in front of each diagram are combinatorial factors counting all possible ways of attaching the external legs. By using Eqs.~\eqref{eq:dclexp} and \eqref{eq:pclexp} we obtain:
\begin{equation}
\label{eq:zeroExpansion1freeSp}
n_1=6\,\delta^2-188\,\delta^3+2385\,\delta^4+\left(\delta^5\right).
\end{equation}
At this point let us study the clusters of two free spins. There are two contributions at order $\delta^3$. The first one is 
\begin{equation}
\label{eq:zeroExp2Free1}
w_2^{(1)}=\,\,18
\begin{gathered}
\scalebox{0.6}{
\begin{tikzpicture}
\node[label=below:{}, shape=circle, scale=1,pattern=north east lines,draw] (0) at (0,0) [right] {};
\node[label=below:{},shape=circle,position=0:{\nodeDist} from 0,scale=1,pattern=north east lines,draw] (dx) {};
\node[label=below:{},shape=circle,position=180:{\nodeDist} from 0,scale=1] (sx) {};
\node[label=below:{},shape=circle,position=-90:{\nodeDist} from 0,scale=1] (dw1) {};
\node[label=below:{},shape=circle,position=90:{\nodeDist} from 0,scale=1] (up) {};
\draw[black,thick,dashed] (0) -- (dw1);
\draw[black,thick,dashed] (0) -- (sx);
\draw[black,thick] (0) -- (dx);
\draw[black,thick] (0) -- (up);
\node[label=below:{},shape=circle,position=90:{\nodeDist} from dx,scale=1] (up1) {};
\node[label=below:{},shape=circle,position=0:{\nodeDist} from dx,scale=1] (dx1) {};
\node[label=below:{},shape=circle,position=-90:{\nodeDist} from dx,scale=1] (dw2) {};
\draw[black,thick,dashed] (dx) -- (dw2);
\draw[black,thick] (dx) -- (up1);
\draw[black,thick] (dx) -- (dx1);
\end{tikzpicture}
}
\end{gathered}
=p^2\,P_{cl}^3\,D_{cl}^3=18\,\delta^3+O\left(\delta^4\right)
\end{equation}
and the other is the following
\begin{equation}
\begin{split}
\label{eq:zeroExp2Free2}
w_2^{(2)}=\,\,18\,
\begin{gathered}
\scalebox{0.6}{
\begin{tikzpicture}
\node[label=below:{}, shape=circle, scale=1,pattern=north east lines,draw] (0) at (0,0) [right] {};
\node[label=below:{},shape=circle,position=0:{\nodeDist} from 0,scale=1,draw] (dx) {};
\node[label=below:{},shape=circle,position=-90:{\nodeDist} from 0,scale=1] (dw1) {};
\node[label=below:{},shape=circle,position=180:{\nodeDist} from 0,scale=1] (sx1) {};
\node[label=below:{},shape=circle,position=90:{\nodeDist} from 0,scale=1] (up) {};
\draw[black,thick,dashed] (0) -- (dw1);
\draw[black,thick,dashed] (0) -- (sx1);
\draw[black,thick] (0) -- (dx);
\draw[black,thick] (0) -- (up);
\node[label=below:{},shape=circle,position=0:{\nodeDist} from dx,scale=1] (dx1) {};
\node[label=below:{},shape=circle,position=90:{\nodeDist} from dx,scale=1] (up2) {};
\node[label=below:{},shape=circle,position=-90:{\nodeDist} from dx,scale=1] (dw2) {};
\draw[black,thick,dashed] (dx) -- (up2);
\draw[black,thick,dashed] (dx) -- (dw2);
\draw[black,thick] (dx) -- (dx1);
\end{tikzpicture}
}
\end{gathered}
&=p\,(1-p)\,P_{cl}^4\,D_{cl}^3=\\&=18\,\delta^3+O\left(\delta^4\right).
\end{split}
\end{equation}
At order $\delta^3$ the density of the number of clusters of two spins is 
\begin{equation}
n_2=\frac{z}{2}\left(w_2^{(1)}+w_{2}^{(2)}\right)+O\left(\delta^4\right),
\end{equation}
where $z/2$ is the number of paths containing two nodes (\emph{i.e.}\ the number of edges) divided by $N$. Note that when considering clusters of more than one spin the configuration space may separate in different ergodic components. For instance in $w_2^{(1)}$ all configurations of $s_1$ and $s_2$ are visitable, while in $w_2^{(2)}$ the configuration $s_1=-1,s_2=-1$ cannot be reached from the initial condition. In the first case the two spins are statistically independent, while in the second they have a non-trivial correlation. At order $\delta^3$ we also have to consider clusters of size three:
\begin{equation}
\begin{split}
w_3^{(1)}=\,\,
\begin{gathered}
\scalebox{0.6}{
\begin{tikzpicture}
\node[label=below:{}, shape=circle, scale=1,pattern=north east lines,draw] (0) at (0,0) [right] {};
\node[label=below:{},shape=circle,position=0:{\nodeDist} from 0,scale=1,draw] (dx) {};
\node[label=below:{},shape=circle,position=90:{\nodeDist} from 0,scale=1] (up) {};
\node[label=below:{},shape=circle,position=180:{\nodeDist} from 0,scale=1] (sx) {};
\node[label=below:{},shape=circle,position=270:{\nodeDist} from 0,scale=1] (dwn) {};
\draw[black,thick] (0) -- (dx);
\draw[black,thick] (0) -- (sx);
\draw[black,thick,dashed] (0) -- (up);
\draw[black,thick,dashed] (0) -- (dwn);
\node[label=below:{},shape=circle,position=0:{\nodeDist} from dx,scale=1,pattern=north east lines,draw] (1) {};
\node[label=below:{},shape=circle,position=90:{\nodeDist} from dx,scale=1] (up1) {};
\node[label=below:{},shape=circle,position=270:{\nodeDist} from dx,scale=1] (dwn1) {};
\node[label=below:{},shape=circle,position=0:{\nodeDist} from dx,scale=1] (dx1) {};
\draw[black,thick,dashed] (dx) -- (up1);
\draw[black,thick,dashed] (dx) -- (dwn1);
\draw[black,thick] (dx) -- (dx1);
\node[label=below:{},shape=circle,position=90:{\nodeDist} from dx1,scale=1] (up2) {};
\node[label=below:{},shape=circle,position=270:{\nodeDist} from dx1,scale=1] (dwn2) {};
\node[label=below:{},shape=circle,position=0:{\nodeDist} from dx1,scale=1] (dx2) {};
\draw[black,thick,dashed] (dx1) -- (up2);
\draw[black,thick,dashed] (dx1) -- (dwn2);
\draw[black,thick] (dx1) -- (dx2);
\end{tikzpicture}
}
\end{gathered}
\,\,&=9\,p^2\,(1-p)\,P_{cl}^6\,D_{cl}^2=\\&=9\,\delta^3+O\left(\delta^4\right).
\end{split}
\end{equation}
At order $\delta^3$ the density of the number of clusters of size three is:
\begin{equation}
n_3=\frac{z(z-1)}{2}w_3^{(1)}+O\left(\delta^4\right),
\end{equation}
where $z(z-1)/2$ is the number of paths containing three nodes, divided by $N$. It is important to note that the contribution to a generic observable $O$ from a given cluster varies depending on the ergodic component, that is selected by the initialization of the spins and the external legs. Indeed the probability distribution of the spins corresponding to different ergodic components are different, implying a different contribution to the observables. 

At a generic order $n$ one should study all possible clusters $C$ of free spins in which the sum of the number of external continuous legs and spins that are up in the initial condition is equal to $n$. These contributions can be counted directly for small $n$. As we are going to see, at order $\delta^4$ one should take into account clusters that have size at most equal to five. Each cluster can be studied by following this scheme:
\begin{enumerate}
    \item choose a connected cluster $C$ of spins. Let us call $\partial C$ the set of spins that do not belong to $C$, but have a neighbor belonging to $C$ (these are the external legs in the previous graphical representations). The spins in $\partial C$ are considered blocked;
    \item consider all possible ways of initializing $n$ spins up in $C\cup \partial C$, and the other spins down, and select those that result in all spins in $C$ being free. Note that the cases in which only a subset $C'\subset C$ of spins is free can be neglected, since those are counted in the analysis of $C'$. Given an initial condition it is easy to distinguish the free spins by the following iterative procedure. Starting from the initial condition suppose to iteratively orient in the up state all the spins that are facilitated (\emph{i.e.}\ the spins that have at least $f$ neighbors in the up state) until there are no spins that can be flipped. At convergence the up spins that are facilitated coincide with the free spins. This simple procedure can be used in bootstrap percolation for finding the $k$-core \cite{iwata2009dynamics};
    \item for each initialization selected at the previous step, compute the set of visitable configurations. To check if a configuration $c_0$ is allowed, one can run the algorithm for finding the free spins, starting with the same configuration in $\partial C$ (the same external legs), but with the spins of $C$ initialized in $c_0$. A configuration $c_0$ is allowed if and only if all the spins in $C$ are free starting from $c_0$;
    \item count all the initializations that produce the same set of allowed configurations (state). We call this number $m$, the multiplicity.  
\end{enumerate}
At this point suppose to choose an order for the set of all possible configurations of $K$ spins:
\begin{equation}
\begin{split}
&c_{1}=(s_1=c_{11},s_2=c_{12},\dots,s_K=c_{1K}),\\
&\vdots\\
&c_{2^K}=(s_1=c_{2^K1},s_2=c_{2^K2},\dots,s_K=c_{2^KK}).
\end{split}
\end{equation}
Let us also define a vector $r=(r_1,\dots,r_{2^K})$, the ``reachability'', such that
\begin{equation}
r_i=
\begin{cases}
&1\quad \text{if $c_i$ is reachable}\\
&0\quad \text{otherwise}.
\end{cases}
\end{equation}
Given a cluster, an initial condition specify an ergodic component, that corresponds to a specific reachability vector. Different ergodic components produce different reachability vectors. With this notation the probability distribution $P_{r}$ of $K$ free spins, given a reachability vector $r$ is:
\begin{equation}
P_{r}(c_i)=\frac{r_i\,\prod_{j=1}^Kp^{\frac{1}{2}(1-c_{ij})}\,(1-p)^{\frac{1}{2}(1+c_{ij})}}{\sum_{k=1}^{2^K}r_k\,\prod_{j=1}^K\,p^{\frac{1}{2}(1-c_{kj})}\,(1-p)^{\frac{1}{2}(1+c_{kj})}}.
\end{equation}
The entropy $S_{r}$ is
\begin{equation}
S_{r}=-\sum_{i=1}^{2^K}P_r(c_i)\,\log{P_r(c_i)}.
\end{equation}
Once the observable of interest $O_{C,r}$ is computed for each cluster $C$ and reachabilitiy $r$ that are relevant at order $n$, the contribution to the average density $\langle O\rangle/N$ of $O$ coming from all the clusters of free spins takes the form:
\begin{equation}
\label{eq:freeSPins}
\sum_{g=1}^n\sum_{C\in\mathcal{C}_{g}}\sum_{r\in\mathcal{R}_{g}(C)}O_{C,r} m_{C,r},
\end{equation}
where $\mathcal{C}_{g}$ is the set of all clusters $C$ for which it is possible to initialize $g$ spins up in $C\cup\partial C$ in such a way as all the spins of $C$ are free. The other quantity we introduced, $\mathcal{R}_g(C)$, is the set of all reachabilities of $C$ given $g$ up spins in $C\cup\partial C$. Of course, depending on the observable, one should also take into account the contribution coming from the blocked spins (like for the self-overlap). 

Following the previous scheme one finds that at order $\delta^4$ the self-overlap (see Sec.~\ref{sec:Overlap}) is
\begin{equation}
\label{eq:expSelfOv}
q\approx 1-96\,\delta^3+700\,\delta^4,
\end{equation}
and the entropy density of the state:
\begin{multline}
\label{eq:expEntr}
s\approx 6\,\Big(1 + 6 \log{2} - \log{\delta}\Big)\,\delta^3+\\+\Big(61 - 1242 \log{2} + 810 \log{3} - 64 \log{\delta}\Big)\,\delta^4.
\end{multline}
For the computation of \eqref{eq:expSelfOv} and \eqref{eq:expEntr} one has to consider two classes of contributions. The first one comes from the cluster of one free node (see Eq.~\eqref{eq:zeroExpansion1freeSp}), and from \eqref{eq:zeroExp2Free1} and \eqref{eq:zeroExp2Free2}, in which $P_{cl}$ and $D_{cl}$ have to be expanded up to order $\delta^4$ (see Eqs.~\eqref{eq:dclexp} and \eqref{eq:pclexp}). The second class  is obtained by considering all possible connected clusters $C$ of free spins that are produced by the presence of four spins up in $C\cup\partial C$ in the initial condition. These terms are listed below, with the following order. For each $C$ we specify an order for the configurations of the spins in $C$, the set of reachabilities $\mathcal{R}(C)$, and the multiplicities $m_{C,r}$. 

\subsubsection{$5$ spins}
The only initial condition of a cluster of five free spins given four spins up is the following:
\begin{equation*}
\begin{gathered}
\scalebox{0.6}{
\begin{tikzpicture}
\node[label=below:{}, shape=circle, scale=1,pattern=north east lines,draw] (0) at (0,0) [right] {};
\node[label=below:{},shape=circle,position=0:{\nodeDist} from 0,scale=1,draw] (dx) {};
\node[label=below:{},shape=circle,position=90:{\nodeDist} from 0,scale=1] (up) {};
\node[label=below:{},shape=circle,position=180:{\nodeDist} from 0,scale=1] (sx) {};
\node[label=below:{},shape=circle,position=270:{\nodeDist} from 0,scale=1] (dwn) {};
\draw[black,thick] (0) -- (dx);
\draw[black,thick] (0) -- (sx);
\draw[black,thick,dashed] (0) -- (up);
\draw[black,thick,dashed] (0) -- (dwn);
\node[label=below:{},shape=circle,position=0:{\nodeDist} from dx,scale=1,pattern=north east lines,draw] (1) {};
\node[label=below:{},shape=circle,position=90:{\nodeDist} from dx,scale=1] (up1) {};
\node[label=below:{},shape=circle,position=270:{\nodeDist} from dx,scale=1] (dwn1) {};
\node[label=below:{},shape=circle,position=0:{\nodeDist} from dx,scale=1] (dx1) {};
\draw[black,thick,dashed] (dx) -- (up1);
\draw[black,thick,dashed] (dx) -- (dwn1);
\draw[black,thick] (dx) -- (dx1);
\node[label=below:{},shape=circle,position=90:{\nodeDist} from dx1,scale=1] (up2) {};
\node[label=below:{},shape=circle,position=270:{\nodeDist} from dx1,scale=1] (dwn2) {};
\node[label=below:{},shape=circle,position=0:{\nodeDist} from dx1,scale=1,draw] (dx2) {};
\draw[black,thick,dashed] (dx1) -- (up2);
\draw[black,thick,dashed] (dx1) -- (dwn2);
\draw[black,thick] (dx1) -- (dx2);

\node[label=below:{},shape=circle,position=90:{\nodeDist} from dx2,scale=1] (up3) {};
\node[label=below:{},shape=circle,position=270:{\nodeDist} from dx2,scale=1] (dwn3) {};
\node[label=below:{},shape=circle,position=0:{\nodeDist} from dx2,scale=1,pattern=north east lines,draw] (dx3) {};
\draw[black,thick,dashed] (dx2) -- (up3);
\draw[black,thick,dashed] (dx2) -- (dwn3);
\draw[black,thick] (dx2) -- (dx3);

\node[label=below:{},shape=circle,position=90:{\nodeDist} from dx3,scale=1] (up4) {};
\node[label=below:{},shape=circle,position=270:{\nodeDist} from dx3,scale=1] (dwn4) {};
\node[label=below:{},shape=circle,position=0:{\nodeDist} from dx3,scale=1] (dx4) {};
\draw[black,thick,dashed] (dx3) -- (up4);
\draw[black,thick,dashed] (dx3) -- (dwn4);
\draw[black,thick] (dx3) -- (dx4);

\end{tikzpicture}
}
\end{gathered}
\end{equation*}
Let us choose the following order for the configurations:
\begin{equation}
c_1=(-1,-1,-1,-1,-1),\quad c_2=(-1,-1, -1, -1, 1),
\end{equation}
\begin{equation}
c_3=(-1, -1, -1, 1, -1),\quad c_4=(-1,-1, 1,-1, -1),
\end{equation}

\begin{equation}
c_5=(-1, 1, -1,-1, -1),\quad c_6=(1, -1, -1, -1, -1),
\end{equation}
\begin{equation}
c_7=(-1, -1, -1, 1, 1),\quad c_8=(-1, -1, 1, -1, 1),
\end{equation}

\begin{equation}
c_9=(-1, 1, -1, -1, 1),\quad c_{10}=(1, -1, -1, -1, 1),
\end{equation}

\begin{equation}
c_{11}=(1, -1, -1, 1, -1),\quad c_{12}=(1, -1, 1, -1,-1),
\end{equation}

\begin{equation}
c_{13}=(1, 1, -1, -1, -1),\quad c_{14}=(-1, 1, 1, -1, -1),
\end{equation}

\begin{equation}
c_{15}=(-1,-1, 1, 1,-1),\quad c_{16}=(-1, 1, -1, 1,-1),
\end{equation}

\begin{equation}
c_{17}=(1, 1, 1, -1, -1),\quad c_{18}=(1, 1, -1, 1, -1),
\end{equation}

\begin{equation}
c_{19}=(1, -1, 1, 1, -1),\quad c_{20}=(-1, 1, 1, 1,-1),
\end{equation}

\begin{equation}
c_{21}=(-1, 1, 1, -1, 1),\quad c_{22}=(-1, 1, -1, 1, 1),
\end{equation}

\begin{equation}
c_{23}=(-1, -1, 1, 1, 1),\quad c_{24}=(1, -1, -1, 1, 1),
\end{equation}

\begin{equation}
c_{25}=(1, 1, -1, -1, 1),\quad c_{26}=(1, -1, 1, -1, 1),
\end{equation}

\begin{equation}
c_{27}=(1, 1, 1, 1, -1),\quad c_{28}=(1, 1, 1, -1, 1),
\end{equation}

\begin{equation}
c_{29}=(1, 1, -1, 1, 1),\quad c_{30}=(1, -1, 1, 1, 1),
\end{equation}

\begin{equation}
c_{31}=(-1, 1, 1, 1, 1),\quad c_{32}=(1, 1, 1, 1, 1)
\end{equation}
Here there is only one reachability that, following the previous order, is 
\begin{widetext}
\begin{equation}
r=(0, 0, 0, 0, 0, 0, 0, 0, 0, 0, 0, 0, 0, 0, 0, 1, 0, 1, 1, 1, 1, 1, 0, 0, 0, 1, 1, 1, 1, 1, 1, 1),\quad m=9.
\end{equation}
\end{widetext}
\subsubsection{Aligned $4$ spins}
There are two contributions from clusters $C$ with four spins. One is discussed here, the other in the next subsection. An example of ``aligned'' $4$-cluster at order $\delta^4$ is:
\begin{equation*}
\begin{gathered}
\scalebox{0.6}{
\begin{tikzpicture}
\node[label=below:{}, shape=circle, scale=1,pattern=north east lines,draw] (0) at (0,0) [right] {};
\node[label=below:{},shape=circle,position=0:{\nodeDist} from 0,scale=1,draw] (dx) {};
\node[label=below:{},shape=circle,position=90:{\nodeDist} from 0,scale=1] (up) {};
\node[label=below:{},shape=circle,position=180:{\nodeDist} from 0,scale=1] (sx) {};
\node[label=below:{},shape=circle,position=270:{\nodeDist} from 0,scale=1] (dwn) {};
\draw[black,thick] (0) -- (dx);
\draw[black,thick] (0) -- (sx);
\draw[black,thick,dashed] (0) -- (up);
\draw[black,thick,dashed] (0) -- (dwn);
\node[label=below:{},shape=circle,position=0:{\nodeDist} from dx,scale=1,pattern=north east lines,draw] (1) {};
\node[label=below:{},shape=circle,position=90:{\nodeDist} from dx,scale=1] (up1) {};
\node[label=below:{},shape=circle,position=270:{\nodeDist} from dx,scale=1] (dwn1) {};
\node[label=below:{},shape=circle,position=0:{\nodeDist} from dx,scale=1] (dx1) {};
\draw[black,thick,dashed] (dx) -- (up1);
\draw[black,thick,dashed] (dx) -- (dwn1);
\draw[black,thick] (dx) -- (dx1);
\node[label=below:{},shape=circle,position=90:{\nodeDist} from dx1,scale=1] (up2) {};
\node[label=below:{},shape=circle,position=270:{\nodeDist} from dx1,scale=1] (dwn2) {};
\node[label=below:{},shape=circle,position=0:{\nodeDist} from dx1,scale=1,pattern=north east lines,draw] (dx2) {};
\draw[black,thick] (dx1) -- (up2);
\draw[black,thick,dashed] (dx1) -- (dwn2);
\draw[black,thick] (dx1) -- (dx2);

\node[label=below:{},shape=circle,position=90:{\nodeDist} from dx2,scale=1] (up3) {};
\node[label=below:{},shape=circle,position=270:{\nodeDist} from dx2,scale=1] (dwn3) {};
\node[label=below:{},shape=circle,position=0:{\nodeDist} from dx2,scale=1] (dx3) {};
\draw[black,thick,dashed] (dx2) -- (up3);
\draw[black,thick,dashed] (dx2) -- (dwn3);
\draw[black,thick] (dx2) -- (dx3);
\end{tikzpicture}
}
\end{gathered}
\end{equation*}
Let us choose the following order: 
\begin{equation}
c_1=(-1,- 1,- 1,- 1),\quad c_2=(-1, -1, -1, 1),
\end{equation}
\begin{equation}
c_3=(-1, -1, 1, -1), \quad c_4=(-1, 1, -1, -1),
\end{equation}
\begin{equation}
c_5=(1,-1,-1,-1),\quad c_6=(1, 1, -1, -1),
\end{equation}
\begin{equation}
\quad c_7=(1, -1, 1, -1),\quad c_8=(1, -1, -1, 1),
\end{equation}
\begin{equation}
c_9=(-1, 1, -1, 1),\quad c_{10}=(-1,-1,1, 1),
\end{equation}
\begin{equation}
c_{11}=(-1, 1, 1, -1),\quad c_{12}=(-1, 1, 1, 1),
\end{equation}
\begin{equation}
c_{13}=(1, -1, 1, 1),\quad c_{14}=(1, 1, -1, 1),
\end{equation}
\begin{equation}
c_{15}=(1,1, 1, -1),\quad c_{16}=(1,1,1,1).
\end{equation}
In this case we have five possible reachabilities:
\begin{equation}
r_1=(0, 0, 1, 0, 0, 0, 1, 0, 1, 1, 1, 1, 1, 1, 1, 1), 
\end{equation}
\begin{equation}
r_2=(0, 0, 0, 0, 0, 0, 1, 0, 1, 0, 1, 1, 1, 1, 1, 1),
\end{equation}
\begin{equation}
r_3=(0, 0, 1, 0, 0, 0, 1, 1, 1, 1, 1, 1,1, 1, 1, 1),
\end{equation}
\begin{equation}
r_4=(0, 0, 0, 1, 0, 1, 1, 1, 1, 0, 1, 1, 1, 1, 1, 1),
\end{equation}
\begin{equation}
r_5=(0,0,0,1, 0, 1, 1, 0, 1, 0, 1, 1, 1, 1, 1, 1).
\end{equation}
The reachabilities have multiplicities:
\begin{equation}
m_1=9,\quad m_2=27,\quad m_3=18,\quad m_4=18,\quad m_5=18.
\end{equation}

\subsubsection{Not-aligned $4$ spins}
The other contribution of $4$-cluster is the ``not-aligned'' case. An example at order $\delta^4$ is:
\begin{equation}
\label{fig:4notalign}
\begin{gathered}
\scalebox{0.6}{
\begin{tikzpicture}
\node[label=below:{}, shape=circle, scale=1,pattern=north east lines,draw] (0) at (0,0) [right] {};
\node[label=below:{},shape=circle,position=0:{\nodeDist} from 0,scale=1,draw] (dx) {};
\node[label=below:{},shape=circle,position=135:{\nodeDist} from 0,scale=1] (up) {};
\node[label=below:{},shape=circle,position=180:{\nodeDist} from 0,scale=1] (sx) {};
\node[label=below:{},shape=circle,position=225:{\nodeDist} from 0,scale=1] (dwn) {};
\draw[black,thick] (0) -- (dx);
\draw[black,thick] (0) -- (sx);
\draw[black,thick,dashed] (0) -- (up);
\draw[black,thick,dashed] (0) -- (dwn);
\node[label=below:{},shape=circle,position=0:{\nodeDist} from dx,scale=1,pattern=north east lines,draw] (1) {};
\node[label=below:{},shape=circle,position=90:{\nodeDist} from dx,scale=1,pattern=north east lines,draw] (up1) {};
\node[label=below:{},shape=circle,position=90:{\nodeDist} from up1,scale=1,] (upup1) {};
\draw[black,thick] (upup1) -- (up1);
\node[label=below:{},shape=circle,position=45:{\nodeDist} from up1,scale=1,] (dxup1) {};
\draw[black,thick,dashed] (dxup1) -- (up1);
\node[label=below:{},shape=circle,position=135:{\nodeDist} from up1,scale=1,] (sxup1) {};
\draw[black,thick,dashed] (sxup1) -- (up1);
\node[label=below:{},shape=circle,position=270:{\nodeDist} from dx,scale=1] (dwn1) {};
\node[label=below:{},shape=circle,position=0:{\nodeDist} from dx,scale=1] (dx1) {};
\draw[black,thick] (dx) -- (up1);
\draw[black,thick,dashed] (dx) -- (dwn1);
\draw[black,thick] (dx) -- (dx1);
\node[label=below:{},shape=circle,position=45:{\nodeDist} from dx1,scale=1] (up2) {};
\node[label=below:{},shape=circle,position=-45:{\nodeDist} from dx1,scale=1] (dwn2) {};
\node[label=below:{},shape=circle,position=0:{\nodeDist} from dx1,scale=1] (dx2) {};
\draw[black,thick,dashed] (dx1) -- (up2);
\draw[black,thick,dashed] (dx1) -- (dwn2);
\draw[black,thick] (dx1) -- (dx2);
\end{tikzpicture}
}
\end{gathered}
\end{equation}
Using the same order for the configurations as for the aligned case, we find the following (unique) reachability:
\begin{equation}
r=(0, 1, 0, 0, 0, 1, 1, 1, 1, 1, 1, 1, 1, 1, 1, 1),\quad m=27.
\end{equation}
This contribution is particularly interesting, indeed it is the only term that at this order does not have the form of a spins chain. This implies a different factorization of the probability distribution of the free spins on this diagram. In particular in this case such probability does not contain only pairwise terms between neighboring spins, but one has also to include interactions between the farthest spins. This fact can be immediately checked. If the distribution contained only pairwise terms between neighboring spins, by conditioning on the spin represented by an empty circle (see Fig.~\ref{fig:4notalign}), their neighbors should become statistically independent. Suppose to condition on the white circle spin to point down. The three neighbors cannot be all pointing down: at least two of them should point up, otherwise there should be some blocked spins. Therefore they are not statistically independent.

\subsubsection{$3$ spins}
An example of cluster of three free spins at order $\delta^4$ is:
\begin{equation}
\begin{gathered}
\scalebox{0.6}{
\begin{tikzpicture}
\node[label=below:{}, shape=circle, scale=1,pattern=north east lines,draw] (0) at (0,0) [right] {};
\node[label=below:{},shape=circle,position=0:{\nodeDist} from 0,scale=1,draw] (dx) {};
\node[label=below:{},shape=circle,position=90:{\nodeDist} from 0,scale=1] (up) {};
\node[label=below:{},shape=circle,position=180:{\nodeDist} from 0,scale=1] (sx) {};
\node[label=below:{},shape=circle,position=270:{\nodeDist} from 0,scale=1] (dwn) {};
\draw[black,thick] (0) -- (dx);
\draw[black,thick] (0) -- (sx);
\draw[black,thick,dashed] (0) -- (up);
\draw[black,thick,dashed] (0) -- (dwn);
\node[label=below:{},shape=circle,position=0:{\nodeDist} from dx,scale=1,pattern=north east lines,draw] (1) {};
\node[label=below:{},shape=circle,position=90:{\nodeDist} from dx,scale=1] (up1) {};
\node[label=below:{},shape=circle,position=270:{\nodeDist} from dx,scale=1] (dwn1) {};
\node[label=below:{},shape=circle,position=0:{\nodeDist} from dx,scale=1] (dx1) {};
\draw[black,thick] (dx) -- (up1);
\draw[black,thick,dashed] (dx) -- (dwn1);
\draw[black,thick] (dx) -- (dx1);
\node[label=below:{},shape=circle,position=90:{\nodeDist} from dx1,scale=1] (up2) {};
\node[label=below:{},shape=circle,position=270:{\nodeDist} from dx1,scale=1] (dwn2) {};
\node[label=below:{},shape=circle,position=0:{\nodeDist} from dx1,scale=1] (dx2) {};
\draw[black,thick,dashed] (dx1) -- (up2);
\draw[black,thick,dashed] (dx1) -- (dwn2);
\draw[black,thick] (dx1) -- (dx2);
\end{tikzpicture}
}
\end{gathered}
\end{equation}
We choose the following order for the configurations:
\begin{equation}
c_1=(-1, -1, -1),\quad c_2=(1, -1, -1), 
\end{equation}
\begin{equation}
c_3=(-1, 1, -1),\quad c_4=(-1, -1, 1),
\end{equation}
\begin{equation}
c_5=(1, 1, -1),\quad c_6=(1, -1, 1),
\end{equation}
\begin{equation}
c_7=(-1,1, 1),\quad c_8=(1, 1, 1).
\end{equation}
The reachabilities are specified by the following set of vectors
\begin{equation}
r_1=(1, 1, 1, 1, 1, 1, 1, 1),\quad r_2=(0, 0, 1, 1, 1, 1, 1, 1)
\end{equation}
\begin{equation}
r_3=(0, 1, 1, 1, 1, 1, 1, 1),\quad r_4=(0, 0, 1, 0, 1, 1, 1, 1)
\end{equation}
\begin{equation}
r_5=(0, 1, 1, 0, 1, 1, 1, 1),
\end{equation}
that are associated with the multiplicities:
\begin{equation}
m_1=54,\quad m_2=18,\quad m_3=54,\quad m_4=27,\quad m_4=18.
\end{equation}

\subsubsection{$2$ spins}
An example of cluster of two free spins at order $\delta^4$ is:
\begin{equation*}
\begin{gathered}
\scalebox{0.6}{
\begin{tikzpicture}
\node[label=below:{}, shape=circle, scale=1,draw] (0) at (0,0) [right] {};
\node[label=below:{},shape=circle,position=0:{\nodeDist} from 0,scale=1,draw] (dx) {};
\node[label=below:{},shape=circle,position=-90:{\nodeDist} from 0,scale=1] (dw1) {};
\node[label=below:{},shape=circle,position=180:{\nodeDist} from 0,scale=1] (sx1) {};
\node[label=below:{},shape=circle,position=90:{\nodeDist} from 0,scale=1] (up) {};
\draw[black,thick,dashed] (0) -- (dw1);
\draw[black,thick,dashed] (0) -- (sx1);
\draw[black,thick] (0) -- (dx);
\draw[black,thick] (0) -- (up);
\node[label=below:{},shape=circle,position=0:{\nodeDist} from dx,scale=1] (dx1) {};
\node[label=below:{},shape=circle,position=90:{\nodeDist} from dx,scale=1] (up2) {};
\node[label=below:{},shape=circle,position=-90:{\nodeDist} from dx,scale=1] (dw2) {};
\draw[black,thick,dashed] (dx) -- (up2);
\draw[black,thick,dashed] (dx) -- (dw2);
\draw[black,thick] (dx) -- (dx1);
\end{tikzpicture}
}
\end{gathered}
\end{equation*}
We choose the following order for the configurations:
\begin{equation}
c_1=(-1, -1),\quad c_2=(-1, 1), 
\end{equation}
\begin{equation}
c_3=(1, -1),\quad c_4=(1, 1).
\end{equation}
The reachabilities are
\begin{equation}
r_1=(1, 1, 1, 1),\quad r_2=(0, 1, 1, 1),
\end{equation}
with multiplicities:
\begin{equation}
m_1=51,\quad m_2=9.
\end{equation}
The cluster with a single spin has already been discussed at the beginning of the section.

\section{Analytical Computation of the Spin-Glass Susceptibility}
\label{sec:AnalyticalCompSGSusc}
In this section we give the details about the computation of the spin-glass susceptibility, $\chi_{SG}$, of the FAM (see \cite{parisi2014diluted} for an analogous computation in case of the Ising SG), that we introduced in Sec.~\ref{sec:IntroSGSusc}. Let us rewrite $\chi_{SG}$ (see Eq.~\eqref{eq:defSpGlSus}) in a more convenient form by noticing that we can perform the sum with respect to an arbitrary reference spin $s_0$ thanks to the translational invariance of the disorder-averaged correlation function, and we can write  the connected correlation function $\langle s_0\,s_i\rangle_c$ as the derivative of the magnetization $m_0$ of $s_0$ w.r.t.\ an external field added \emph{after} the extraction of initial condition (see Eq.~\eqref{eq:addExtField}), and that affects therefore only the free spins. In this way we obtain:
\begin{equation}
\chi_{SG}=\mathds{E}\sum_{i}\langle s_0 s_i\rangle_{c}^2=\frac{1}{\beta^2}\mathds{E}\sum_{i}\left(\frac{\dd m_0}{\dd H_i}\right)^2.
\end{equation}
On a tree, we can define by $T_{i\rightarrow j}$ the set of nodes that can be reached starting from node $i$ without passing through node $j$. We have that:
\begin{widetext}
\begin{equation}
\label{eq:ChiSG}
\frac{1}{\beta^2}\sum_{i}\left(\frac{\dd m_0}{\dd H_i}\right)^2=\big(1-m_0^2\big)^2+\frac{4\,e^{-4\beta}}{\big(e^{-2\beta}+1-R_0\big)^4}\sum_{j\in\partial 0}\left(\frac{\dd R_0}{\dd R_{j\rightarrow 0}}\right)^2\sum_{k\in T_{j\rightarrow 0}}\left(\frac{\dd R_{j\rightarrow 0}}{\dd H_k}\right)^2,
\end{equation}
where the first term on the RHS is the contribution coming from the case with $i=0$. At this point let us define the susceptibility $\chi(\eta,\mu,R)$ on a single branch of the tree conditioned to a triplet $(\eta,\mu,R)$:
\begin{equation}
\label{eq:OnebranchAvSusc}
\chi(\eta,\mu,R)=\mathds{E}\left\{\delta_{\eta,\eta_{i\rightarrow j}}\delta_{\mu,\mu_{i\rightarrow j}}\delta(R-R_{i\rightarrow j})\sum_{k\in T_{i\rightarrow j}}\left(\frac{\dd R_{i\rightarrow j}}{\dd H_k}\right)^2\right\}.
\end{equation}
Note that $\chi(1,1,R)$ is always zero, indeed if $(\eta,\mu)_{i\rightarrow j}=(1,1)$, $s_i$ is blocked down, and since the external fields $H_i$'s are added after the extraction of the initial condition, they don't have effect on the blocked spins. Note also that $\chi(\eta,\mu,R)$ does not depend on $i\rightarrow j$ due to the average over disorder. Substituting \eqref{eq:OnebranchAvSusc} into \eqref{eq:ChiSG} it is possible to express $\chi_{SG}$ in terms of $\chi(\eta,\mu,R)$, leading to Eq.~\eqref{eq:LinearEqChiSg},
where
\begin{equation}
\Gamma_0=\int\dd R\,P_{site}(R) \left[1-\bigg(\frac{e^{-2\beta}-1+R}{e^{-2\beta}+1-R}\bigg)^2\right]^2,
\end{equation}
\begin{equation}
\mathbf{G}\chi=z\,\mathds{E}_{s_n}\sum_{\eta_1\mu_1}\,\int\dd R_1 \left[\left(\prod_{i=2}^{z}\,\sum_{\eta_i\mu_i}\int \dd P(\eta_i,\mu_i,R_i)\right)\Omega_{s_n}^{site}\big(\{(\eta_i,\mu_i)\}_i^z\big)\,W_{site}(\{R_i\}_i^z)\right]\,\chi(\eta_1,\mu_1,R_1),
\end{equation}
and
\begin{equation}
W_{site}(\{R_i\}_i^z)=\frac{4\,e^{-4\beta}}{\big(e^{-2\beta}+1-R_{site}(\{R_i\}_i^z)\big)^4}\left(\frac{\dd R_{site}(\{R_i\}_i^z)}{\dd R_1}\right)^2.
\end{equation}
As anticipated in Sec.~\ref{sec:IntroSGSusc} the susceptibility on one branch $\chi(\eta,\mu,R)$ satisfies a fixed-point equation (see Eq.~\eqref{eq:LinearEqChi}). In order to show that, observe that Eq.~\eqref{eq:OnebranchAvSusc} can be written as the sum of two contributions:
\begin{equation}
\chi=\xi_0(\eta,\mu,R)+\xi_1(\eta,\mu,R).
\end{equation}
The first one, $\xi_0(\eta,\mu,R)$, corresponds to the case $k=i$:
\begin{multline}
\label{eq:chi1}
\xi_0(\eta,\mu,R)\equiv\mathds{E}\left\{\delta_{\eta,\eta_{i\rightarrow j}}\delta_{\mu,\mu_{i\rightarrow j}}\delta(R-R_{i\rightarrow j})\left(\frac{\dd R_{i\rightarrow j}}{\dd H_i}\right)^2\right\}=\\=\mathds{E}_{s_n}\left(\prod_{i}^{c}\sum_{\eta_i\mu_i}\,\int \dd P(\eta_i,\mu_i,R_i)\right)\,\left(\frac{\dd R_{iter}\big(\{R_i\}_{i}^{c}|H\big)}{\dd H}\Big|_{H=0}\right)^2\,\delta^{(k)}\Big(\eta-T_{s_n}(\{\eta_i,\mu_i\}_{i}^{c})\Big)\times\\ \times\delta^{(k)}\Big(\mu-M_{s_n}(\{\eta_i,\mu_i\}_{i}^{c})\Big)\,\delta\Big(R-F_{s_n}(\{\eta_i,\mu_i,R_i\}_{i}^{c})\bigg),
\end{multline}
where in the case $(4,2)$ we have:
\begin{equation}
R_{iter}\big(R_1,R_2,R_3|H\big)=\frac{R_1 R_2 +R_1 R_3+R_2 R_3-3R_1 R_2 R_3}{e^{-2\beta(1-H)}+1-R_1 R_2 R_3}.
\end{equation}
The second contribution, $\xi(\eta,\mu,R)$, is given by the cases $k\neq i$. Suppose that $k\in T_{q\rightarrow i}$ for some $q\in\partial i\setminus j$. By writing:
\begin{equation}
\label{eq:obs1}
\frac{\dd R_{i\rightarrow j}}{\dd H_k}=\frac{\dd R_{i\rightarrow j}}{\dd R_{q\rightarrow i}}\frac{\dd R_{q\rightarrow i}}{\dd H_k},
\end{equation}
we find:
\begin{multline}
\label{eq:chi2}
\xi_1(\eta,\mu,R)\equiv\mathds{E}\left\{\delta_{\eta,\eta_{i\rightarrow j}}\delta_{\mu,\mu_{i\rightarrow j}}\delta(R-R_{i\rightarrow j})\sum_{k\in T_{i\rightarrow j}\setminus i}\left(\frac{\dd R_{i\rightarrow j}}{\dd H_k}\right)^2\right\}=\\=c\,\mathds{E}_{s_n}\sum_{\eta_1\mu_1}\,\int\dd R_1  \Bigg[\left(\prod_{i=2}^{c}\,\sum_{\eta_i\mu_i}\int \dd P(\eta_i,\mu_i,R_i)\right)\,W_{iter}(\{R_i\}_i^c) \,\delta^{(k)}\Big(\eta-T_{s_n}(\{\eta_i,\mu_i\}_i^c)\Big)\times\\\times\delta^{(k)}\Big(\mu-M_{s_n}(\{\eta_i,\mu_i\}_i^c)\Big)\,\delta\Big(R-F_{s_n}(\{\eta_i,\mu_i,R_i\}_i^c)\bigg)\Bigg]\,\chi(\eta_1,\mu_1,R_1),
\end{multline}
where
\begin{equation}
W_{iter}(\{R_i\}_i^c)=\left(\frac{\dd R_{iter}(\{R_i\}_i^c)}{\dd R_1}\right)^2.
\end{equation}
As anticipated in Sec.~\ref{sec:IntroSGSusc}, Eq.~\eqref{eq:chi2} takes the form of a linear operator applied to $\chi$, namely $\xi_{1}(\eta,\mu,R)=(\mathbf{F}\chi)(\eta,\mu,R)$.
\end{widetext}
\section{Numerical Estimation of the Spin-Glass Susceptibility}
\label{sec:NumericalCompSGSusc}
The SG susceptibility can be measured in numerical experiments by evolving different replicas of the same system according to different thermal histories starting from the same realisation of the disorder (that in this case is the initial condition). The overlap between such replicas allows to estimate different correlation functions between the spins. For example we have \cite{parisi2013critical}:
\begin{equation}
\label{eq:fluctOverl1}
M_1\equiv N\langle\delta q_{12}^2\rangle=\frac{1}{N}\sum_{ij}\big(\,\overline{\langle s_is_j\rangle^2}-q^2\big),
\end{equation}
\begin{equation}
\label{eq:fluctOverl2}
M_2\equiv N\langle\delta q_{12}\,\delta q_{23}\rangle=\frac{1}{N}\sum_{ij}\big(\,\overline{\langle s_is_i\rangle\langle s_i\rangle\langle s_j\rangle}-q^2\big),
\end{equation}
\begin{equation}
\label{eq:fluctOverl3}
M_3\equiv N\langle\delta q_{12}\,\delta q_{34}\rangle=\frac{1}{N}\sum_{ij}\big(\,\overline{\langle s_i\rangle^2\langle s_j\rangle^2}-q^2\big),
\end{equation}
where $q$ is the overlap and $\delta q_{\alpha\beta}$ is the shifted overlap between replicas $\alpha$ and $\beta$:
\begin{equation}
q=\frac{1}{N}\sum_i\overline{\langle s_i\rangle^2},\quad \delta q_{\alpha\beta}=\frac{1}{N}\sum_i s_i^{\alpha} s_i^{\beta}-q.
\end{equation}
Note that $\langle\delta q_{12}^2\rangle$, $\langle\delta q_{12}\delta q_{23}\rangle$ and $\langle\delta q_{12}\delta q_{34}\rangle$ can be easily computed numerically, while the quantities on the RHS of Eqs.~\eqref{eq:fluctOverl1}, \eqref{eq:fluctOverl2}, \eqref{eq:fluctOverl3} cannot. The SG susceptibility is immediately obtained by combining the previous expressions:
\begin{equation}
\chi_{SG}=\frac{1}{N}\sum_{ij}\overline{\langle s_i s_j\rangle^2_c}=M_1-2M_2+M_3. 
\end{equation}
Clearly the expressions Eqs.~\eqref{eq:fluctOverl1}, \eqref{eq:fluctOverl2}, \eqref{eq:fluctOverl3} are invariant under a permutation of the replica indexes. Therefore it is convenient to define the estimators $S_{M_1},S_{M_2}$ and $S_{M_3}$ of, respectively, $M_1$,$M_2$ and $M_3$, that are constructed by averaging over all equivalent permutations:
\begin{equation}
S_{M_1}=N\frac{2}{n(n-1)}\sum_{(\alpha\beta)}\langle\delta q_{\alpha\beta}^2\rangle
\end{equation}
\begin{equation}
S_{M_2}=N\frac{6}{n(n-1)(n-2)}\sum_{(\alpha\beta)(\beta\gamma)}\langle\delta q_{\alpha\beta}\delta q_{\beta\gamma}\rangle
\end{equation}
\begin{equation}
S_{M_3}=N\left(\frac{2}{n(n-1)}\right)^2\sum_{(\alpha\beta)(\gamma\delta)}\langle\delta q_{\alpha\beta}\delta q_{\gamma\delta}\rangle,
\end{equation}
where $n\geq 4$ is the number of replicas, and the sums are over the distinct pairs: $\sum_{(\alpha\beta)}=\sum_{\alpha<\beta}$. The previous expressions can be also rewritten in terms of unrestricted sums. Indeed by defining: 
\begin{equation}
A_1=N\sum_{\alpha\beta}\langle\delta q_{\alpha\beta}^2\rangle,\quad A_2=N\sum_{\alpha\beta\gamma}\langle\delta q_{\alpha\beta}\delta q_{\beta\gamma}\rangle,
\end{equation}
\begin{equation}
A_3=N\sum_{\alpha\beta\gamma\delta}\langle\delta q_{\alpha\beta}\delta q_{\gamma\delta}\rangle,
\end{equation}
we obtain
\begin{equation}
S_{M_1}=\frac{A_1}{n(n-1)}
\end{equation}
\begin{equation}
S_{M_2}=\frac{1}{n(n-1)(n-2)}(A_2-A_1)
\end{equation}
\begin{equation}
S_{M_3}=\frac{4}{n(n-1)(n-2)(n-3)}\left(\frac{A_1}{2}-A_2+\frac{A_3}{4}\right).
\end{equation}
Therefore for $n=4$ we obtain the following relation for $\chi_{SG}$
\begin{equation}
\chi_{SG}=\frac{1}{24} (6 A_1 - 6 A_2 + A_3),
\end{equation}
where $A_1,A_2$ and $A_3$ can be directly computed in numerical simulations. 
\clearpage
\bibliography{bibliografia.bib}
\end{document}